\documentclass{aa}  

\usepackage{graphicx}
\usepackage{lipsum}
\usepackage{txfonts}

\usepackage{float}
\usepackage{newfloat}
\usepackage{hyperref}
\hypersetup{colorlinks, linkcolor={blue}, citecolor={blue}, urlcolor={blue}}
\usepackage{afterpage}
\usepackage{amsmath}
\usepackage{subcaption}
\usepackage{siunitx}
\usepackage{booktabs}
\usepackage{multirow}
\usepackage{color}
\let\mc\multicolumn
\definecolor{raspberry}{rgb}{0.7,0.,0.3}
\definecolor{magenta}{rgb}{0.8,0,0.8}
\definecolor{purple}{rgb}{0.5,0,0.5}
\definecolor{gray}{rgb}{0.5,0.6,0.7}

\usepackage{silence}
\begin{document}

   \title{AXES-2MRS: A new all-sky catalogue of extended X-ray galaxy groups}

   \author{H. Khalil \inst{1}
          \and
          A. Finoguenov \inst{1}
        \and
        E. Tempel \inst{2,3}
        \and
        G. A. Mamon \inst{4}
          }

   \institute{Department of Physics,  University of Helsinki, Gustaf Hällströmin katu 2A, Helsinki, FI-00014, Finland 
         \and
   Tartu Observatory, University of Tartu, Observatooriumi 1, 61602 T\~oravere, Estonia
\and
Estonian Academy of Sciences, Kohtu 6, 10130 Tallinn, Estonia
             \and
Institut d’Astrophysique de Paris (UMR 7095: CNRS \& Sorbonne Université), F-75014 Paris, France             
             }

   \date{Received March 21, 2024}

 
  \abstract
   {Understanding baryonic physics at the galaxy-group level is a prerequisite for cosmological studies of the large-scale structure. One poorly understood aspect of galaxy groups is related to the properties of their hot intragroup medium. The well-studied X-ray groups have strong cool cores by which they were selected, so expanding the selection of groups is currently an important avenue in uncovering the diversity within the galaxy group population.}
   {We present a new all-sky catalogue of X-ray-detected groups (AXES-2MRS) based on the identification of large X-ray sources found in the ROSAT All-Sky Survey (RASS) with the Two Micron Redshift Survey (2MRS) Bayesian Group Catalogue. We studied the basic properties of these galaxy groups to gain insights into the effect of different group selections on the properties.}
   {In addition to X-ray luminosity from shallow survey data of RASS, we obtained detailed X-ray properties of the groups by matching the AXES-2MRS catalogue to archival X-ray observations by \textit{XMM-Newton} and complemented this by adding the published \textit{XMM-Newton} results on galaxy clusters in our catalogue. We analysed temperature and density to the lowest overdensity accessible by the data, obtaining hydrostatic mass estimates {at a uniform overdensity of 10000 times the critical, $M_{10000}$,} and comparing them to the velocity dispersions of the groups. We explored the relationship between X-ray and optical properties of AXES-2MRS groups through the $\sigma_{\mathrm v}-L_{\mathrm X}$, $\sigma_{\mathrm v}-kT$, $kT-L_{\mathrm X}$, $\sigma_{\mathrm v}-M$, and  $c_{200}-L_{\rm X}$ scaling relations.}
   {We find a large spread in the central mass {($M_{10000}$), measured by \textit{XMM-Newton},} to virial mass   {($M_{200}$), inferred by the velocity dispersion,} ratios for galaxy groups. This can either indicate that large non-thermal pressure of galaxy groups affects our X-ray mass measurements or the effect of a diversity of halo concentrations on the X-ray properties of galaxy groups. Previous catalogues based on detecting the peak of the X-ray emission preferentially sample the high-concentration groups. In contrast, our new catalogue uncovered many low-concentration groups, completely revising our understanding of X-ray groups.}
   {}


   \maketitle
%

\section{Introduction}

The hot intergalactic medium of galaxy groups plays an important role in galaxy evolution and reflects the energetics of galactic outflows and metal production. Several studies have suggested a direct link between the baryonic content of galaxy groups and the shape of matter power spectrum on spatial scales below 10 Mpc \citep{Debackere20}. 
Deep X-ray surveys have enabled significant advances in the understanding of galaxy groups, as they have discovered a large population of X-ray emitting groups down to masses below $10^{13} M_\odot$ and reaching redshifts above two on high-mass groups \citep{Gozaliasl2019}. However, there is still not a full understanding of the low-redshift ($z<0.1$) population of galaxy groups. Nevertheless, this population is a main source of knowledge regarding the detailed properties of galaxy groups.
Previous catalogues of X-ray-selected local groups and clusters of galaxies have primarily been based on identifying sources encompassing the emitting zone of $2^\prime$. This has been shown to account for only a fraction of galaxy groups that consist of relaxed groups with luminous central objects \citep{mulchaey00}. A large population of sources is lacking in these catalogues \citep{Xu18},  which has been confirmed by the dedicated consideration of galaxy group emission by \citet{Kaefer2019}. In this paper, we continue the investigation of such sources, considering the spatially resolved X-ray emission down to the lowest signal-to-noise ratio of the ROSAT All-Sky Survey (RASS) and detected on virial spatial scales. 

This paper is organised as follows: In Sect. \ref{sectiondata}, we present the construction and basic properties of the new X-ray source catalogue, we describe the 2MRS optical group catalogue used for the identification, and we introduce a representative subsample observed by \textit{XMM-Newton}. The analysis of X-ray and optical properties of X-ray-detected groups in our catalogue is provided in Sect. \ref{sectionresults}. In Sect. \ref{scalingrelations}, we present the scaling relations and include a comparison with the literature. We summarise our results in Sect. \ref{conclusion}. In this study, we adopted a flat $\Lambda$CDM cosmology with the parameters $H_{0}$ = 70 km s$^{-1}$ Mpc$^{-1}$, $\Omega_{\rm m}$ = 0.3, and $\Omega_{\Lambda} = 1 - \Omega_{\rm m}$. Unless otherwise stated, errors represent standard $1\sigma$ uncertainties (drawn at the 68$\%$ confidence level).  
For radii, masses, and concentrations, the suffixes 200, 500, and 10000 correspond to the {encompassed} densities relative to the critical density of the Universe at the redshift of the group.

\section{Data}
\label{sectiondata}
\subsection{AXES: A new catalogue of X-ray sources from ROSAT All Sky Survey}

The ROSAT all-sky survey (RASS) has been an enormous legacy for X-ray astronomy \citep[see][for a review]{Truemper93}. Of particular importance are the all-sky catalogues of sources \citep{Voges99}, which formed the base of X-ray studies in the last three decades. Exploration of the RASS data down to its faint limits has recently become an active field  (e.g. \citealp{Finoguenov20}). In the present paper, we report a new study of RASS data. We have produced a new catalogue of RASS sources, All-sky X-ray Extended Sources (AXES), found using 0.5--2.0 keV band images. For the source detection and determination of the flux extraction regions, we employ the wavelet scales of 12 and 24 arcmin after removing the emission detected on scales of 6 arcmin and below.   {The image reconstruction on scales of 12 and 24 arcmin is combined before running the source extraction, so the sources detected by both or either scale would be considered. We use ellipses to characterize the source, define the flux extraction region, and use the geometric average of major and minor axes as the source extent, $R_{\rm E}$.} This is a multiscale detection, unaffected by the emission on scales smaller than the scale of interest.   {Further details on the wavelet decomposition can be found in \cite[][\footnote{\url {https://github.com/avikhlinin/wvdecomp}}]{vikhlinin98}.}
We construct the experiment to scale with the baryonic content of galaxy groups near $R_{500}$, which is different from the point of finding spatially resolved X-ray sources on scales of the point spread function, which forms a base of the \cite{Xu18} catalogue.

By reducing the dependence of the detection on the shape of X-ray emission in the centre, the modelling of the source detection becomes feasible through currently available hydrodynamical simulations. While there's a disagreement between the behaviour of the observed gas in the central group regions and the expected gas profiles from simulations \citep{borgani04}, this mismatch disappears towards the outskirts \citep{roncarelli}.

AXES contains over six thousand unique X-ray sources, with a large concentration of sources towards the galactic centre, with many of them identified with supernova remnants. Therefore, to report on a new galaxy group, external identification of sources is required.

\begin{figure*}
\centering
   \includegraphics[width=\textwidth]{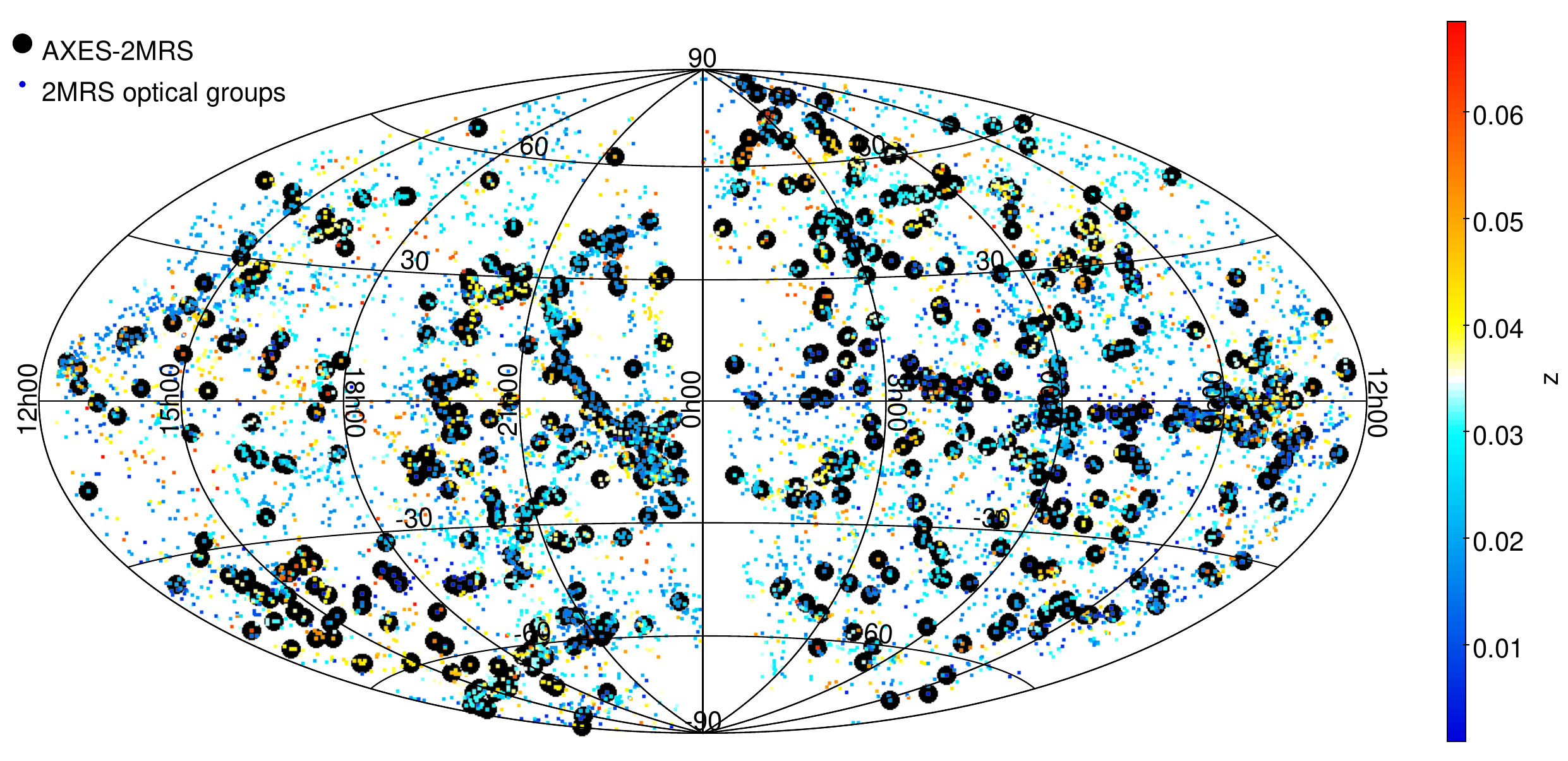}
     \caption{Distribution of AXES-2MRS X-ray sources and 2MRS optical groups represented with supergalactic coordinates on the sky. The small points are 2MRS optical groups bar coloured according to their (CMB rest frame corrected) redshift, while larger black circles denote the full AXES-2MRS X-ray catalogue of 558 groups.}
     \label{sgc}
\end{figure*}

\subsection{  {AXES-2MRS: Matching AXES with the 2MRS optical group catalogue}}
\label{section:optical}

The choice of the angular scales for our X-ray detection is designed to cover the virial radius of groups at $z<0.1$. To identify the AXES sources, we consider a group catalogue from the 2MASS spectroscopic survey \citep[2MRS]{tempel18}, selecting the groups that contain at least three spectroscopic members.
  {The \citet{tempel18} group catalogue is derived from the 2MRS dataset, described in \citet{2012ApJS..199...26H}. This dataset includes galaxies brighter than 11.75~mag in the $K_\mathrm{S}$ band and is highly complete above the Galactic plane (Galactic latitudes $|b| > 5^{\circ}$). The 2MRS galaxy sample becomes sparse at farther distances, so we limited our study to galaxies within 300~Mpc. This selection results in a sample of 42\,620 galaxies and 1\,933 groups, each with at least three spectroscopic members. For group detection, \citet{tempel18} employs a probabilistic approach, modelling the groups within a Bayesian framework using a marked point process model. In practice, this probabilistic algorithm produces groups very similar to those identified by the widely used Friends of Friends algorithm \citep{2016A&A...588A..14T}.}
The advantage of this group catalogue is that it is all-sky and extends into the galactic disc, allowing us to improve on the studies of the local dynamics. In assigning the X-ray sources to the optical group, we computed the radius of 200 kpc using the redshift of the group and used it to find an X-ray counterpart within this radius. Using the redshift of the group and the \ion{H}{i} absorption-corrected flux of the X-ray source, we then computed the source rest-frame X-ray luminosity in the 0.1--2.4 keV band, with K-corrections obtained iteratively using the $L-T$ relation.  
Our choice of using 2MRS groups down to 3 spectroscopic members is an attempt to improve the completeness of the 2MRS towards $10^{13} M_\odot$ groups expected to emit X-rays while avoiding the inclusion of a large population of low-mass groups (with masses extending down to $10^{12} M_\odot$), present in the two-member catalogue \citep[for a discussion of tracing group mass with a few members, see e.g.][]{knobel09}. Our choice of a minimum of 3 members compares well with the results of REFLEX spectroscopic identification \citep{reflex}, which made the largest contribution to the exhaustive X-ray cluster catalogue MCXC \citep{MCXC2011}. Given the importance of the low-z systems to the studies of the local dynamics, we did not cut the catalogue to the extragalactic areas, which was uniquely possible given our choice of source identification using the 2MRS catalogue. Most groups in the 2MRS catalogue have no mass estimates, given they have just a few members. Our catalogue improves this situation by providing an X-ray luminosity estimate, which is a mass proxy, and marks the massive parts of the local cosmic web.

\begin{figure}[hbt!]
\centering
   \includegraphics[width=\columnwidth]{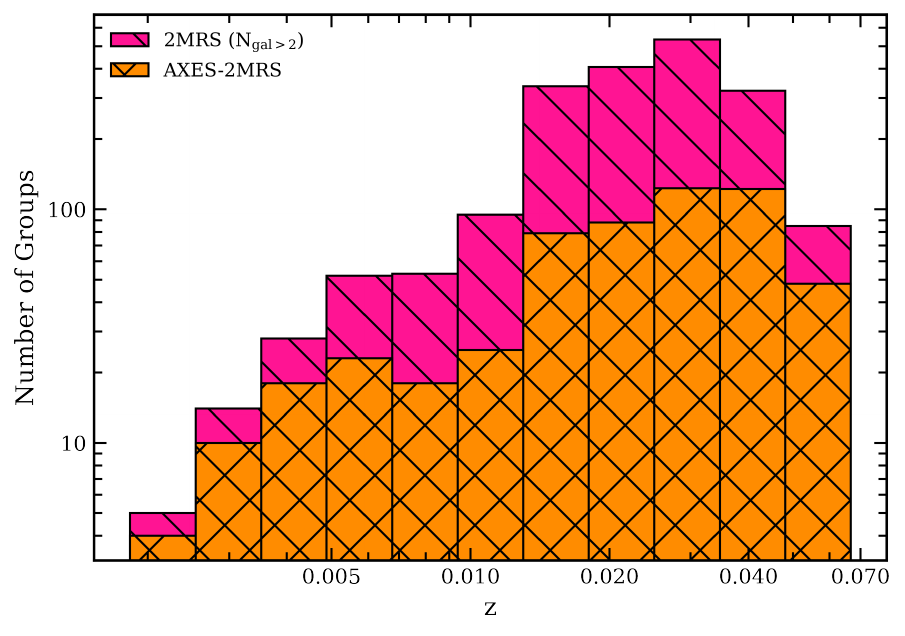}
     \caption{  {Distribution of AXES-2MRS group redshifts (orange) overlaid with that of the 2MRS optical group catalogue with at least three members (pink)}.}
     \label{dist_z}
\end{figure}

\begin{figure}[hbt!]
\centering
   \includegraphics[width=\columnwidth]{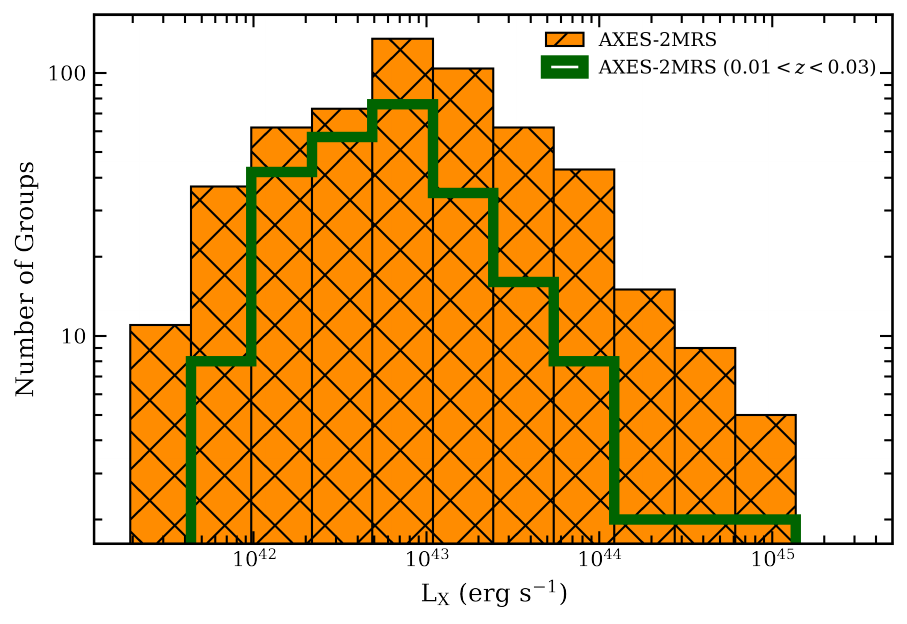}
     \caption{{Distribution of the RASS X-ray luminosities of AXES-2MRS groups (orange) overlaid with that of a subset defined by the best redshift range ($0.01 < z < 0.03$) for the L$_{\rm X}$ completeness (green).}}
     \label{dist_lx}
\end{figure}

\begin{figure}[hbt!]
\centering
   \includegraphics[width=\columnwidth]{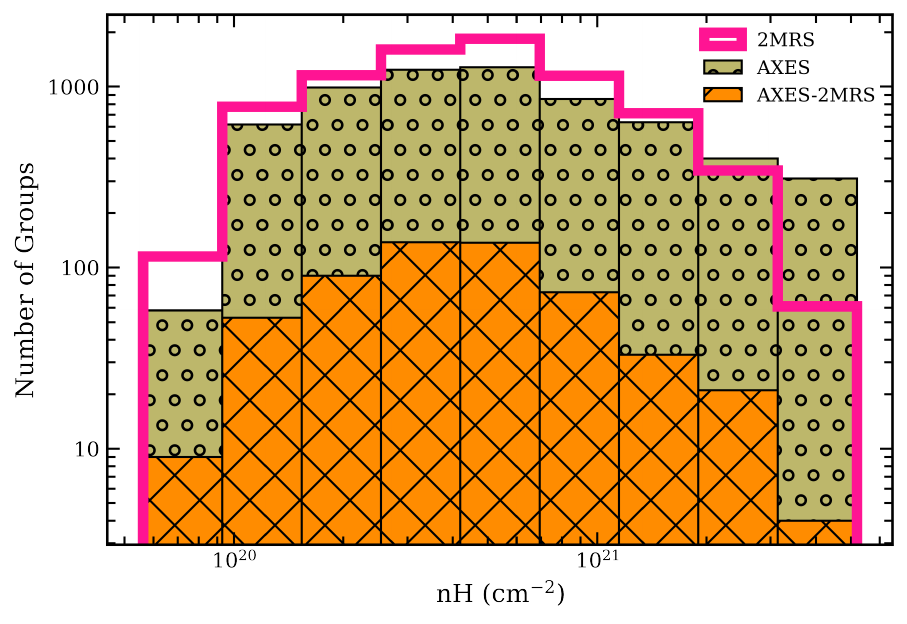}
     \caption{  {Distribution of nH of AXES-2MRS (orange) overlaid with that of the 2MRS optical catalogue (pink) and that of the AXES X-ray source catalogue (khaki)}.}
     \label{dist_nh}
\end{figure}

\begin{figure}[hbt!]
\centering
   \includegraphics[width=\columnwidth]{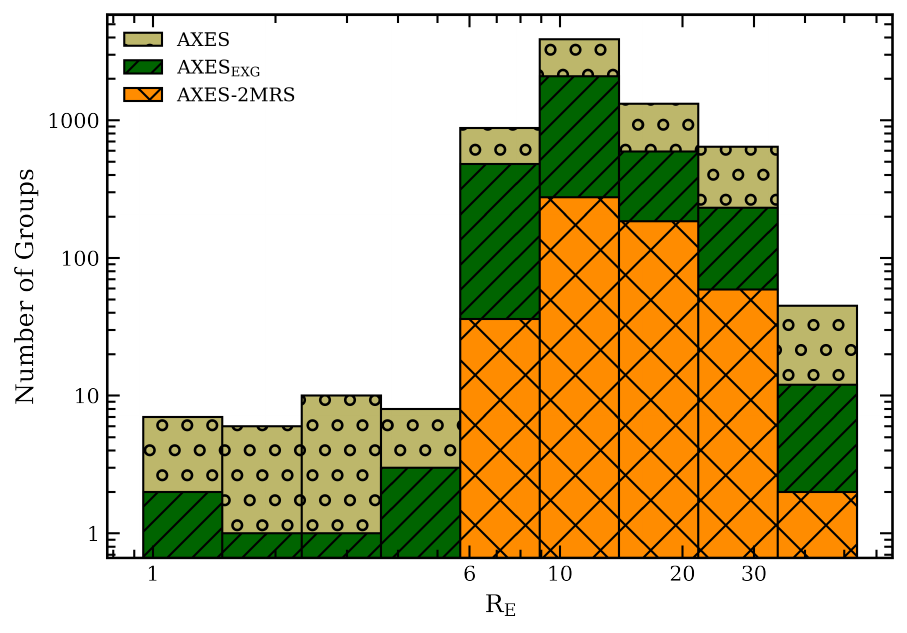}
     \caption{  {Distribution of the apparent radial extent of the X-ray emission of AXES-2MRS groups (orange) overlaid with that of the full AXES X-ray sources (khaki) and that of the extragalactic (nH $< 5 \cdot 10^{20}$ cm$^{-2}$) AXES sources (green).}}
     \label{dist_RE}
\end{figure}

\begin{figure}[hbt!]
\centering
   \includegraphics[width=\columnwidth]{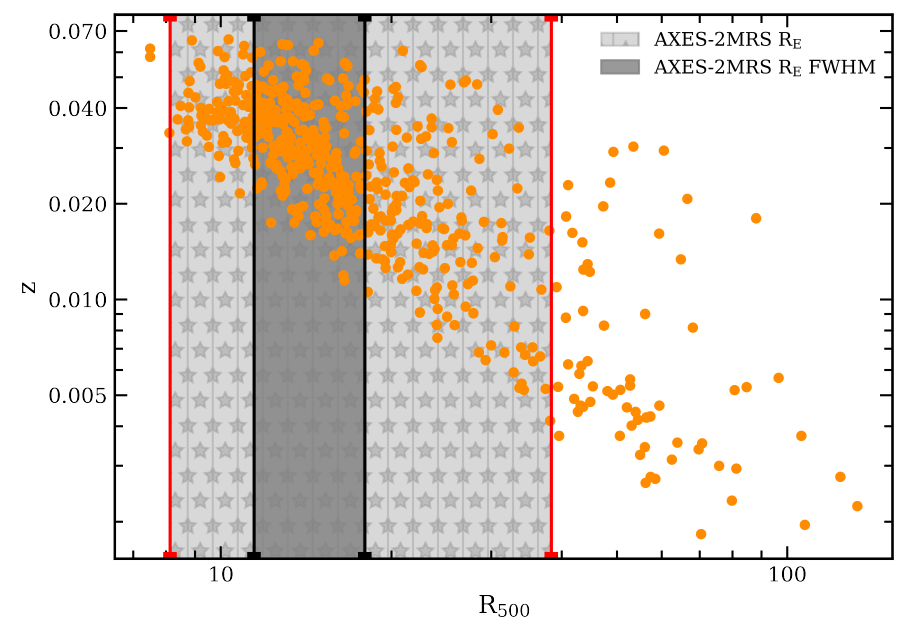}
     \caption{  {Redshift--$R_{500}$ relation for the AXES-2MRS groups. Vertical lines show the range of the X-ray emission, $R_{\rm E}$ (region defined with red lines) and its FWHM (region defined with black lines, see Fig. \ref{dist_RE}).}}
     \label{zr500}
\end{figure}

  {With 558 groups, AXES-2MRS has a high level of purity of 97\% given the small number of sources and the high fraction of matches.} In Fig.\ref{sgc} we show the sky distribution of the groups in AXES-2MRS using a supergalactic coordinate system. We use the symbol's colour to illustrate the group's redshift. 
X-ray sources identified with the groups are marked with large black-filled circles.   {Figure \ref{dist_z} shows the redshift distribution of AXES-2MRS and that of the 2MRS optical group catalogue with at least 3 spectroscopic members. Without selection effects, the number of sources should just increase with redshift, as the volume increases. We see that it is the parent catalogue of optical groups that stops increasing with redshift first at a $z\sim0.03$ and then the X-ray counterparts. We also notice that the fraction of X-ray emitting groups is high at $z<0.005$ and then levels off at a typical value of 25\% observed in deep surveys \citep{knobel09}, and improves as $z>0.03$ due to optical catalogue sampling more massive systems. 
In Fig. \ref{dist_lx}, we show the RASS X-ray luminosity distribution of AXES-2MRS groups. We see that the catalogue is incomplete at $L_{\rm X} < 10^{43}$ erg s$^{-1}$, and to remove the optical incompleteness from the consideration, we also show a histogram of X-ray luminosity at $0.01<z<0.03$, which reveals the incompleteness at $L_{\rm X} < 4 \times 10^{42}$ erg s$^{-1}$, which is due to the depth of RASS data. Figure \ref{dist_nh} shows the distribution of nH of AXES, 2MRS, and AXES-2MRS. We see that the fraction of optical groups identified with X-ray does not change much with nH,  {the column density of neutral hydrogen}. Still, the fraction of unidentified X-ray sources increases towards the large nH, which we ascribe to a larger fraction of galactic sources. Figure \ref{dist_RE} shows the distribution of the radial extend ($R_{\rm E}$) of the X-ray emission of AXES-2MRS compared to that of AXES X-ray sources and the extragalactic subset of AXES sources defined at nH $< 5 \times 10^{20}$ cm$^{-2}$. We see the onset of wavelet source filtering, suppressing the number of sources with an extent below $12^{\prime}$.  The few sources with $R_{\rm E} < 6^\prime$ are artefacts of source extraction and have no optical counterparts. Galactic sources prevail at spatial scales exceeding half a degree. Figure \ref{zr500} shows the relation between the redshift and $R_{500}$ of AXES-2MRS groups. It also shows the extent of the X-ray detection scale. We see that $R_{500}$ increases with decreasing redshift, and the scales of X-ray detections are appropriate for detecting $R_{500}$ except at $z<0.01$.} 

\begin{figure}[hbt!]
\centering
   \includegraphics[width=\columnwidth]{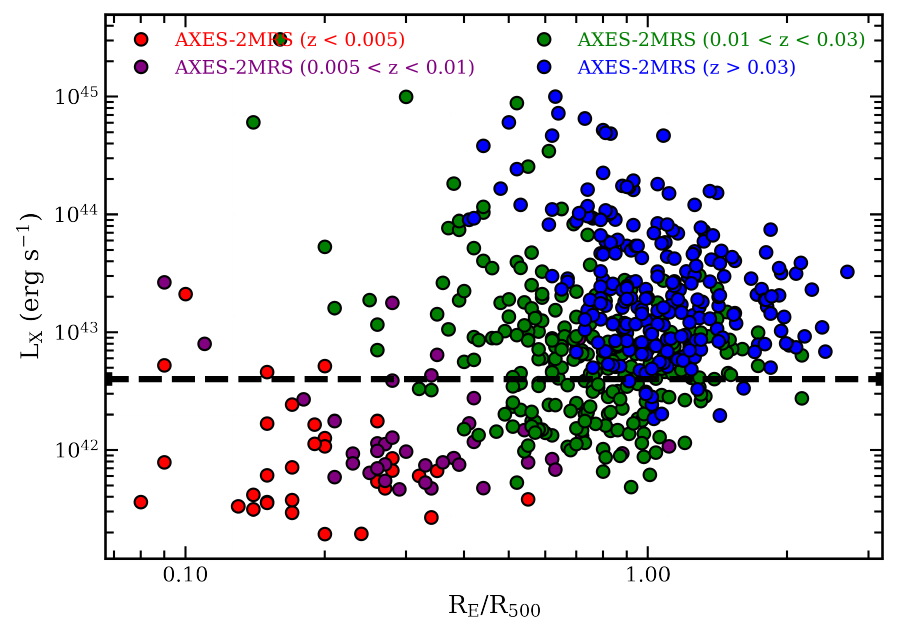}
     \caption{  {Ratio between the detected extent of the X-ray emission $R_{\rm E}$ and $R_{500}$ of AXES-2MRS groups versus the RASS X-ray luminosity split into four redshift bins:  $z < 0.005$ (red), $0.005< z<0.01$ (\textit{purple}), $0.01 < z < 0.03$ (green), and  $z > 0.03$ (blue). Horizontal dashed line represents the $5 \times 10^{42}$ erg s$^{-1}$ AXES-2MRS completeness limit.}}
     \label{re500lx}
\end{figure}

  {In Fig. \ref{re500lx} we test whether the ratio of the observed extent to the group size changes. We see that most AXES-2MRS groups have $R_{\rm E}/$$R_{500}$ around 1. The previously noticed onset of incompleteness at $L_{\rm X} < 10^{43}$  erg s$^{-1}$ is also associated with several effects. At $z>0.03$, there is a clear lack of detections, which we associate with large masses of optical groups of 2MRS 3+ member catalogue. At $z<0.01$, we see that the detection changes towards the core of the emission, so a comparison is not very meaningful. $z<0.01$ sources occupy a larger area in the sky compared to what we could trace with our fixed angular scales of the detection. The best redshift range for the $L_{\rm X}$ completeness is therefore $0.01<z<0.03$, where indeed the completeness is slightly better, reaching $4\times10^{42}$ erg s$^{-1}$. We report an increase in the size of source detection to twice the $R_{500}$ radius for sources with $L_{\rm X} < 5 \times 10^{43}$ erg s$^{-1}$, which potentially manifests the effect of elevating the hot baryons beyond the virial scales, predicted by the AGN feedback models (Mark Voit, private communication 2024). In those models, the reduced baryon fraction of galaxy groups is explained by a large extent of the X-ray emitting gas, so observation of this effect serves as clear evidence in favour of these models.}

\begin{figure}
    \centering
    \includegraphics[width=0.45\textwidth]{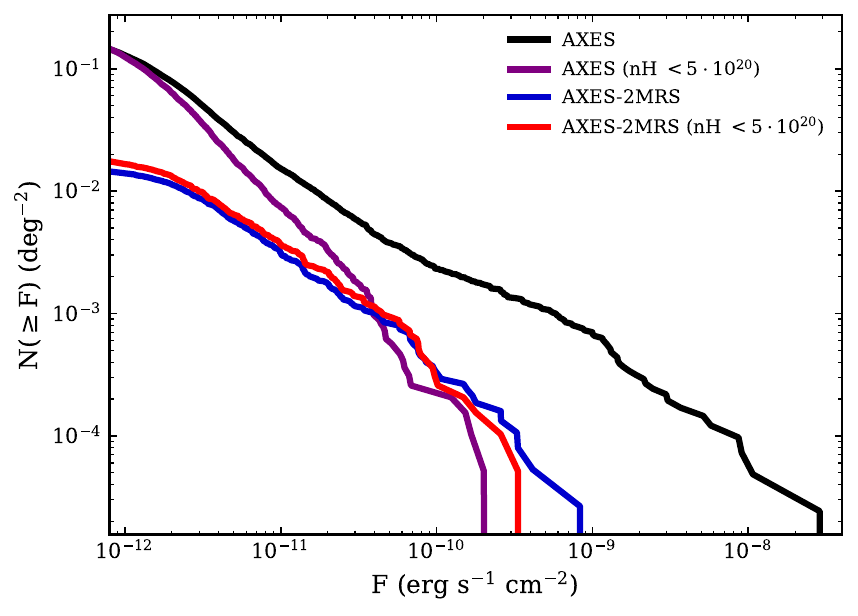}
    
    {\includegraphics[width=0.45\textwidth]{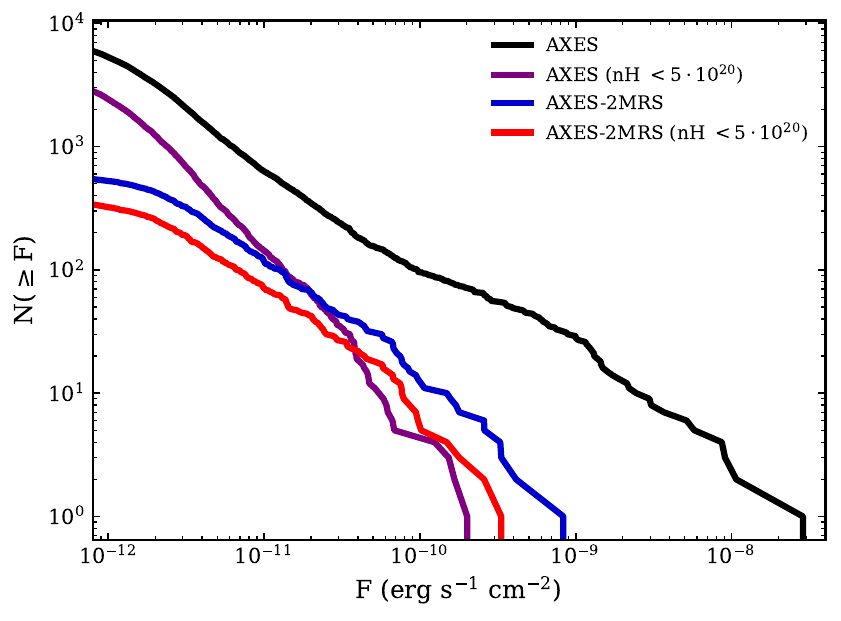}}
    
     \caption{{Sky density of AXES X-ray sources and AXES-2MRS groups.} \textit{Top panel:} Sky-density of sources as a function of X-ray flux in the 0.5--2.0 keV band ($\log N(>S)-\log S$) for AXES X-ray sources (black), and AXES-2MRS groups (blue). AXES X-ray sources in the extragalactic region (nH $< 5\times 10^{20}$ cm$^{-2}$) are shown in purple, while extragalactic AXES-2MRS groups are shown in red. The flux of the AXES-2MRS sources has been extrapolated to $R_{500}$, while for AXES we show the measured flux.   {\textit{Bottom panel:} Cumulative distribution for the X-ray flux in the 0.5--2.0 keV band. Other details are the same as the top panel.}}
     \label{flux_cdf}
\end{figure}

\begin{figure}[hbt!]
\centering
   \includegraphics[width=\columnwidth]{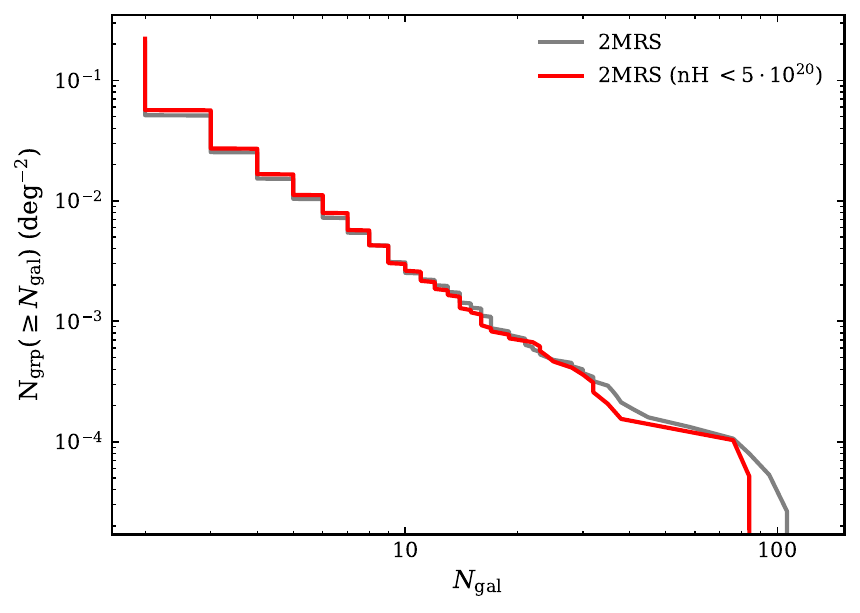}
     \caption{  {Cumulative distribution for the number of member galaxies per sky area for the 2MRS group survey (grey) and its extragalactic (nH $< 5 \cdot 10^{20}$) subset (red).}}
     \label{2mrs_ngrp}
\end{figure}

In {the top panel of} Fig. \ref{flux_cdf}, we show the normalised cumulative number count of AXES sources and AXES-2MRS groups (see Sect. \ref{section:optical}) as a function of the flux in the 0.5--2.0 keV band ($\log N-\log S$). {While we show the unnormalised version in the bottom panel for comparison.} We do not attempt to restore the original $\log N-\log S$, but rather to look for indications of the catalogue completeness.  On the adopted spatial scales the detection is background limited and completeness as a function of flux indicates a completeness limit of $2\times 10^{-12}$ ergs s$^{-1}$  cm$^{-2}$, which is characteristic of the extragalactic areas. The 2MRS survey stops where X-ray sensitivity drops by a factor of two due to foreground absorption. When comparing the $\log N-\log S$ distribution inside and outside the zone of avoidance, we see an excess of the bright sources in the zone of avoidance, which we attribute to additional sources coming from the galactic plane. 
  {We note that AXES-2MRS sources (blue curve in the top panel of Fig. \ref{flux_cdf}) and their extragalactic subset (red curve) are offset in the faint end. To explore this, we show the cumulative distribution of the number of member galaxies normalised by the sky area for the 2MRS group survey and its extragalactic subset in Fig. \ref{2mrs_ngrp}. We show that the completeness of the full 2MRS optical survey is similar to its extragalactic subset, with a slight excess of rich systems, characterising the contribution of local large-scale structure. Thus, we conclude that the small mismatch of AXES-2MRS curves in the faint end of  {the top panel of} Fig. \ref{flux_cdf} is due to the dependence of the X-ray completeness on nH and that the RASS data is deeper in the area of low nH.}

 In the application of X-ray studies to the Galactic areas, it is important to acknowledge technological differences in X-ray detectors. ROSAT PSPC detector, used in RASS, is based on the photon interaction with gas, which is not sensitive to stellar light, the latter being a serious issue for X-ray CCDs. On \textit{XMM-Newton}, this problem is addressed by selecting an appropriate filter for the observation. However, this cannot be done for eROSITA, resulting in an additional source of contamination, absent in RASS. Thus, the AXES catalogue provides a reliable list of sources, against which to compare CCD detections on comparable spatial and flux scales.

\begin{figure}[hbt!]
\centering
   \includegraphics[width=\columnwidth]{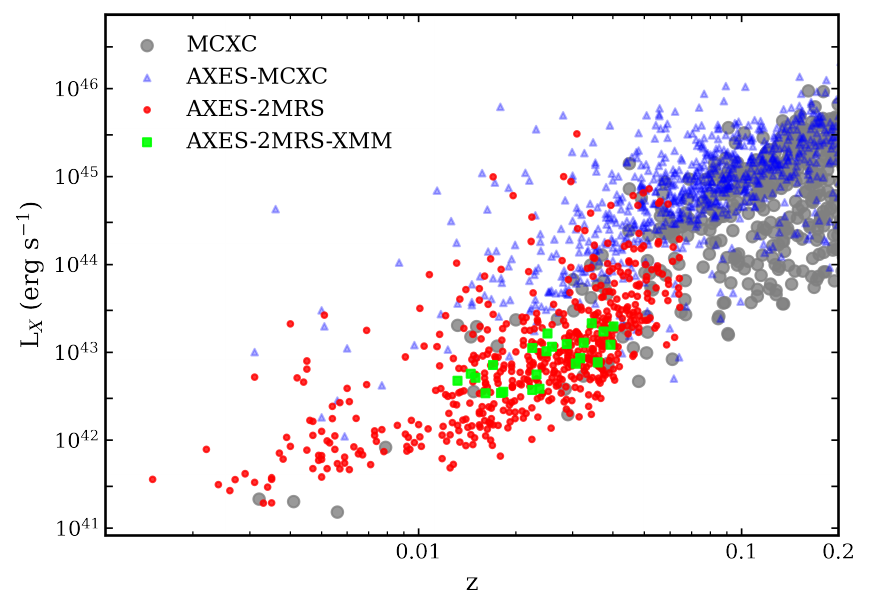}
     \caption{Full AXES-2MRS catalogue (558 groups, \textit{red} circles) compared to overlapping systems between MCXC and AXES 
     (blue triangles) and unique MCXC systems (grey circles) in the $L_{\rm X}-z$ plane. The \textit{XMM-Newton}-observed subset with at least eight members selected from AXES-2MRS is shown with neon green squares.
     }
     \label{lxz}
\end{figure}

To demonstrate the state of our new catalogue relative to the literature, we compare it with the MCXC in the $L_{\rm X}-z$ plane in Fig. \ref{lxz}. The MCXC consists of 1743 unique cluster detections representing a compilation of different catalogues based on RASS data, and ROSAT pointed data.   
Using MCXC mass estimates we find 905 AXES sources inside $R_{200}$ of 838 unique clusters.  In Fig.\ref{lxz} we plot the full MCXC catalogue, zooming on the low-redshift ($z<0.2$) subspace, its overlap with AXES sources, and present the AXES-2MRS catalogue. MCXC, being a literature compilation catalogue, reveals clear signatures of different completeness below and above $10^{44}$ ergs s$^{-1}$. Nevertheless, for our purpose of illustrating the limitations imposed by using 2MRS as follow-up data, it is sufficient. We note that for the overlapping systems in AXES and MCXC (blue triangles in Fig. \ref{lxz}), we quote $L_{\rm X}$ from the MCXC catalogue. AXES-2MRS sources are limited to $z<0.08$ with incompleteness signatures appearing at $z>0.04$, which match the expectations on the completeness of the 2MRS catalogue. At $z>0.1$, the extent of X-ray emission reaches AXES detection scales only for the most massive clusters, so the selection effects become noticeable close to a redshift of 0.2.  At $z<0.04$ AXES-2MRS luminosities are located well on the extrapolation of the MCXC trends towards lower redshift, while MCXC itself does not have many systems in the same redshift range.  We also cross-matched AXES-2MRS with the newly released X-ray-selected extended galaxy cluster catalogue (RXGCC, \citealt{RXGCC}), which is based on RASS and contains 944 systems. Within a 5$'$ radius (positional uncertainty of large X-ray sources at lowest detection significance), we identified 162 overlapping systems, out of a total number of 558 AXES-2MRS groups.
To further illustrate the role of AXES-2MRS in enhancing the completeness of X-ray group catalogues, we compared our catalogue to the extensive eROSITA-based catalogue of galaxy clusters and groups \cite[eRASS1]{erass1_new}. With over 12,200 systems, eRASS1 covers a total of 13,116 deg${^2}$ in the western Galactic hemisphere of the sky and has a redshift range of $0.003 < z < 1.32$. Despite the majority (68$\%$) of eRASS1 systems being new identifications with no counterparts in the literature, we could match only 73 overlapping groups with AXES-2MRS within a 5$'$ radius and a redshift tolerance of 0.01. The published eRASS1 catalogues reach much fainter fluxes, but the source detection is limited to $3^{\prime}$, which is not optimal for low-redshift groups \citep{Kaefer2019}. In addition, the red sequence identification, employed in the identification of eRASS1 sources, is incomplete for groups \citep{rykoff14}.

\begin{figure*}
\subcaptionbox{$0^{\circ} < \text{Dec} \leq 90^{\circ}$}{
    \centering
    \begin{subfigure}[b]{0.5\textwidth}
            \hspace{0.1in}
            \includegraphics[height=0.42\textheight]{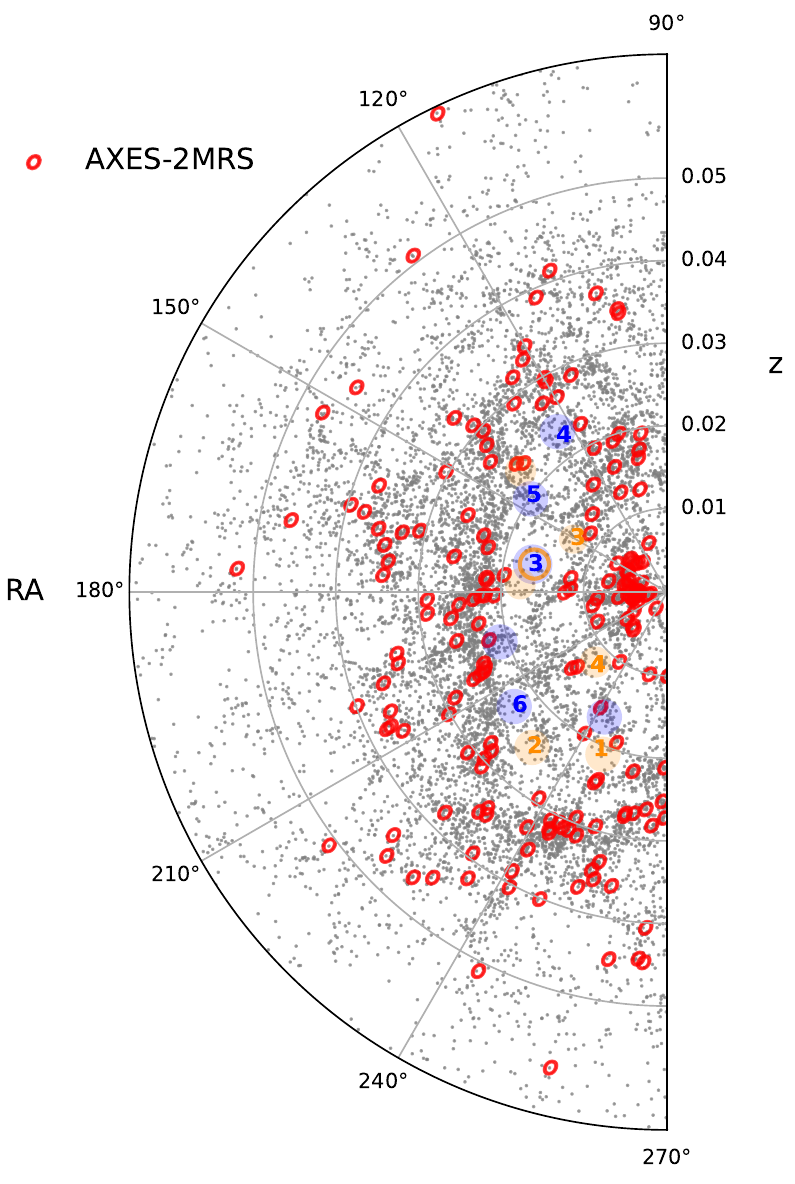}
    \end{subfigure}
    \begin{subfigure}[b]{0.5\textwidth}
            \hspace{-0.6in}
            \includegraphics[height=0.42\textheight]{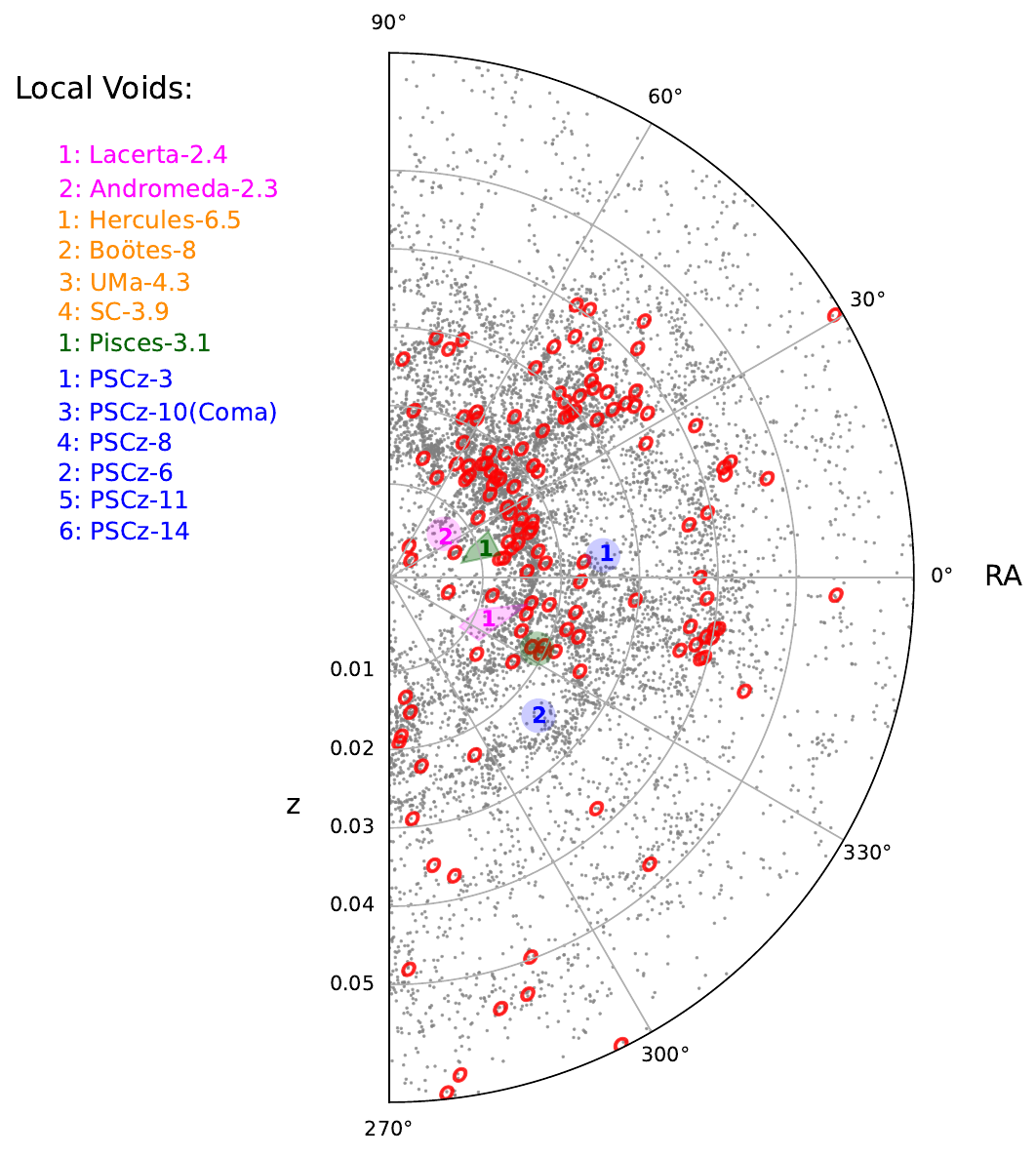}        
    \end{subfigure}
    }
\subcaptionbox{$-90^{\circ} \leq \text{Dec} \leq 0^{\circ}$}{   
    \begin{subfigure}[b]{0.5\textwidth}
            \hspace{0.1in}
            \includegraphics[height=0.42\textheight]{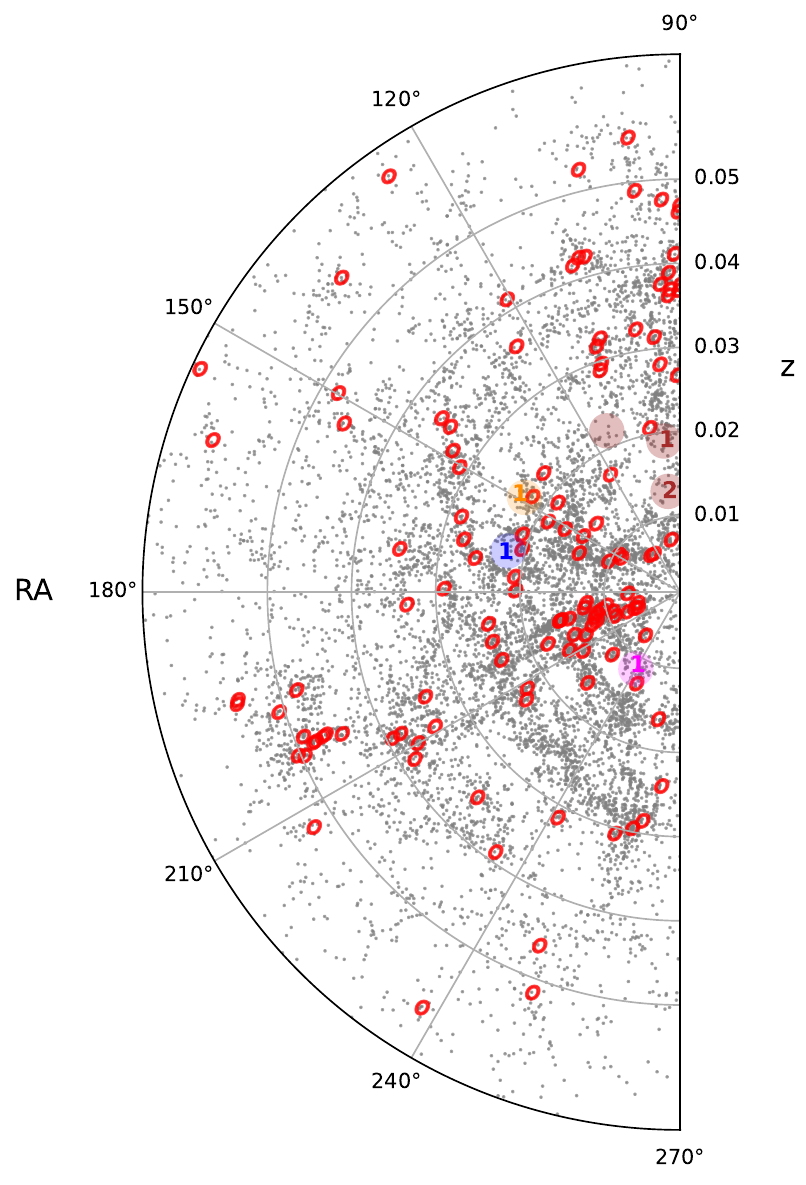}
    \end{subfigure}
    \begin{subfigure}[b]{0.5\textwidth}
            \hspace{-0.6in}
            \includegraphics[height=0.42\textheight]{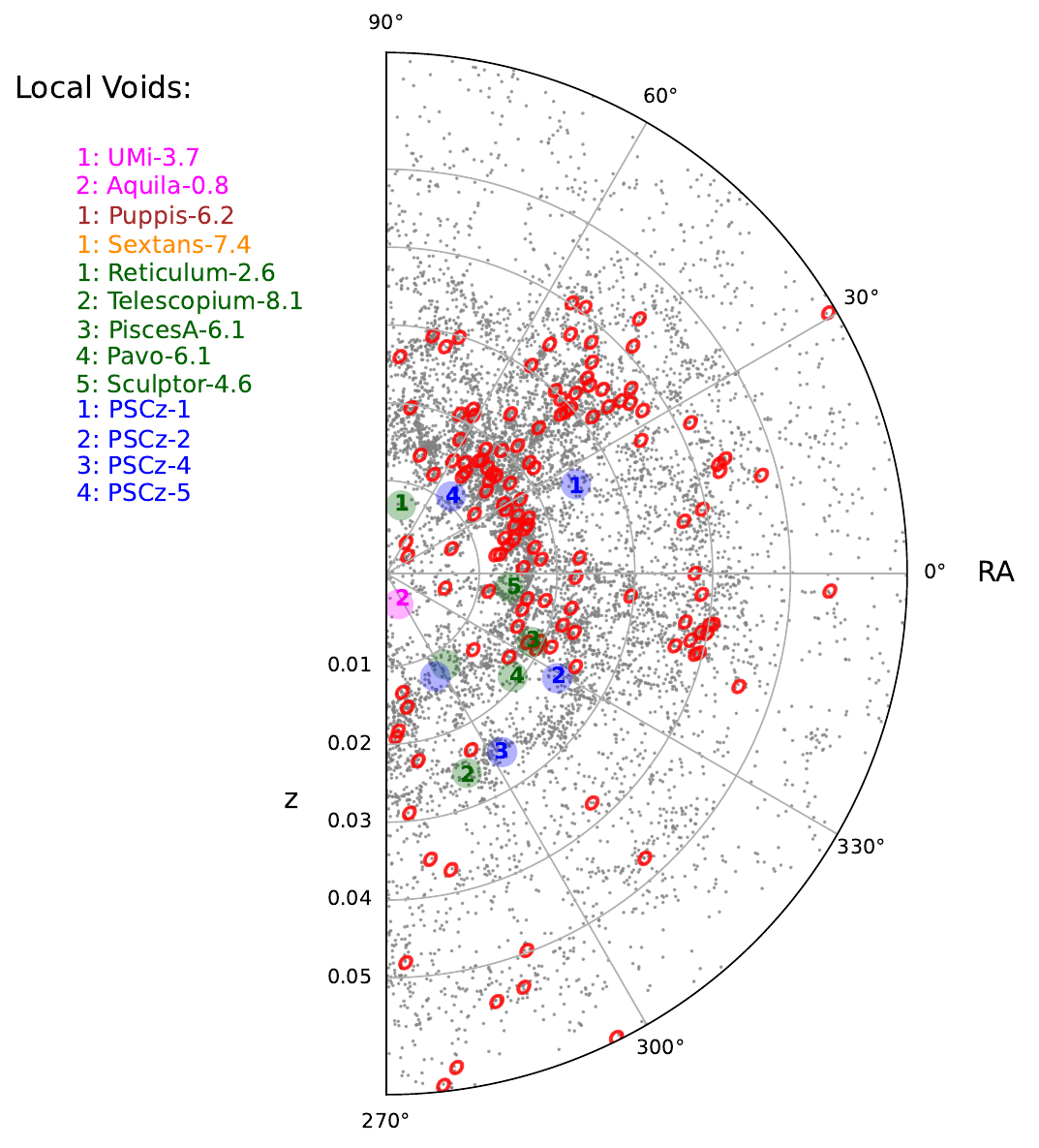}        
    \end{subfigure}
}
    \caption{Projection of the three-dimensional distribution of local cosmic structure onto the RA-redshift plane, illustrating the correspondence of the AXES-2MRS groups (red ellipses) to large-scale structure traced by the 2MRS galaxies (grey dots), with the location of the largest local voids marked. The coloured regions denote void centres, but their sizes do not reflect void dimensions. Pink regions and labels indicate density minima in the Local Void. Orange is the Hercules Void, while green and brown represent the Sculptor and Eridanus Voids, respectively. The blue colour denotes the PSCz-IRAS survey data \citep{voids}, while the other voids are based on \citet{cosmicflows}. Empty circles denote significantly smaller voids. (a) Northern hemisphere. Lacerta-2.4 and Pisces-3.1 are two prominent voids with their projected orientations traced by the shapes. (b) Southern hemisphere.}
    \label{lss}
\end{figure*}

The all-sky distribution of the identified AXES-2MRS  X-ray groups overlaid with the galaxy compilation used to create the 2MRS optical catalogue is shown in Fig. \ref{lss}. The figure is split about the equator into two declination regions: the top panel shows the northern [$0^{\circ} < \text{Dec} \leq 90^{\circ}$] range with 316 AXES-2MRS groups and 21484 galaxies. While the bottom panel shows the southern [$-90^{\circ} \leq \text{Dec} \leq 0^{\circ}$] range with 242 AXES-2MRS groups and 20727 galaxies. The maps show the location of the largest local structures in the Universe, on which 2MRS-AXES sources mark the most massive virialised systems. In addition, we mark the locations of the local underdensity regions (voids) taken from \citet{voids} and \citet{cosmicflows}. The former work detected voids in the PCSz redshift survey \citep{pscz}, while the latter modelled the morphology of the local voids using peculiar velocities.
 
\subsection{\textit{XMM-Newton} data}
\label{sectionxmm}

For a subset of AXES-2MRS groups, we obtained a much more detailed picture of the group emission using \textit{XMM-Newton} observations.
We have searched the archival \textit{XMM-Newton} data on the sample, limiting the study to the groups having at least 8 spectroscopic members, retained after cleaning the membership using {\it Clean} \citep{Mamon2013}. 
  {We constrain our analysis to 8 members because we want to probe the relation between the X-ray and optical properties of AXES-2MRS (see Sect. \ref{scalingrelations}) through the \textit{XMM-Newton} subsample and to demonstrate a detailed match of X-ray source to the optical counterpart. This comparison sample is extended in  \citet{xgap}, which presents a complete follow-up on AXES groups with 10 or more spectroscopic members based on the SDSS \citep{axes-sdss}.}
 Our \textit{XMM-Newton} data reduction pipeline is described in \citet{finoguenov07}. We used the XMMSAS version 21.0.0. 
In the imaging analysis, we used the 0.5--2 keV band, combining EMOS and EPN data after removing the energy intervals strongly affected by the instrument lines, as described in \citet{finoguenov07}. Point source detection and removal were performed following \cite{finoguenov10}. We apply similar wavelet decomposition as for the RASS data but for the extended emission on scales from half an arcminute to 4 arcmin. These scales provide insight into the central part of the object, unresolved by RASS. Larger scales are analysed using a symmetrical beta model. 
In cases where several optical groups are present, the availability of \textit{XMM-Newton} data helped in selecting the correct counterpart of the emission, and often we found more than one extended X-ray source. Spectroscopic group membership for galaxies is also complicated in those cases, and the catalogues list the probability of being a member of several adjacent groups. In these cases, we considered all the galaxies associated with the main optical counterpart with a probability above 10\%.
Looking at the target selection in the archival data, we see that some studies are follow-ups of radio sources, and some are follow-ups of X-ray sources. The optically driven survey with X-ray follow-up, CLOGS \citep{clogs}, which is analogous to our approach here, occupies a smaller redshift range compared to our data.

After screening the data to eliminate failed observations and   {selecting the ones with the best S/N in case several observations were available for the same field of view (FOV),}
we retained 25 distinct and usable \textit{XMM-Newton} observations, detailed in Table \ref{table: Optical Results} and Table \ref{table: Observations}. The uniquely identified  X-ray groups are shown in Fig.~\ref{zoo1}, where we compare the wavelet-filtered X-ray images cleaned from background and point sources with the location of group galaxies. The spatial scales shown in Fig.\ref{zoo1} range from 0.5 to 8 arcmin. In most cases, X-ray emission can be unambiguously identified with a single galaxy group. 

We detected a merging behaviour in three systems   {with AXES-2MRS Group IDs (see Table. \ref{tab:catalogue_columns}): 361, 5089, 6407, superscripted "M" in Table \ref{table: Observations})}, where the X-ray emission could be linked to more than one galaxy group (see Fig. \ref{zoo1}), while positional and velocity difference between the optical groups is small. We established a correspondence of X-ray emission to optical groups in these merging systems based on the dominant representation of optical sources near the centre of the X-ray emission regions. Nonetheless, in one case (Group ID: 5089), an extreme merging behaviour (with Group ID: 5084) was observed in which galaxies from the two optical sources were tightly packed in the centre. We selected the optical system 5089 because it had the closest galaxy member to the core of X-ray emission. This system was omitted from the scaling relations analysis due to its notably non-relaxed behaviour, significantly surpassing the other systems in the \textit{XMM-Newton} subsample in terms of dynamical complexity. Furthermore, four systems were identified as over-split (superscripted "OS" in Table \ref{table: Observations}), having double X-ray emission components. These double X-ray component groups were examined separately, and a single component was selected as a representative for each system determined by the X-ray peak best associated with the optical group. The coordinates of those X-ray peaks are listed in Table \ref{table: Optical Results}.

\section{Results}
\label{sectionresults}
  {Next, we study the optical and X-ray properties of AXES-2MRS groups and its \textit{XMM-Newton} subsample (AXES-2MRS-XMM). Section \ref{xraytemp} discusses the spectral modelling and gas temperature calculation for AXES-2MRS-XMM. We perform the surface brightness profile analysis and the hydrostatic mass estimates for AXES-2MRS-XMM in Sects. \ref{sectionsb} and \ref{mass}, respectively. In Sects. \ref{sec:velocity_dispersion} and \ref{substructure}, we present the results of our study of line-of-sight optical velocity dispersion and the velocity substructure split,  respectively. The dark matter halo concentration calculation for AXES-2MRS-XMM is shown in Sect. \ref{concentration}.}

\subsection{X-ray temperatures}
\label{xraytemp}

For the spectral extraction, we applied the \texttt{evselect} task in XMMSAS to filter out bright pixels and hot columns (\texttt{FLAG == 0}). We selected only single and double patterns (\texttt{PATTERN $<=$ 4}) for the EPIC pn camera. Source and background spectra were created from the same FOV using the same criteria. Redistribution matrix files (rmfs) and ancillary files (arfs) were generated using XMMSAS's \texttt{rmfgen} and \texttt{arfgen} tasks, respectively. Point sources identified in detector images were visually checked and excluded from the event files.
To ensure Gaussian statistics in both background and source spectra for the $\chi^{2}$ minimisation used in temperature modelling, a channel binning scheme was applied using the \texttt{grppha} task from HEASARC’s \texttt{FTOOLS}\footnote{\url{https://heasarc.gsfc.nasa.gov/ftools/}} package. Background spectra were re-binned to achieve at least 30, 60, or 200 counts per bin, depending on the source brightness and the observation's S/N.

Intragroup medium (IGrM) X-ray temperatures were estimated based on fitting the spectra with an absorbed APEC thermal plasma model \citep{apec} using the redshift of each group ($z_{\rm med}$ in Table \ref{table: Observations}), and allowing the normalisation and metal abundances to vary. The Galactic absorption component was fixed using the emission centres coordinates and the online HEASARC's web tool \texttt{nH}\footnote{\url{https://heasarc.gsfc.nasa.gov/cgi-bin/Tools/w3nh/w3nh.pl}} which is based on the HI4PI Survey \citep{hi4pi}.   {We employed the ellipsoidal quadratic mean radius to associate a radius with the temperature extraction region:
\begin{eqnarray}
  R_{kT} = \sqrt{\frac{a_{\rm spec}^{2}+b_{\rm spec}^{2}}{2}},
\end{eqnarray}
with $a_{\rm spec}$ and $b_{\rm spec}$ as  semi-major and semi-minor axes, respectively. These parameters together with the position angle of the elliptical region $\theta$ are listed in Table. \ref{table: Observations}.}
To ensure consistent temperature measurements, we used a fixed spectral fitting range of [0.4-3.0] keV. The lower limit is set to avoid energies with a high background-to-signal ratio.  The [1.45-1.6] keV range was excluded due to the strong instrumental Al line, which dominates the background and its exclusion improves the S/N of the data. We checked that the number of spectral bins left after channel trimming was larger than the degrees of freedom of the model. In one case (Group ID: 5089), an inspection of the background spectrum showed strong peaks in the soft band ($ > 0.6$ keV) indicating the presence of an extra background component due to soft protons \citep{kuntz}. In that case, we increased the lower fitting limit to 0.6 keV.

To better constrain our analysis, the temperature modelling was repeated using a wider energy range of [0.4-7.0] keV. The systematic error, defined as the difference in $kT$ between the two energy ranges, was always checked and found to be less than 10$\%$. In all of the observations analysed, the reduced $\chi^{2}$ values were in the range of [0.7-1.8]. The obtained temperature estimates for the \textit{XMM-Newton} subsample are summarised in Table \ref{table: Observations}.

\subsection{Surface brightness profiles}
\label{sectionsb}

The \textit{Chandra} Interactive Analysis of Observations (CIAO) v4.15.1 \citep{ciao} was used in the surface brightness profile extraction and fitting. Assuming circular symmetry, a set of concentric circular annuli centred on the temperature extraction region was used for each observation. Depending on the location of each centre of extended emission on the detector, the outer radius of the annuli R$_{kT}$ was set to fully encompass the outskirts of the galaxy groups and reach the background level. In the four over-split systems, the same component used for temperature estimation was also used in the surface brightness profile extraction, and the other component was manually masked. In one case (Group ID: 6116), the centre of the source was on the edge of the detector, and since we have no data available on scales outside \textit{XMM-Newton}'s FOV, no usable radial profile fit could be obtained. Thus, it is excluded from the surface brightness, mass, and X-ray luminosity analyses; however, we kept it in the $\sigma_{\rm v}-kT$ study, as its temperature and velocity dispersion are well constrained.
Point sources were given special attention due to the faint nature of the extended X-ray emission from our group sample. Therefore, any left-over emission from interlopers, failure to detect faint-emission interlopers, or both, was found to affect the radial profile fit to a large extent. Accordingly, point source regions were re-examined by eye and often modified manually to fully cover their emission area. CIAO’s \texttt{dmcopy} and \texttt{dmextract} tasks were used in the point source subtraction and radial profile extraction, respectively. A variance map ($\sigma^{2}$) was produced by squaring the error on counts and fed to \texttt{dmextract} along with the background subtracted and point source corrected image. Surface brightness profiles were then calculated by dividing the number of counts in each concentric ring by its area, and errors were estimated by dividing the variance of each ring by its area.

A one-dimensional beta model was fitted to the radial profiles of each galaxy group using the \texttt{beta1d} model of the \texttt{Sherpa} package. The \texttt{beta1d} model has the form

\begin{eqnarray}
   \Sigma(r) = \Sigma_{0} \left[ 1+ \left( \frac{r}{r_{c}} \right)^{2} \right]^{-3\beta + 0.5},
\end{eqnarray}
where $\Sigma(r)$ is the surface brightness at radius $r$, r$_{c}$ is the core radius, and $\beta$ is the slope parameter of the profile.

We are only interested in the slope of the profiles at large radii; hence, a simple power-law relation with one slope parameter $\beta$ fitting the outskirts of the groups was our goal. Good fits, with reduced $\chi^{2}$ in the range $0.8<\chi^{2}<2$, were obtained in which the core radius parameter was fixed at artificially small values. The extreme merger case (Group ID: 5089) was the only system that did not have robust statistical results ($\chi^{2}> 2$). The parameters used in the radial profile extraction and fitting are listed in Table \ref{table: SB-Mass Results}.

\begin{figure}[hbt!]
\centering
   \includegraphics[width=\columnwidth]{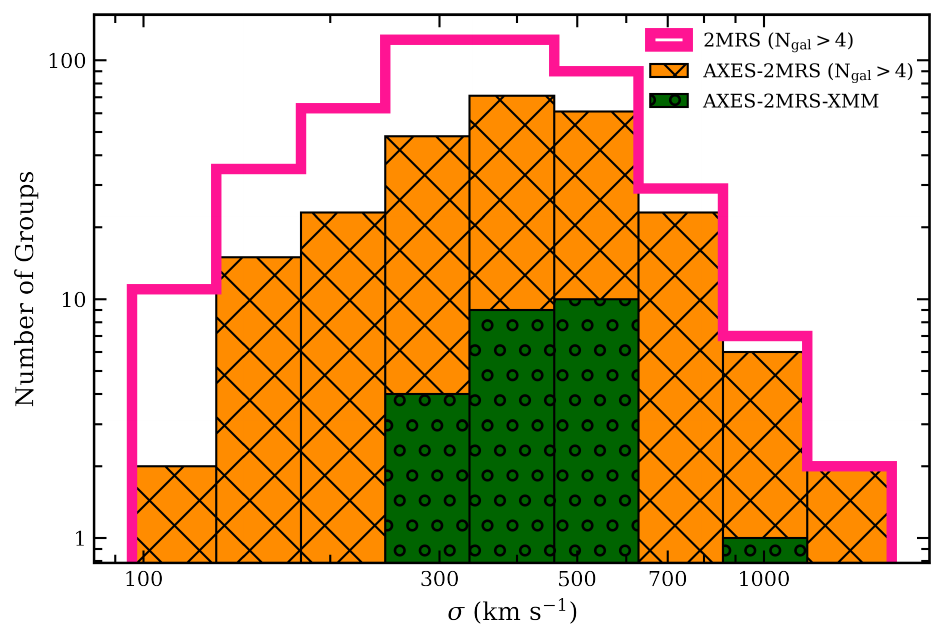}
     \caption{  {Distribution of the velocity dispersion of the cleaned 5+ member 2MRS groups (pink) overlaid with that of the 5+ member AXES-2MRS groups (orange) and that of the AXES-2MRS-XMM subset (green).}}
     \label{sigma_dist}
\end{figure}

\begin{figure}
\centering
   \includegraphics[width=\columnwidth]{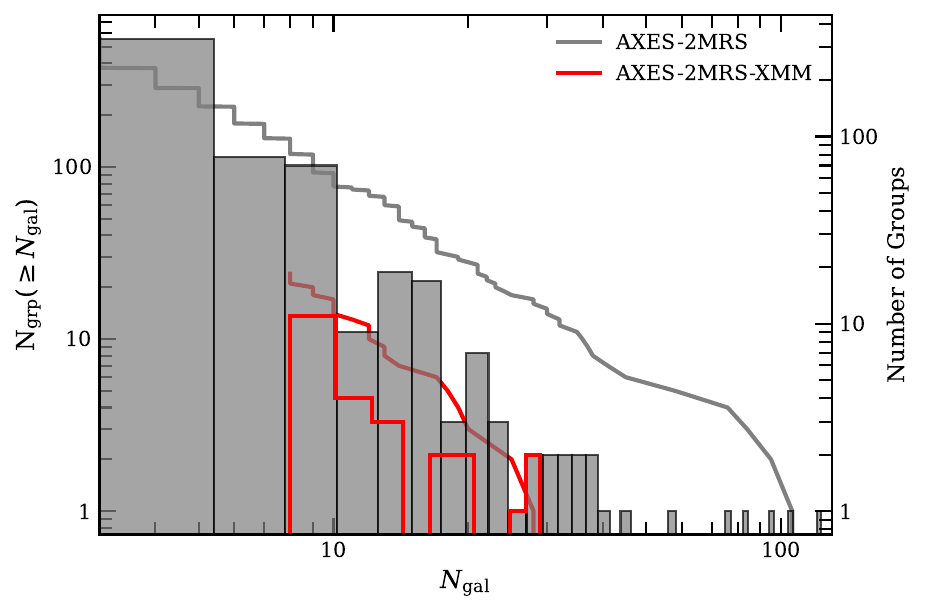}
     \caption{Cumulative distributions (curves) and frequency distributions (histograms) of the number of member galaxies (which also traces the group richnesses) for the full AXES-2MRS groups (grey) and the \textit{XMM-Newton} subsample (red). The left Y-axis shows the values of the cumulative distribution curves, while the right Y-axis shows the values of the frequency distribution curves.}
     \label{ngalhist}
\end{figure}

\subsection{Hydrostatic mass estimates}
\label{mass}

The total X-ray-deduced mass inside a radius $r$ assuming hydrostatic equilibrium, a $\beta$-model shape of the gas density, and a polytropic temperature profile ($T(r) \propto n_{\rm e}^{\gamma -1}$, where $n_{\rm e}$ is the electron density) is \citep{finoguenov01} 
\begin{eqnarray}
   M(r) = 3.7 \times 10^{13} \ T(r) \ r \ \frac{3 \ \beta \ \gamma \ \left( \frac{r}{r_{\rm c}} \right)^{2}}{1+  \left( \frac{r}{r_{\rm c}} \right)^{2}}\, M_{\odot} \ ,
\end{eqnarray}
where $\gamma = 1.1$. The X-ray masses were calculated inside the temperature extraction radius R$_{kT}$, and thus $T(r)$ is just the spectroscopic temperature. {While we used elliptical regions in the temperature extraction, we assumed a circular symmetry in the mass analysis.} 
  The overdensity of the measurement was estimated as
\begin{eqnarray}
   \Delta = \frac{M(r)}{(4\pi/3)\, r^{3} \rho_{c}(z)},
\end{eqnarray}
where $\rho_{c}(z)$  = $1.37\times10^{11} E^{2}(z) M_{\odot}/(\rm Mpc)^{3}$ is the critical density of the Universe at redshift $z$. Table \ref{table: SB-Mass Results} includes the calculated X-ray masses and their overdensities as well as the rescaled masses at $\Delta = 10000$. In rescaling the masses, we used the mean expected concentration following the prescription of \citet{hukrastov}.   {To calculate the concentration, we used the code provided in the appendix of \citet{hukrastov}, which also presents the relevant equations.} In addition, we explore our constraints on the concentration of individual halos coming from comparing direct mass estimates at two different overdensities{: the central mass   {($M_{10000}$), measured by \textit{XMM-Newton}} and the virial mass ($M_{200}$), inferred by the velocity dispersion.}

\subsection{Velocity dispersion}
\label{sec:velocity_dispersion}

We choose the gapper velocity dispersion estimator \citep{beers90} as it is preferred in case of a low number of member galaxies. In particular, \citet{beers90} favours this method for clusters with fewer than 15 members.    {In Fig. \ref{sigma_dist}, we show the distribution of the measured line-of-sight velocity dispersion of both the cleaned 5+ member optical 2MRS groups and the cleaned 5+ member X-ray AXES-2MRS groups. We also show the distribution of the velocity dispersion of the AXES-2MRS-XMM subset. The values in Fig. \ref{sigma_dist} are used in the scaling relations in Sect. \ref{scalingrelations}. The fraction of the X-ray emitting groups is steadily increasing with velocity dispersion, from 20-30\% in the 100--200 km s$^{-1}$ range to $>70$\% at $\sigma_v>500$ km s$^{-1}$. } Figure \ref{ngalhist} shows the cumulative distribution of the number of groups  ($N_{\rm grp}$) in the full AXES-2MRS, and the \textit{XMM-Newton} subsample, as a function of the number of member galaxies $N_{\rm gal}$, as well as the corresponding frequency distribution.

The velocity dispersion for the groups is calculated as
\begin{eqnarray}
   \sigma_{\rm gap} = \frac{c}{{1+\langle z\rangle}}\,\left(\frac{\sqrt{\pi}}{N(N-1)} \sum_{i=1}^{N-1} w_{i}g_{i}\right),
\end{eqnarray}
where $c$ is the speed of light, $N$ is the number of members, $w_{i}$ = $i(N-i)$,  $g_{i}$ = $z_{i+1} - z_{i}$ is the pairwise difference between member redshifts, {and $\langle z\rangle$ is the mean redshift}. The uncertainty on $\sigma_{\rm gap}$ is calculated as
\begin{eqnarray}
   \Delta\sigma_{gap} = \frac{C\sigma_{gap}}{\sqrt{N-1}},
\end{eqnarray}
where C=0.91, based on the modelling of  \citet{Ruel14}. We adopted the gapper velocity dispersion within the {\it Clean}, which removes interlopers and only uses the galaxies inside the $R_{200}$ in the calculation. 
We report our velocity dispersion measurements of the \textit{XMM-Newton} subsample in Table \ref{table: Optical Results}, and for the full AXES-2MRS catalogue in Table \ref{tab:catalogue_columns}.

\subsection{Halo concentration}
\label{concentration}

Our X-ray mass measurements and velocity dispersions constrain the masses of the groups at very different overdensities: the X-ray measurements cover the central part of the group, while spectroscopic members used for the velocity dispersion estimates extend to the virial radius. We take the optical mass measurement from the calibrations of \citet{munari13}, which linked the $M_{200}$ to the measured $\sigma_v$. We seek a value of concentration that describes both X-ray and optical mass measurements. The largest contribution to the error is the statistical error on the velocity dispersion, which is of the order 30\% for many groups. This is much larger than any bias associated with the assumption of the hydrostatic equilibrium (HSE, \citealp{hse1};\citealp{hse2}; \citealp{hse3}), especially taking into account the previously found good performance of the HSE technique at the overdensities we use for X-ray mass measurements. The resulting constraints on the concentration are rather uncertain (see Sect. \ref{cresult} and Table \ref{table: SB-Mass Results}) with no group clearly showing deviations from the expected concentration of galaxy groups of 2--7 \citep{neto07}.

\subsection{Velocity substructure}
\label{substructure}

We used the Anderson-Darling (AD) normality test \citep{AD} to split the AXES-2MRS catalogue and the \textit{XMM-Newton} subsample into Gaussian (G) and non-Gaussian (NG) groups and study the effect of velocity substructure on the scaling relations. We followed the procedure outlined in \citet{adtest} for applying the AD test. In particular, we calculated the test statistic $A^{2}$ and its sample-size-weighted modification $A^{2*}$, from the ordered velocities of the galaxy group members {$x_{i}$}, as:
\begin{align}
    & A^{2} = -N - \frac{1}{N} \sum_{i=1}^{N} (2i-1)\,\left\{\ln\Phi(x_{i})+\ln\left[1-\Phi\left(x_{N+1-i}\right)\right]\right\}\\
    & A^{2*} = A^{2}\left( 1+ \frac{0.75}{N} + \frac{2.25}{N^{2}} \right) \ ,
\end{align}
where $x_{i} \leq x < x_{i+1}$ and $\Phi(x_{i})$ is the cumulative distribution function of the hypothetical underlying Gaussian distribution given as
\begin{eqnarray}
    \Phi(x_{i}) = \frac{1}{2} \left[ 1+ \text{erf} \left( \frac{x_{i}-\mu}{\sqrt{2}\sigma_{\rm v}} \right) \right] \ ,
\end{eqnarray}
where $\mu$ is the mean velocity of the group and $\sigma_{\rm v}$ is the velocity dispersion. The term $A^{2*}$ is then used to compute the significance level $\alpha_{AD}$, which is used to assess the Gaussianity assumption, as
\begin{eqnarray}
    \alpha_{AD} = a \exp{\left ( -\frac{A^{2*}}{b} \right)},
\end{eqnarray}
where $a = 3.6789468$ and $b = 0.1749916$ are numerical fitting parameters taken from \citet{nelson}. The percentage of the G groups in AXES-2MRS with at least eight spectroscopic members is $\sim{77\%}$, consistent with \citet{damsted}, who studied the dynamics of CODEX clusters, which overlap in X-ray luminosity but are located at higher redshift.

\section{Scaling relations}
\label{scalingrelations}

\subsection{{Statistical methods}}
\label{sec:statmeth}

Primarily, the Python package \texttt{linmix}\footnote{\url{https://github.com/jmeyers314/linmix}} \citep{kelly07} was used in the scaling relations of interest in this paper. It was shown by \citet{kelly07} that it performs better than other regression estimators (e.g. OLS, BCES(Y|X), and FITEXY) when the measurement errors are large and the sample size is small. Optical $\sigma_{\rm v}$ values naturally exhibit a relatively large uncertainty ($\sim$20-30$\%$), and we perform the scaling relations on a limited-size group sample. The \texttt{linmix} package is based on a hierarchical Bayesian model that approximates a distribution function of the input data points using a mixture of several Gaussian components ($K$). Except for $c_{200}-L_{\rm X}$ relation which has $K = 4$ (see Sect. \ref{cresult}), all the scaling relations presented in this section use $K = 3$. 
To better illustrate \texttt{linmix}'s structure, we take the $\sigma_{\rm v}-kT$ relation as an example (see Sect. \ref{sectionsigt} and figures therein). Essentially, $kT$ and $\sigma_{\rm v}$ are supposed to follow a bivariate log-normal distribution $\mathcal{N}_{2}(\mu, \Sigma)$ with the mean $\mu = (\xi, \eta)$. These values $(\xi, \eta)$ represent the true, yet unobservable {means of} $\ln{kT}$ and $\ln{\sigma_{\rm v}}$, and the covariance matrix $\Sigma$ contains the errors observed in the logarithmic values of kT and $\sigma_{\rm v}$. The connection between $\xi$ and $\eta$ is established through the conditional probability distribution $P(\eta\,|\,\xi)$ = $\mathcal{N}(A + B\,\xi, S^{2})$. Here, $A$ is the intercept, $B$ is the slope, and $S$ is the Gaussian intrinsic scatter of $\eta$ around the regression line. The exact equation used in the fitting is
\begin{eqnarray}
    \eta_i = A + B \,\xi_{i} + S,
\end{eqnarray}
where $x_i = \xi_{i} + x_{\rm err, i}$ is the predictor vector of the data points + errors, and $y_{i} = \eta_i + y_{\rm err, i}$ is the target vector. The exact fitting formulas used in the scaling relations are listed in Table \ref{table: formulas}. We use 100 000 Markov chain Monte Carlo (MCMC) iterations within {\tt linmix} and report the mean of the posterior distribution of the best-fit parameters in Table \ref{tab:allresults}.

\begin{figure}
\centering
   \includegraphics[width=\columnwidth]{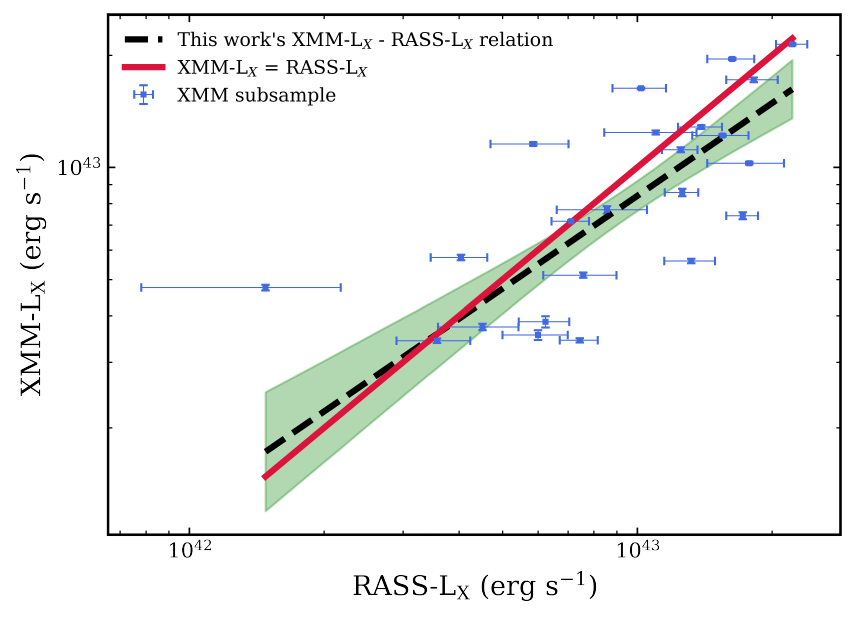}
     \caption{Comparison of AXES-2MRS X-ray luminosity measurements between \textit{XMM-Newton} and RASS. The dashed line {is the best fit}. The solid line gives the one-to-one relation between \textit{XMM-Newton} and RASS. The shaded region is the 1$\sigma$ uncertainty of the slope and intercept.
     }
     \label{lxxmmrass}
\end{figure}

\begin{figure}
\centering
   \includegraphics[width= \columnwidth]{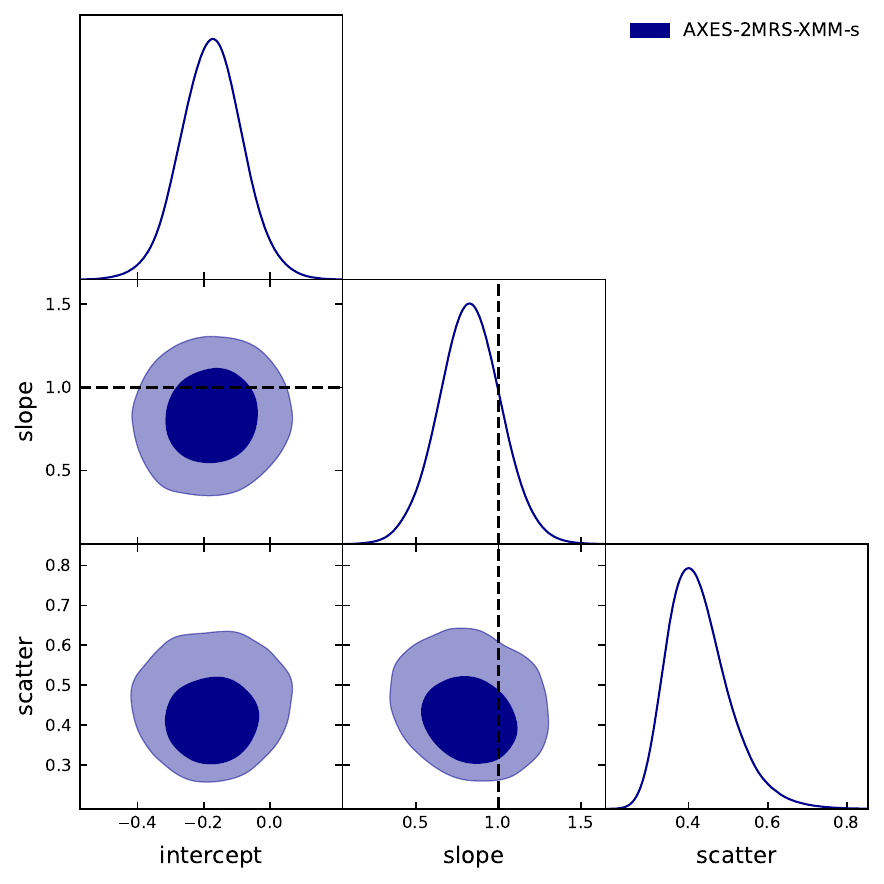}
     \caption{MCMC fitting results with one- and two-dimensional projections of the posteriors of the XMM-$L_{\rm X}$--RASS $L_{\rm X}$ scaling relation for the AXES-2MRS-XMM sample. The vertical and horizontal lines represent the one-to-one relation and correspond to the red line in Fig. \ref{lxxmmrass}. Dark and light contours represent 68$\%$ and 95$\%$ confidence levels, respectively.} 
     \label{lx_rassxmm_getdist}
\end{figure}

Given that we introduce a new sample of galaxy groups and use Bayesian methods to analyse the scaling relations, it is important to separate the contribution of the sample from the difference in the analysis. So, we also apply the 
orthogonal distance regression \citep[ODR]{odr} through its SciPy wrapper \texttt{ODRPACK}\footnote{\url{https://docs.scipy.org/doc/scipy/reference/odr.html}}.
  The main ODR problem can be represented as the minimisation of the residual sum of the squares of the orthogonal distances between each data point ($x_{i}$,$y_{i}$) and the curve representing the model equation $y = \alpha x + \beta$. The ODR algorithm can be expressed as
\begin{eqnarray}
    \mathrm{min} \sum_{i=1}^{n} \left(\left[ \frac{\alpha(x_{i} + \delta_{i})+\beta - y_{i}}{\sigma_{\epsilon_{i}}}\right]^{2} + \left[\frac{\delta_{i}}{\sigma_{\delta_{i}}}\right]^{2}\right),
\end{eqnarray}
where $\alpha$ and $\beta$ are the true, yet unknown values of the model parameters, $x_{i}$ and $y_{i}$ are the data vectors, $\delta_{i}$ is the true, yet unknown error on $x_{i}$ (an equivalent for $y_{i}$ is $\epsilon_i$, but it does not affect the solution), $\sigma_{\delta_i}$ and $\sigma_{\epsilon_i}$ are the weights to differentiate the contribution of different points to the solution, which here are selected to be the estimated statistical errors on $x_{i}$ and $y_{i}$, respectively. 
As Bayesian methods typically produce wider confidence intervals than frequentist ones, uncertainties of the \texttt{linmix} parameters are somewhat larger. In the following subsections, we present the scaling relations between the \textit{XMM-Newton} and ROSAT X-ray luminosities ($L_{\rm XMM}-L_{\rm RASS}$), the optical velocity dispersion and each of the X-ray luminosity ($\sigma_{\rm v}-L_{\rm X}$), the X-ray gas temperature ($\sigma_{\rm v}-kT$), and the mass ($\sigma_{\rm v}-M_{10000}$) as well as between the X-ray temperature and luminosity ($kT-L_{\rm X}$) and between the dark matter-halo concentration and luminosity ($c_{200}-L_{\rm X}$).\\

\subsection{Characterisation of AXES X-ray luminosities}

\begin{figure*}
    \centering
    \begin{subfigure}[b]{0.6\textwidth}
            \hspace{-0.1in}
            \includegraphics[height=0.28\textheight]{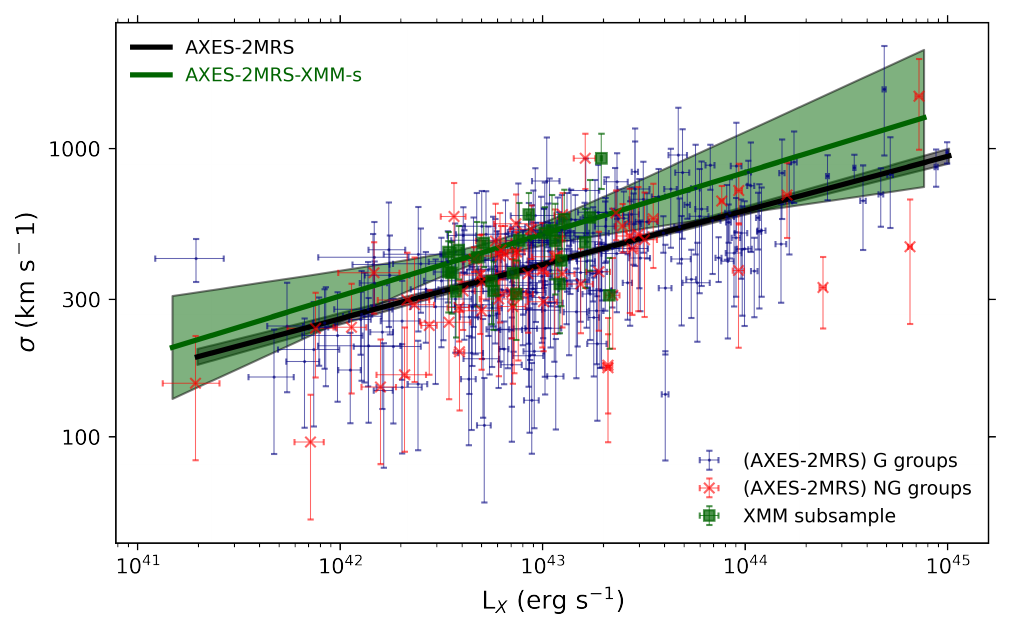}
    \end{subfigure}
    \begin{subfigure}[b]{0.39\textwidth}
            \hspace{-0.1in}
            \includegraphics[height=0.28\textheight]{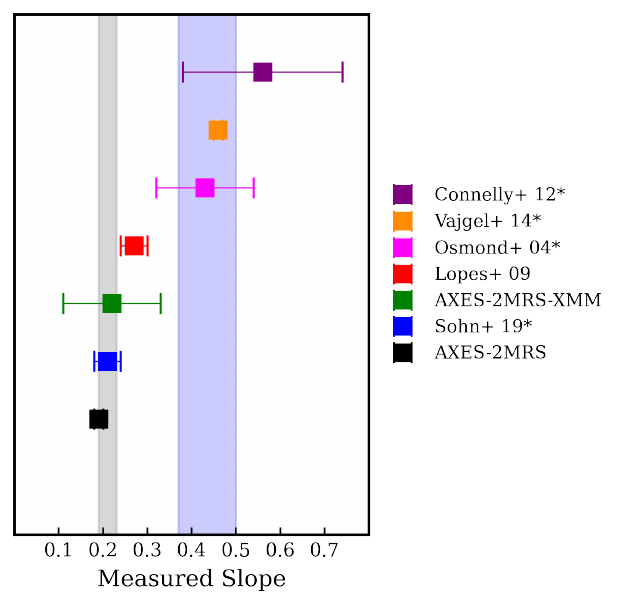}
    \end{subfigure}
    \caption{Velocity dispersion versus X-ray luminosity. \textit{Left panel:} 
    Velocity dispersion versus X-ray luminosity for the full AXES-2MRS systems with at least five members and our \textit{XMM-Newton} subsample. The G and NG groups are marked with blue circles and red crosses, respectively. The \textit{XMM-Newton} subsample is marked with filled green squares. Shaded regions are 1$\sigma$ uncertainties. \textit{Right panel:} Slope comparison for the $\sigma_{\rm v}-L_{\rm X}$ relation with the literature. The grey-shaded region is the self-similar expectation calibrated according to the X-ray emissivity in the band-limited {range of $0.1-2.4$ keV for groups with temperatures in the $0.7-3.0$ keV interval}, while the blue-shaded region is for the bolometric luminosity. (* refers to relations originally expressed as $L_{\rm X}-\sigma_{\rm v}$ and was inverted for comparability). } 
    \label{sigmalx}
\end{figure*}

\begin{figure}
\centering
   \includegraphics[width= \columnwidth]{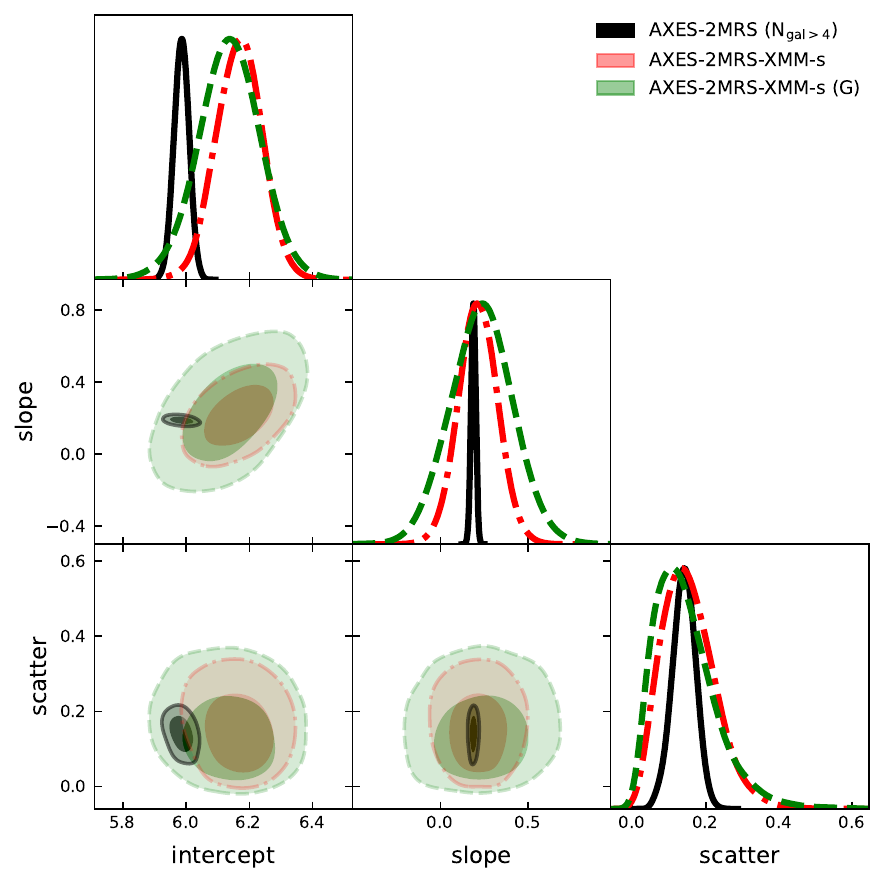}
     \caption{MCMC fitting results with one- and two-dimensional projections of the posteriors of the $\sigma_{\rm v}-L_{\rm X}$ scaling relation for the AXES-2MRS groups with more than four galaxies (black solid curves and contours), AXES-2MRS-XMM-s (red dash-dot curves and contours), and its Gaussian subset (green dashed curves and contours). We refer to Table \ref{tab:allresults} for details about each sample. Other details are the same as Fig. \ref{lx_rassxmm_getdist}.
     }
     \label{sigmalx_getdist}
\end{figure}

Since AXES presents new estimates of X-ray luminosities, measured using large sky areas, it is important to assess the quality of these estimates, using better X-ray observations available with \textit{XMM-Newton}. 
In the computation of X-ray luminosities, we select the band to report the luminosity to be consistent with other studies, 0.1--2.4 keV. The flux measurement is performed using {the} 0.5--2.0 keV band for both  ROSAT and \textit{XMM-Newton} data.
In the calculation of the luminosity, a K-correction using the redshift of the source, temperature estimate using the same $L-T$ relation, and a band difference between observed 0.5--2 keV (used for the flux measurements) and the rest-frame 0.1-2.4 keV (used for $L_{\rm X}$) is made. There are some subtle differences, such as in the RASS data (we used the full flux) and in \textit{XMM-Newton} (we removed the contribution of point sources). Also, with \textit{XMM-Newton} data, the flux apertures capture more precisely the contribution of the group, thus avoiding an extra source of scatter due to confusion. Thus, even with a smaller sample size, \textit{XMM-Newton} data on luminosities deliver improvements on the association of X-ray emission with the galaxy group. Both RASS and \textit{XMM-Newton} fluxes are extrapolated to the estimated $R_{500}$ radius following the procedure described in \citet{finoguenov07}.  The RASS data require no extrapolation, while these aperture corrections to \textit{XMM-Newton} fluxes are within 20\%.   {The flux extrapolation in \textit{XMM-Newton} $L_{\rm X}$ estimates are higher than in measuring RASS luminosity, which is a source of additional scatter.}


The correspondence between the X-ray luminosity measurements between \textit{XMM-Newton} and RASS is illustrated in Fig. \ref{lxxmmrass}, Fig. \ref{lx_rassxmm_getdist}, and detailed in Table \ref{tab:allresults}. The relation is close to the one-to-one relation, and the main effect we see is a 40\% scatter in the luminosity estimates, which we attribute to be a characteristic of the quality of survey-type measurement of RASS. 

\subsection{Velocity dispersion--X-ray luminosity relation}

\begin{figure*}
    \centering
    \begin{subfigure}[b]{0.6\textwidth}
            \includegraphics[height=0.28\textheight]{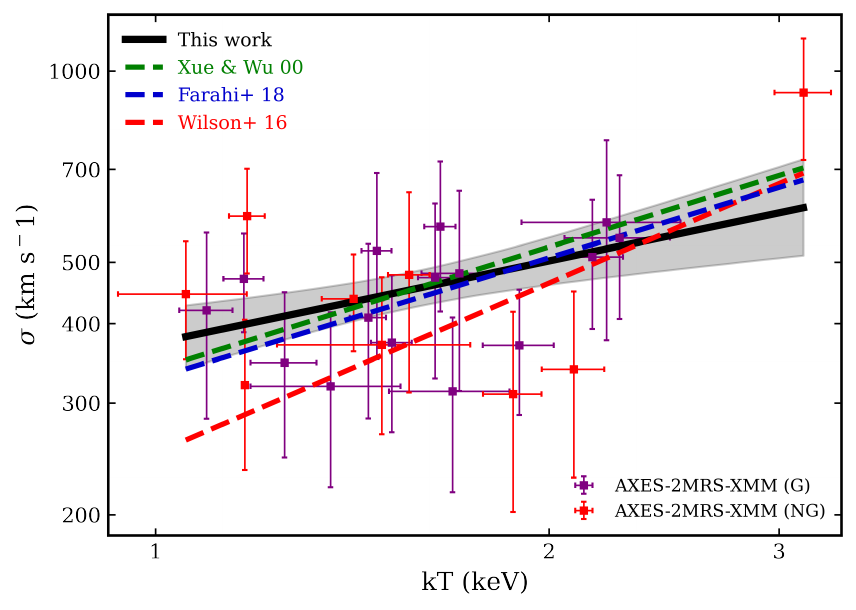}
    \end{subfigure}
    \begin{subfigure}[b]{0.39\textwidth}
            \hspace{-0.3in}
            \includegraphics[height=0.28\textheight]{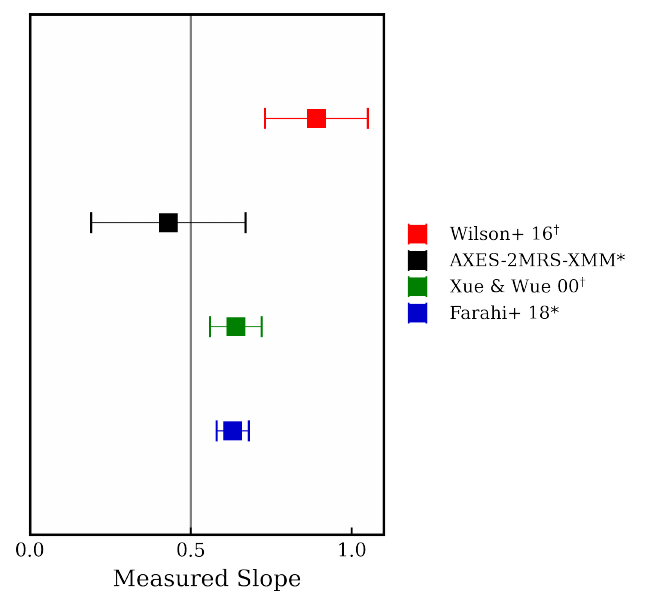}
    \end{subfigure}
    \caption{Velocity dispersion versus X-ray temperature. \textit{Left panel:} 
    {Velocity dispersion versus temperature} for the \textit{XMM-Newton} subsample used in this work in comparison with the literature. The G and NG groups in our AXES-2MRS-XMM sample are represented by purple and red squares, respectively. The grey transparent band is 1$\sigma$ confidence interval for the slope and intercept. \textit{Right panel:} Comparison of the slope of the $\sigma_{\rm v}-kT$ relation with the literature. The grey solid line is the theoretical self-similar expectation. The asterisk mark (*) refers to works using Bayesian linear regression, while the dagger mark ($\dagger$) refers to works using ODR. The slope of \citet{wilson16} is subject to a potential bias, as claimed in \citet{farahi18}.}
    \label{sigmat}
\end{figure*}

\begin{figure}
\centering
   \includegraphics[width= \columnwidth]{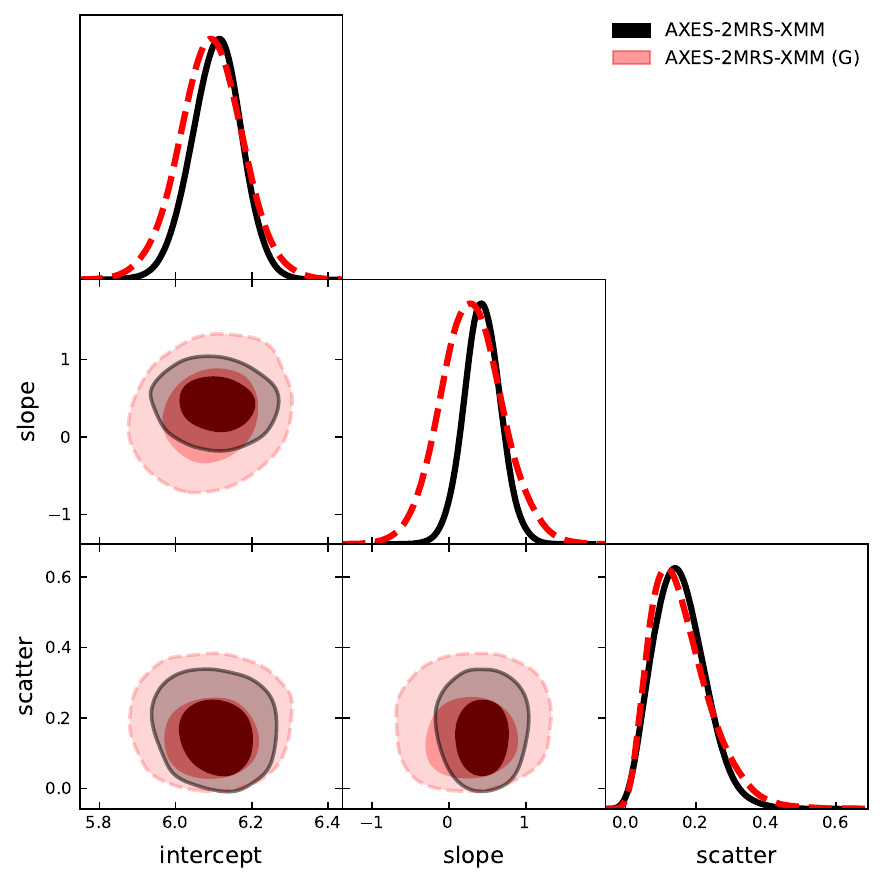}
     \caption{MCMC fitting results with one- and two-dimensional projections of the posteriors of the \textit{$\sigma_{\rm v}-kT$} scaling relation for the full AXES-2MRS-XMM sample (black solid curves and contours) and AXES-2MRS-XMM (G; red dashed curves and contours). Refer to Table \ref{tab:allresults} for details about each sample. Other details are the same as Fig. \ref{lx_rassxmm_getdist}. 
     }
     \label{tsgima_getdist}
\end{figure}

Since no selection criteria were applied on $\sigma_{\rm v}$ in our \textit{XMM-Newton} subsample, we examine the $\sigma_{\rm v}-L_{\rm X}$ relation, instead of $L_{\rm X}-\sigma_{\rm v}$, and compare it to the relation of the full AXES-2MRS catalogue \citep[for a detailed discussion on the choice, see][]{kelly07}.
We are particularly interested in comparing the constrained intrinsic scatters to assess the effect of sources of contamination (point sources, nearby groups, etc.) on the $\sigma_{\rm v}-L_{\rm X}$ relation, due to  \textit{XMM-Newton}’s higher sensitivity to those compared to ROSAT. We present our scaling relation work and compare it to the literature results on the group scale in Fig. \ref{sigmalx}, with the one- and two-dimensional projections of the posteriors of the parameters shown in Fig. \ref{sigmalx_getdist}. Our result for the full  XMM-sample ($\sigma_{\rm v} \propto$ $L_{\rm X}^{0.22 \pm 0.11}$) is in agreement with the calibration of the self-similar expectation based on the X-ray emissivity behaviour in the 0.1-2.4 keV energy band and 0.7-3.0 keV temperature range (see the right panel in Fig. \ref{sigmalx} for our result in comparison with the literature). \citet{Lovisari21} presents an argument about the truly expected slope of this relation based on the behaviour of the X-ray emissivity in the low-temperature regime. They claim that instead of $L_{\rm X} \propto \sigma_{\rm v}^{4}$, the scaling should follow $L_{\rm X} \propto \sigma_{\rm v}^{3+2\gamma}$, where $\gamma$ is a constant determined based on the temperature range, energy band, and metallicity (see Table 1 in \citealp{Lovisari21}). 
Indeed, the most deviating results are obtained for galaxy groups at lower X-ray luminosities, compared to our sample. 
We detail the parameters of the $\sigma_{\rm v}-L_{\rm X}$ relation in Table \ref{tab:allresults}. We notice almost no change in the scatter of both \textit{XMM-Newton} and ROSAT relations, which we attribute to the level of intrinsic scatter being much larger than the scatter between \textit{XMM-Newton} and RASS luminosities. Although there is a noticeable difference in the normalisation of the scaling relation between RASS and \textit{XMM-Newton} samples, once we retain only the high-quality measurements of velocity dispersion by limiting the sample to that of at least 8 spectroscopic members, the difference disappears. This means that our \textit{XMM-Newton} subsample is representative of the eight or more member AXES groups, while the full catalogue is somewhat different.
Moreover, we investigated the effect of the number of spectroscopic members on the relation, and we found that adding poor galaxy groups (${N_{\rm gal}} < 8$) does not affect the quality of the fit in terms of the statistical errors and the intrinsic scatter (on the 1-$\sigma$ level). Up to our level of precision, there is no clear evidence of substructure (NG) deviations in the $\sigma_{\rm v}-L_{\rm X}$ relation as shown in Table \ref{tab:allresults}. Additionally, we matched the AXES-2MRS sample and the \textit{XMM-Newton} subsample with The Third Cambridge Catalog (3C) and its revised version (3CR) within 5 arcmin to test whether including galaxy groups with an active radio AGN changes the scaling relations. We removed a total of 6 systems from the full AXES-2MRS sample and 1 system from the \textit{XMM-Newton} subsample, and we did not find any noticeable change in the scaling relations, concluding that ongoing AGN activity is not immediately seen in the group properties. In the subsequent analysis, we no longer split the sample based on the radio properties.

\begin{figure*}
    \centering
    \begin{subfigure}[b]{0.65\textwidth}
            \includegraphics[height=0.32\textheight]{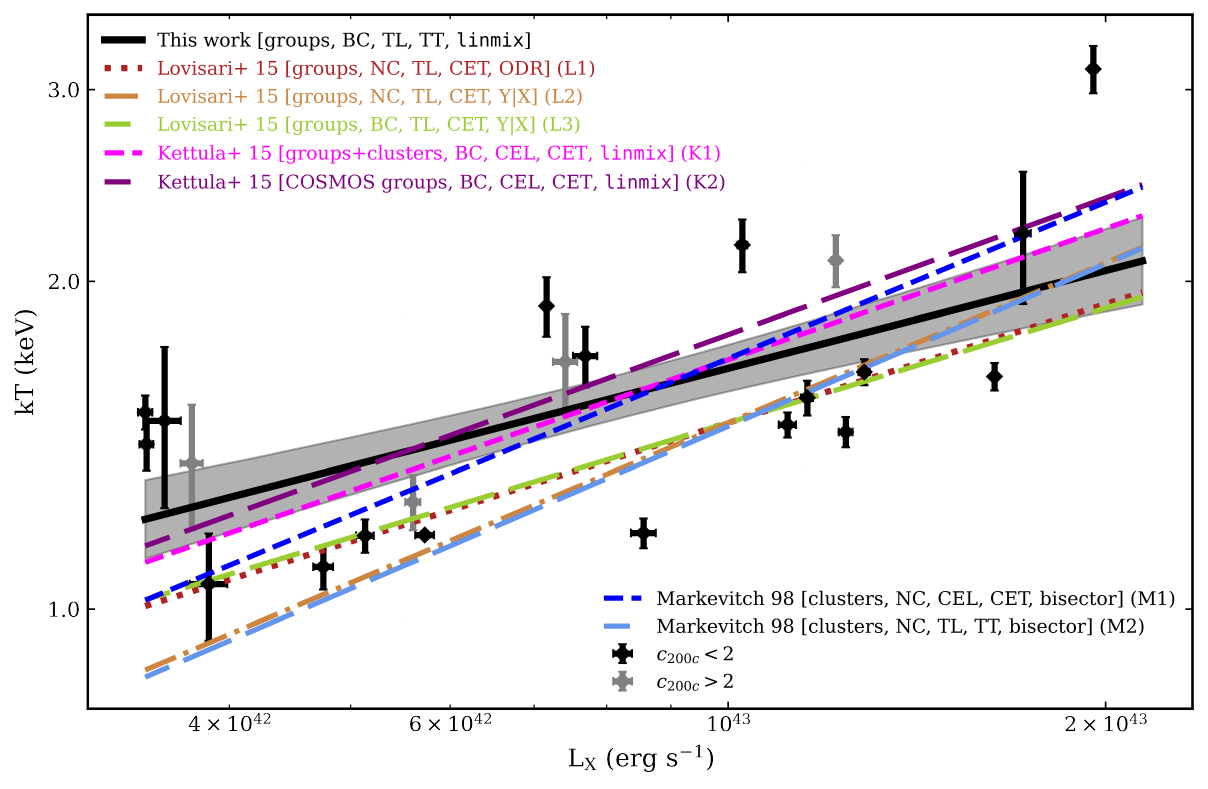}
    \end{subfigure}
    \begin{subfigure}[b]{0.3\textwidth}
            \includegraphics[height=0.32\textheight]{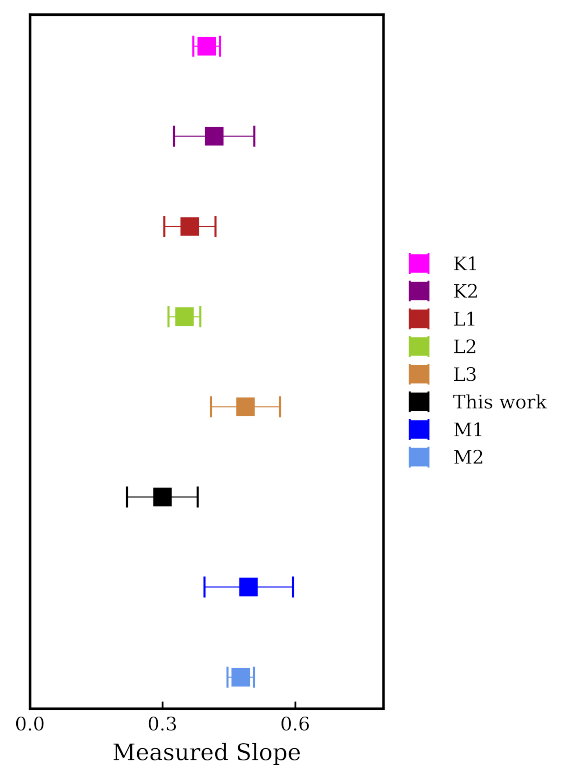}
    \end{subfigure}
    \caption{{X-ray temperature versus luminosity.} \textit{Left panel:} 
    {Temperature versus X-ray luminosity}  for the \textit{XMM-Newton} subsample used in this work (black solid line) in comparison with the literature. The terms BC and NC indicate bias-corrected and non-corrected relations, respectively, while TL and CEL are total and core-excised $L_{\rm X}$, respectively. The abbreviations TT and CET are for total and core-excised $kT$, respectively. The type of the sample used and the regression method are also indicated (see Sect. \ref{tlx_sec} for more details). Grey and black points are high and low concentration groups, respectively, defined at $c_{200} = 2$. The shaded region is the 1-$\sigma$ uncertainty. \textit{Right panel:} Comparison of the slope of the $kT-L_{\rm X}$ relation with the literature. Slope labels are the same as relation labels from the left panel. Only band-limited $L_{0.1-2.4}$ relations are considered.
        }
    \label{tlx}
\end{figure*}

\begin{figure}
\centering
   \includegraphics[width= \columnwidth]{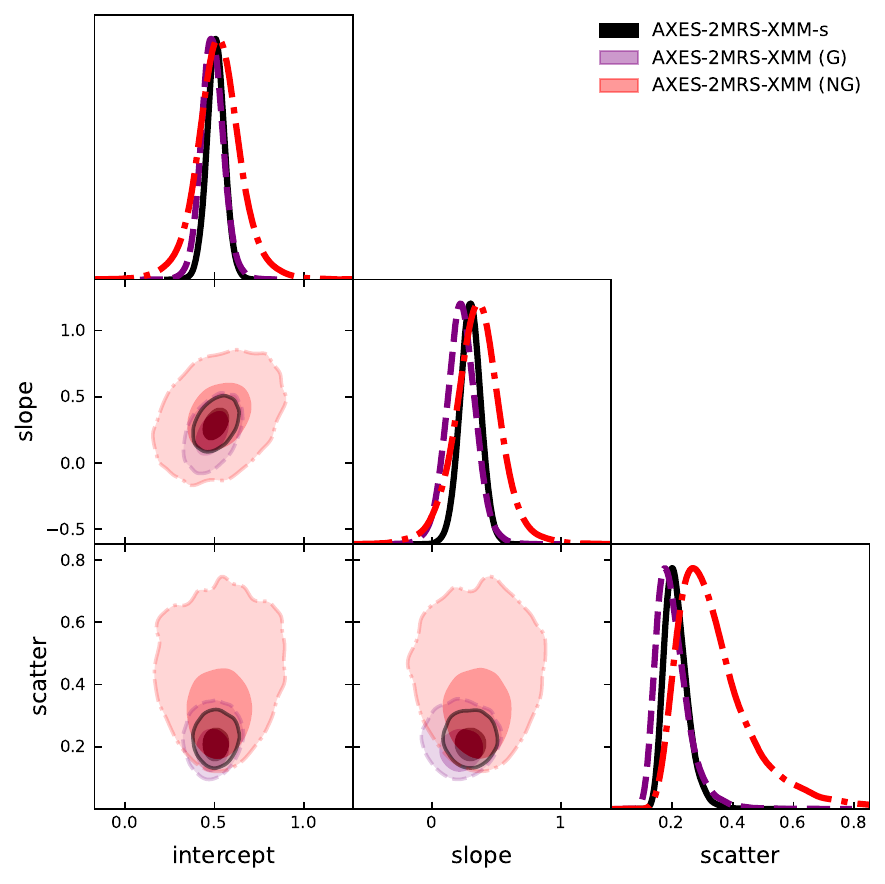}
     \caption{MCMC fitting results with one- and two-dimensional projections of the posteriors of the $kT-L_{\rm X}$ scaling relation for AXES-2MRS-XMM-s (black solid curves and contours), its Gaussian subset (purple dashed curves and contours), and its non-Gaussian subset (red dash-dot curves and contours). Refer to Table \ref{tab:allresults} for details about each sample. Other details are the same as Fig.~\ref{lx_rassxmm_getdist}.
     }
     \label{tlx_getdist}
\end{figure}

Tracking the efforts done studying the $\sigma_{\rm v}-L_{\rm X}$ relation, we find that \citet{sohn19} used a sample of 74 groups from the COSMOS survey \citep{finoguenov07, Gozaliasl2019} and reported a scaling of the form $\sigma_{\rm v} \propto L_{\rm X}^{0.21 \pm 0.03}$. Their sample suffered from anomalously low velocity dispersion values, so they reduced their sample to seven velocity dispersion bins and fitted them to get the above result. Notably, they stated that the intrinsic scatter of their relation was high; however, they did not report a specific value. \citet{lopes2019b} used a (group + cluster) sample of 97 systems from the NoSOCS-SDSS survey \citep{lopes2019a} and reported a result of the form $\sigma_{\rm v} \propto L_{\rm X}^{0.27 \pm 0.03}$ with an intrinsic scatter ($S_{\mathrm{ln}Y|\mathrm{ln}X}$) of $0.29 \pm 0.03$. \citet{osmond04} was one of the earliest efforts in constraining this relation on the group scale. They used ROSAT data for a group sample of 60 systems and reported a result of the form $\sigma_{\rm v} \propto L_{\rm X}^{0.43 \pm 0.11}$. Next, \citet{vajgel14} used a small sample of 14 groups from the X-Boötes survey \citep{murray05} and reported a relation of the form $\sigma_{\rm v} \propto L_{\rm X}^{0.46 \pm 0.01}$ and intrinsic log-normal scatter of 0.3. Lastly, \citet{connelly12}, using a sample of 38 high redshift X-ray groups with luminosities around $10^{42}$ ergs s$^{-1}$ measured using deep \textit{XMM-Newton} observations, derived a scaling of the form $\sigma_{\rm v} \propto L_{\rm X}^{0.56 \pm 0.18}$, with $L_{\rm X}$ estimates similar to those obtained in the COSMOS field, so the difference is mainly in the covered luminosity range. We report the least intrinsic scatter in the $\sigma_{\rm v}-L_{\rm X}$ relation in the literature with a $0.15 \pm 0.07$ value. We note that the original relations reported in \citet{sohn19}; \citet{osmond04}; \citet{vajgel14}; \citet{connelly12} were in the form $L_{\rm X}-\sigma_{\rm v}$, while we inverted them for the slope comparison.

\subsection{Velocity dispersion--X-ray temperature relation}
\label{sectionsigt}

We studied the $\sigma_{\rm v}-kT$ for the full \textit{XMM-Newton} sample of 24 groups that we have (the effect of adding the dynamically complex 2MRS ID: 5089 system is explored later in this section). The importance of these two observables arises as they are independent baryonic tracers for the depth of the group potential. In the absence of non-gravitational heating,  one expects the scaling to go as $\sigma_{\rm v} \propto T^{0.5}$. We report a consistent relation of the form $\sigma_v \propto T^{0.42 \pm 0.24}$. Detailed parameter estimations for the $\sigma_{\rm v}-kT$ relation are summarised in Table \ref{tab:allresults}.

We considered both the group and the group plus cluster samples when comparing our fit to the literature, with the results on the slope summarised in Fig.~\ref{sigmat}. The role of the velocity substructure in enhancing the scatter is marginally supported by the data with the Gaussian (G) groups and the full sample having 40\% lower intrinsic scatter than the non-Gaussian (NG) groups. Figure \ref{tsgima_getdist} shows the one- and two-dimensional projections of the posteriors of the relation. 

\citet{xue00} studied a low-redshift sample of 274 groups and clusters with $kT <$ 10.1 keV, and found a relation of the form $\sigma_{\rm v} \propto T^{0.64 \pm 0.08}$ using the ODR method with no characterisation of intrinsic scatter. Our data are more uniform because all the X-ray observations used were obtained from the same instrument, and we have more control over the velocity dispersion values as they were all computed using the same (gapper) estimator. In contrast, they used 20 different literature sources to assemble their X-ray and spectroscopic sample (see Table. 1 in \citealp{xue00}). Using a total of 19 groups plus clusters with $kT <$ 5.5 keV taken from the XMM Cluster Survey \citep{mehrtens}, \citet{wilson16} reported a steep relation of the form $\sigma_{\rm v} \propto T^{0.86 \pm 0.16}$. Lastly, \citet{farahi18} used a 138 group plus cluster sample selected from the XXL survey. X-ray temperatures ($kT <$ 6.5 keV) were taken from the XXL analysis of \citet{xxl18}, and spectroscopic data for galaxies were obtained from seven different literature sources. They report a relation in the form $\sigma_{\rm v} \propto T^{0.63 \pm 0.05}$. Despite the formal difference in the slope values, as illustrated in Fig.\ref{sigmat}, all the fits are broadly consistent with our data points. Therefore, obtaining a different slope of the scaling relation might be a result of a different sampling of the data. Exploring larger datasets and understanding the covariance of the scatter with the group properties affecting the selection is needed to establish a full picture.
\subsection{X-ray temperature--luminosity  relation}
\label{tlx_sec}

The soft-band X-ray luminosity and temperature are two X-ray observables obtained with minimal covariance. In our analysis, X-ray temperatures were obtained from spectral modelling of the X-ray observations. In contrast, X-ray luminosities were derived from the flux measurements in the 0.5--2 keV energy band, which, at the redshifts of this study, are primarily sensitive to gas density and metallicity. Accordingly, the $kT-L_{\rm X}$ relation is one of the most studied relations in the literature. The gravitational self-similar expectation is $L_{\rm bol} \propto T^{2}$, where $L_{\rm bol}$ is the bolometric luminosity \citep[e.g.][and references therein]{Boeh}. However, the expected behaviour of the band-limited [0.1-2.4 keV] relation is $L_{0.1-2.4} \propto T^{1.5 + \gamma}$ with $\gamma = -0.04$ for a typical metal abundance of $Z = 0.3\,Z_{\odot}$ and a temperature range of 0.4--3.0 keV (\citealp{Lovisari21}). Thus, the slope of the $kT-L_{\rm X}$ relation is expected to be $\sim$ 0.68. We report a relation of the form $kT \propto L_{\rm X}^{0.3 {\pm} 0.08}$ for the \textit{XMM-Newton} subsample representing AXES-2MRS. We show our relation, together with that of \citet{Kettula15, Lovisari15, maxim} in Fig. \ref{tlx}, while the MCMC fitting parameters of the relation are shown in Fig. \ref{tlx_getdist}.   {We split our sample on high and low concentration groups defined at $c_{200}$ in Fig. \ref{tlx} only to show the location of systems with extremely low concentration values.} To ensure an in-depth comparison, we list several properties of the literature relations in the left panel of the figure. In particular, we specify whether the relation is based on a group, cluster, or group + cluster sample. We also indicate whether it employs a selection bias correction (BC) or not (NC), and specify whether the central regions were excluded in estimating $L_{\rm X}$ (core excised, CEL) or the total $L_{\rm X}$ was used (TL). Furthermore, the total (TT) versus core-excised (CET) temperature is clarified together with the regression method. These acronyms are summarised in Table. \ref{table: acronyms}. To correct for the sampling bias of our \textit{XMM-Newton} subsample, we provide a detection vector to \texttt{linmix} to take into account the distribution of source over the luminosity, where we set temperatures for the systems without \textit{XMM-Newton} temperature measurements to its upper limit of 20 keV. We do not find differences in the results obtained without a detection vector, which illustrates that the sampling bias is negligible.

\begin{table}[hbt!]
\caption{Acronyms used in the $kT-L_{\rm{X}}$ relation.}
\label{table: acronyms}
\centering
\begin{tabular}{c c}
\hline\hline
Acronym & Definition\\
\hline
BC & Selection bias corrected\\
NC & Selection-bias non-corrected\\
CEL & Core-excised X-ray luminosity\\
TL & Total X-ray luminosity\\
CET & Core-excised X-ray temperature\\
TT & Total X-ray temperature\\
\hline
\end{tabular}
\end{table}

The literature scaling relations shown in Fig. \ref{tlx} are detailed as follows: firstly, \citet{Lovisari15} compiled a group sample by applying a flux limit of $F_{0.1-2.4} = 5 \times 10^{-12}$ erg s$^{-1}$cm$^{-2}$ and two redshift cuts ($0.01 < z < 0.035$) to the northern ROSAT all-sky galaxy cluster survey \citep[NORAS]{noras} and the ROSAT-ESO flux-limited X-ray
galaxy cluster survey \citep[REFLEX]{reflex} catalogues. The resulting 23-group sample span a similar temperature ($0.85 < kT < 2.8$ keV) and band-limited luminosity range ($0.4 < L_{\rm X} < 5.3 \times 10^{43}$ erg s$^{-1}$) as our XMM-subsample, and uses total luminosities and core-excised temperatures. Moreover, multiple regression methods were used, and in Fig. \ref{tlx} we show the relation using two of them (ODR and BCES Y|X) as well as the BC and NC relations. On the other hand, \citet{Kettula15} produced the $L_{\rm X}-kT$ relation based on a compiled sample of 12 low-mass clusters from the XMM-CFHTLS survey \citep{xmmcfhtls}, 10 low-mass systems from the COSMOS field \citep{Kettula13b}, and 48 high-mass clusters from the Canadian Cluster Comparison Project \citep[CCCP]{cccp1, cccp2}. The combined sample has a temperature and luminosity range of $1-12$ keV, and $10^{43}-10^{45}$ erg s$^{-1}$, respectively. In Fig. \ref{tlx}, we show their relation with the combined group plus cluster sample and that with only the low-mass COSMOS systems since the latter shows greater similarity in the parameter ranges with our study. Nevertheless, we note that they exclusively used CEL and CET. As a high-mass reference, \citet{maxim} used a low-redshift ($0.04 < z < 0.09$) 35-cluster sample selected from RASS that excludes systems with $kT \leq 3.5$ keV and has a mean TL of $\sim 3.86 \times 10^{44}$ erg s$^{-1}$. We show their relation using CEL and CET as well as that using TL and TT. In the right panel of Fig. \ref{tlx}, we separately compare the slope of the relation with the mentioned works. 

Figure \ref{tlx} {indicates} that all the {previous studies of the $T-L_{\rm X}$ relation in groups are consistent with our}  data and differences in the slope can be attributed to the selection effects or the fitting method used. All these slopes are shallower than the cluster slope of \citet{maxim},  indicating that the groups are hotter {than expected from the extrapolation of the cluster relation}, which can be attributed to the effect of feedback, to higher concentration, or an earlier formation epoch. The effect of feedback on the scaling relations is largely removed when using total mass measurements, instead of temperature, while the effect of the difference in the concentration would remain in those studies. This motivates us to consider the scaling relations using the total mass and to measure the concentration for our sample.

\begin{table*}
\caption{Fitting formulas of the scaling relations.}
\label{table: formulas}
\centering
\begin{tabular}{c c}
\hline\hline
Relation & Fitting formula\\
\hline
$L_{\rm XMM}-L_{\rm RASS}$ & $\ln(L_{\rm XMM} / 10^{43}$ erg s$^{-1}) = A + B \ln(L_{\rm RASS} / 10^{43}$ erg s$^{-1}) + S_{\mathrm{ln}Y|\mathrm{ln}X}$\\
$\sigma_{\rm v}-L_{\rm X}$ & $\ln(\sigma_{\rm v}$ / km s$^{-1}) = A + B \ln(L_{\rm X} / 10^{43}$ erg s$^{-1}) + S_{\mathrm{ln}Y|\mathrm{ln}X}$\\
$\sigma_{\rm v}-kT$ & $\ln(\sigma_{\rm v}$ / km s$^{-1}) = A + B \ln(kT / 1.5$ keV) + $S_{\mathrm{ln}Y|\mathrm{ln}X}$\\
$kT-L_{\rm X}$ & $\ln(kT$ / keV$^{-1}) = A + B \ln(L_{\rm X} / 10^{43}$ erg s$^{-1}) + S_{\mathrm{ln}Y|\mathrm{ln}X}$\\
$\sigma_{\rm v}-M_{10000}$ & $\ln(\sigma_{\rm v}$ / km s$^{-1}) = A + B \ln(M_{10000} / 10 ^{13} M_{\odot}) + S_{\mathrm{ln}Y\mathrm{ln}|X}$\\
$c_{200}-L_{\rm X}$ & $\ln(c_{200})$ = $A$ + $B$ $\ln(L_{\rm X} / 8 \times 10^{42}$\tablefootmark{$\dagger$} erg s$^{-1}$) + $S_{\mathrm{ln}Y|\mathrm{ln}X}$\\
\hline
\end{tabular}
\tablefoot{$A$, $B$, and $S_{\mathrm{ln}Y|\mathrm{ln}X}$ are the normalisation, slope, and intrinsic scatter of the scaling relations, respectively (see Sect. \ref{sec:statmeth} for more details). \tablefoottext{$\dagger$}{$8 \times 10^{42}$ erg s$^{-1}$ is the median L$_{\mathrm{X}}$ of the concentration sample and is found to decrease the slope uncertainty by a factor of 2.}}

\end{table*}

\begin{table*}

    \caption{Scaling relations.}
    \label{tab:allresults}
    
    \centering
\begin{tabular}{c c ccc cc}
\hline\hline 
\toprule
 \multirow{2}[3]{*}{Relation}
 & \multirow{2}[3]{*}{Sample}
 & \mc3c{\texttt{linmix}} 
 & \mc2c{ODR}\\
 \cmidrule(lr){3-5}
 \cmidrule{6-7} 
    & & intercept & slope & scatter & intercept & slope \\
   & & $A$ & $B$ & $S_{\mathrm{ln}Y|\mathrm{ln}X}$ & $A$ & $B$ \\
\midrule
$L_{\rm XMM}-L_{\rm RASS}$ & AXES-2MRS-XMM-s (23) & -0.18 $\pm$ 0.1 & 0.83 $\pm$ 0.19 & 0.43 $\pm$ 0.08 & -0.31 $\pm$ 0.12 & 1.56 $\pm$ 0.31\\
$\sigma_{\rm v}-L_{\rm X}$ & AXES-2MRS ($N_{\rm gal}> 4$) (252) & 5.99 $\pm$ 0.02 & 0.19 $\pm$ 0.01 & 0.14 $\pm$ 0.03 & 6.01 $\pm$ 0.02 & 0.19 $\pm$ 0.01\\
 & AXES-2MRS ($N_{\rm gal}> 7$) (134) & 6.03 $\pm$ 0.03 & 0.17 $\pm$ 0.02 & 0.17 $\pm$ 0.04 & 6.1 $\pm$ 0.03 & 0.17 $\pm$ 0.01\\
 & AXES-2MRS ($N_{\rm gal}>$ 7, G) (102) & 6.07 $\pm$ 0.03 & 0.17 $\pm$ 0.02 & 0.16 $\pm$ 0.04 & 6.1 $\pm$ 0.03 & 0.17 $\pm$ 0.01\\
 & AXES-2MRS ($N_{\rm gal}>$ 7, NG) (32) & 6.06 $\pm$ 0.06 & 0.17 $\pm$ 0.05 & 0.23 $\pm$ 0.08 & 6.09 $\pm$ 0.05 & 0.16 $\pm$ 0.04\\
 &AXES-2MRS-XMM-s (23) & 6.17 $\pm$ 0.07 & 0.21 $\pm$ 0.11 & 0.15 $\pm$ 0.07 & 6.19 $\pm$ 0.06 & 0.22 $\pm$ 0.09\\
 & AXES-2MRS-XMM-s (G) (14) & 6.14 $\pm$ 0.1 & 0.23 $\pm$ 0.18 & 0.14 $\pm$ 0.09 & 6.14 $\pm$ 0.05 & 0.23 $\pm$ 0.09\\
 & AXES-2MRS-XMM-s (NG) (9) & 6.17 $\pm$ 0.2 & 0.17 $\pm$ 0.27 & 0.41 $\pm$ 0.24 & 6.25 $\pm$ 0.12 & 0.23 $\pm$ 0.16\\
$\sigma_{\rm v}-kT$ & AXES-2MRS-XMM (24) & 6.11 $\pm$ 0.07 & 0.43 $\pm$ 0.24 & 0.15 $\pm$ 0.07 & 6.12 $\pm$ 0.05 & 0.44 $\pm$ 0.18\\
& AXES-2MRS-XMM (G) (15) & 6.09 $\pm$ 0.08 & 0.29 $\pm$ 0.41 & 0.16 $\pm$ 0.09 & 6.1 $\pm$ 0.05 & 0.28 $\pm$ 0.2\\
& AXES-2MRS-XMM (NG) (9) & 6.1 $\pm$ 0.18 & 0.48 $\pm$ 0.54 & 0.39 $\pm$ 0.23 & 6.16 $\pm$ 0.11 & 0.52 $\pm$ 0.31\\
$kT-L_{\rm X}$ & AXES-2MRS-XMM-s (23) & 0.51 $\pm$ 0.05 & 0.3 $\pm$ 0.08 & 0.21 $\pm$ 0.04 & 0.4 $\pm$ 0.04 & 0.36 $\pm$ 0.07\\
 & AXES-2MRS-XMM-s (G) (14) & 0.49 $\pm$ 0.07 & 0.22 $\pm$ 0.12 & 0.2 $\pm$ 0.06 & 0.42 $\pm$ 0.04 & 0.16 $\pm$ 0.08\\
 & AXES-2MRS-XMM-s (NG) (9) & 0.53 $\pm$ 0.14 & 0.35 $\pm$ 0.19 & 0.35 $\pm$ 0.15 & 0.43 $\pm$ 0.07 & 0.47 $\pm$ 0.12\\
$\sigma_{\rm v}-M_{10000}$ & Full mass sample (30) & 6.1 $\pm$ 0.06 & 0.13 $\pm$ 0.05 & 0.12 $\pm$ 0.06 & 6.15 $\pm$ 0.04 & 0.12 $\pm$ 0.04\\
& AXES-2MRS-XMM-s (G) (14) & 6.15 $\pm$ 0.12 & 0.29 $\pm$ 0.31 & 0.16 $\pm$ 0.09 & 6.13 $\pm$ 0.07 & 0.13 $\pm$ 0.11\\
& AXES-2MRS-XMM-s (NG) (9) & 6.23 $\pm$ 0.18 & 0.31 $\pm$ 0.26 & 0.34 $\pm$ 0.22 & 6.31 $\pm$ 0.11 & 0.34 $\pm$ 0.13\\
$c_{200}-L_{\rm X}$ & Full $c_{200}$ sample (24) & --0.12 $\pm$ 0.96 & 0.15 $\pm$ 5.37 & 0.76 $\pm$ 0.31 & 9.51 $\pm$ 17.75 & --1.51 $\pm$ 3.55\\
\bottomrule
\end{tabular}
\tablefoot{Column 1 gives the scaling relation (the exact fitting formulas are shown in Table. \ref{table: formulas}). Column 2 provides the sample and the number of systems used in the relation (in parenthesis). Columns 3 to 7 show the parameters of the scaling relations using the Bayesian regression package \texttt{linmix}, and the orthogonal distance regression (ODR), respectively. See Sect. \ref{sec:statmeth} and Table~\ref{table: formulas} for details. AXES-2MRS-XMM-s denotes the AXES-2MRS-XMM subsample excluding the system with the Group ID: 6116 discussed in Sect. \ref{sectionsb}.}

\end{table*}

\subsection{Velocity dispersion--mass relation}
\label{msigmasection}

\begin{figure}
\centering
   \includegraphics[width=\columnwidth]{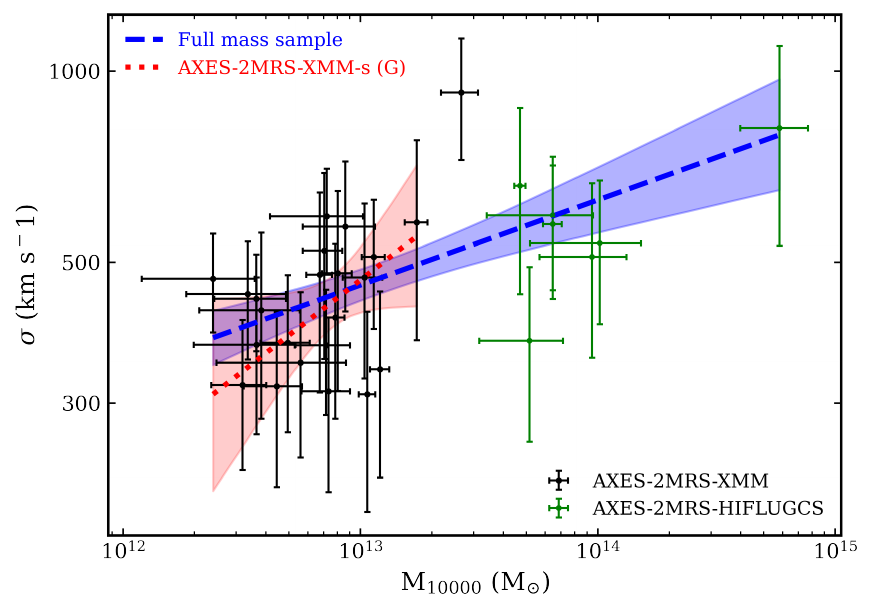}
     \caption{ 
     Velocity dispersion versus mass within a sphere of overdensity $10^4$ relative to the critical density of the Universe. Blue dashed and red dotted lines represent the scaling relation of our full mass sample (see Sect. \ref{msigmasection}) and the AXES-2MRS-XMM-s Gaussian subset, respectively. Black points are the full AXES-2MRS-XMM groups while green points are AXES-2MRS published groups.}
     \label{msigma}
\end{figure}

\begin{figure}
\centering
   \includegraphics[width= \columnwidth]{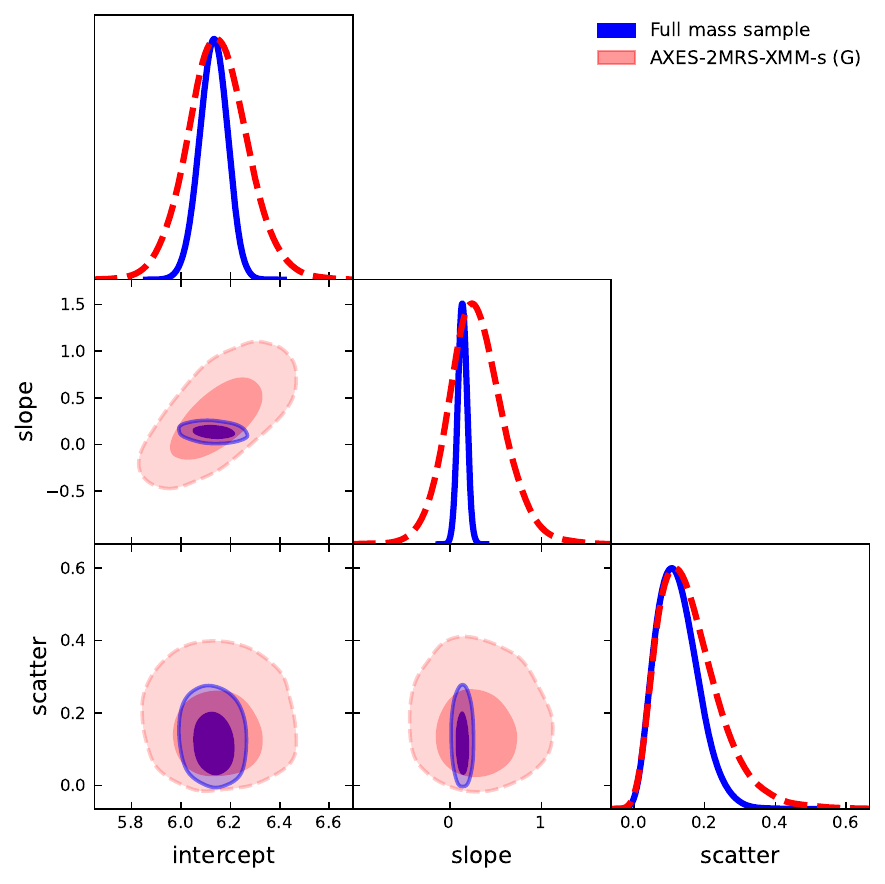}
     \caption{MCMC fitting results with one- and two-dimensional projections of the posteriors of the $\sigma_{\rm v}-M_{10000}$ scaling relation for the full mass sample (blue solid curves and contours) and the Gaussian subset of AXES-2MRS-XMM (red dashed curves and contours). Refer to Table \ref{tab:allresults} for details about each sample. Other details are the same as Fig.~\ref{lx_rassxmm_getdist}. 
     }
     \label{sigmam_getdist}
\end{figure}

As most of our mass measurements correspond to a high overdensity, finding a correspondence to our measurement of velocity dispersion, which traces virial mass, is sensitive to halo concentration. In this section we consider the scatter in the estimate of the central mass this effect introduces and include an effect of complexity in the velocity dispersion profile into consideration.  
The obtained relation for our mass sample is shown in Fig.~\ref{msigma} while the details of the fitting parameters and the one- and two-dimensional posteriors are presented in Table \ref{tab:allresults} and Fig. \ref{sigmam_getdist}, respectively. While we have concentrated on the \textit{XMM-Newton} analysis on galaxy group scales, an analysis of AXES-2MRS clusters can be readily found in the literature \citep[e.g.][]{HICOSMO}. To improve on the mass coverage, we included previously published results from the HIFLUGCS cluster sample \citep{HICOSMO}, in overlap with our velocity dispersion measurements. We report a nearly flat relation for the full mass sample of $\sigma_{\rm v} \propto M_{10000}^{0.13 \pm 0.05}$. The deviations of the slope from the $M^{1/3}$ expectation can be explained either if the contribution of the non-thermal pressure to the mass estimate at the overdensities of 10000 is high in galaxy groups, or if our sample is dominated by the low concentration systems, which we explore below. 
Splitting the systems, based on the Gaussianity of velocity dispersion, does not lead to any detectable changes in the parameters of this scaling relation. 

\subsection{Concentration--X-ray luminosity relation}
\label{cresult}

\begin{figure}
\centering
   \includegraphics[width= \columnwidth]{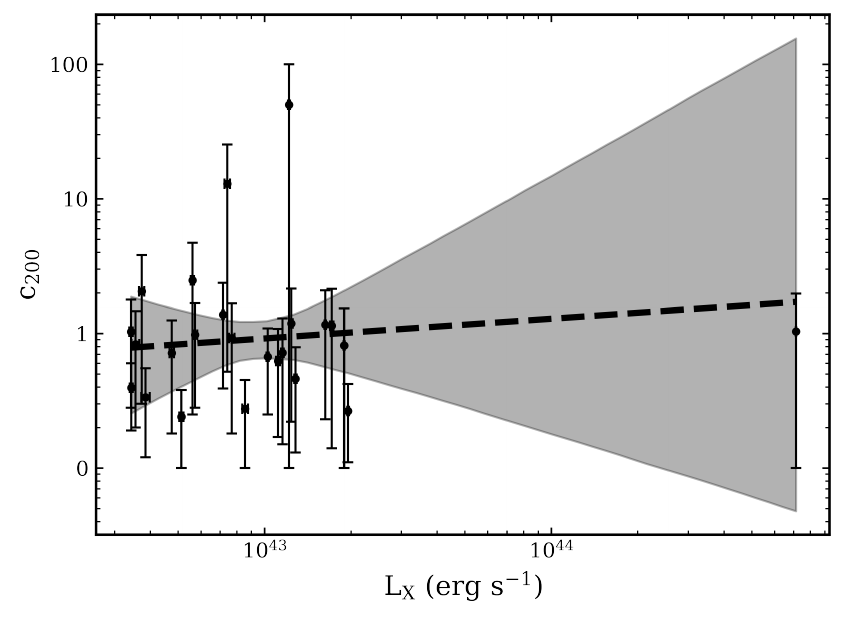}
     \caption{Halo concentration versus X-ray luminosity for the $c_{200}$ sample used in this work.}
     \label{c-lx}
\end{figure}

\begin{figure}
\centering
   \includegraphics[width= \columnwidth]{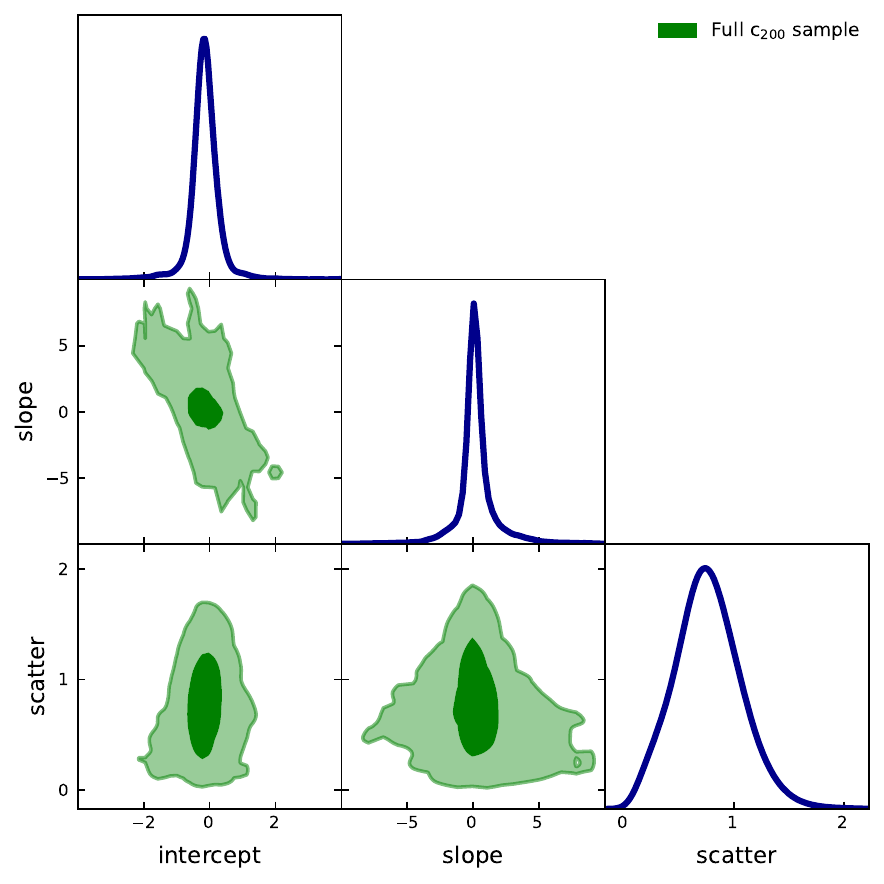}
     \caption{MCMC fitting results with one- and two-dimensional projections of the posteriors of the $c_{200}-L_{\rm X}$ scaling relation for the full $c_{200}$ sample. Other details are the same as Fig.~\ref{lx_rassxmm_getdist}. 
     }
     \label{clx_getdist}
\end{figure}

Owing to the hypersensitivity of the concentration calculation to variations in the velocity dispersion, coupled with the presence of relatively large errors in our velocity dispersion measurements, approximately of the order of 20-30\%, the determination of the concentration parameter poses a significant challenge. As a result, we were only able to confidently constrain the concentration of 22 systems within our \textit{XMM-Newton} subsample (see Table~\ref{table: SB-Mass Results}). Additionally, due to the intricate nature of this analysis, we could extend our concentration determinations to only two systems from previously published datasets (Group IDs: 6041 and 6093).
The result of the $c_{200}-L_{\rm X}$ scaling relation is shown in Figs.~\ref{c-lx} and \ref{clx_getdist} and the details of the parameters are shown in Table \ref{tab:allresults}. We used $K = 4$ Gaussian components in fitting this relation with \texttt{linmix} instead of 3 (see Sect. \ref{sec:statmeth}) as it was found to increase the significance level of the parameters by a factor of 2. Remarkably, we measure a low halo concentration across the majority of systems within our \textit{XMM-Newton} subsample, with values below 1. In performing the linear regression analysis, as an input on concentration we described the probability distribution of concentrations in the expected range between 1 and 10 using a log-normal distribution, as most points have asymmetric profiles of the probability distribution function, and of relevance is the correct estimate of this probability in the range of expected solution. Our result on the slope of the scaling of $-0.12 \pm 0.96$ broadly agrees with the expectations based on the numerical modelling of the $c_{\rm 200} - M$ relation on the scales of groups and clusters of galaxies. 

We report a logarithmic intrinsic scatter on the concentration of $S(\ln c_{200}) = 0.76 \pm 0.31$. Within errors, our mean value is slightly higher than the results on the $c_{\rm 200} - M$ relation from cosmological simulations of  $S(\ln c_{200}) \approx 0.33$ \citep{wechsler02, neto07, maccio07, maccio, bhattacharya13, child18}. Observational efforts on the $S(\ln c_{200})$ parameter, despite being limited and mostly inferred from the $c_{\rm 200} - M$ relation, show a fair agreement with the simulations (e.g. \citealp{wojtak, amodeo, umetsu}). In particular, \citet{wojtak} assembled a low-redshift ($z < 0.1$) sample of 41 relaxed and rich ($N_{\rm gal} \geq 70$) clusters from the NASA/IPAC Extragalactic Database (NED) and the WIde field Nearby Galaxy cluster Survey \citep[WINGS]{wings}. The median mass of the sample is $5 \times 10^{14} M_{\odot}$ and the median concentration is 7. They report an intrinsic scatter of $S(\ln c_{200}) = 0.35$. On the other hand, \citet{amodeo} used a high-redshift ($z \geq 0.4$) sample of 47 clusters collected from the \emph{Chandra} public archive and representing the high-mass end of the cluster population with a median mass of $1.3 \times 10^{15} M_{\odot}$. They report a mean log-normal scatter of $S(\ln c_{200}) = 0.32$. Finally, \citet{umetsu} reported a low upper limit on $S(\ln c_{200})$ of 0.24 using the X-ray-selected XXL cluster sample \citep{xxl18}. They claim that this low scatter can be attributed to a selection bias related to the dynamical state of the clusters or a result of overestimation of the $c_{200}$ measurement errors.

It is important to note that \citet{amodeo} has shown that the intrinsic scatter 
decreases with increasing mass. The weak but systematic anti-correlation of mass and intrinsic scatter is also supported by the simulation work of \citet{neto07}. Given the low masses of our \textit{XMM-Newton} subsample, this effect could potentially explain the relatively high mean value of the scatter. Additionally, a larger sample with higher luminosity and concentration ranges can improve the significance of the scatter measurement. 

\section{Conclusions and summary}
\label{conclusion}

In this work, we have presented a new catalogue of AXES-2MRS X-ray galaxy groups that has a selection based on the baryonic content at $M_{500}$, and we examined its properties. We have significantly enhanced the representation of the under-explored low-redshift, low-luminosity galaxy groups. In addition, our sample is prevalent in low-mass X-ray systems ($<$ 10$^{14} M_{\odot}$, see Table \ref{table: SB-Mass Results}), which enhances the completeness of the galaxy group catalogues, potentially addressing the longstanding problem of missing faint low-mass systems.

The main parameter in common between the various subsamples studied here is velocity dispersion. We find that the main sample exhibits a comparable scaling relation between the X-ray luminosity and the velocity dispersion, and in particular, it exhibits a similar scatter. The value of the scatter is high, which is in agreement with the conclusion of \cite{damsted} that the scatter at $z<0.15$ is very large. We observed that using our measurements of intragroup medium temperatures does not resolve this problem, and the scatter is still large. Our use of masses, as opposed to temperature, reduces the scatter, which indicates that feedback effects contribute significantly to the scatter. 
Our reported $\sigma_{\rm v}-kT$ relation is also marginally flatter than the self-similar expectation, which also points to either the importance of non-gravitational heating or the effect of halo concentration. 
Our analysis of the $c_{200}-L_{\rm X}$ relation reveals a large intrinsic scatter that we deem representative of galaxy groups. 
Thus, we conclude that both feedback and halo concentration are at the root of the large scatter of properties of X-ray groups.
We believe that a combination of large scatter and group selection can explain differences in the mean scaling relations for galaxy groups, as all published relations pass through some of the points presented in this study. The main question that remains open is what is the right balance of groups {to be included in the samples of X-ray properties}. 

We note that there is a similarity in the slope and normalisation of the scaling relation between the velocity dispersion and the X-ray luminosity between the AXES-2MRS galaxy group sample and the distant COSMOS sample, which was obtained using a similar X-ray detection technique. In addition, the \textit{XMM-Newton}-observed subsample is comparable to the full AXES-2MRS catalogue in this relation and can be safely held as representative of the full AXES-2MRS sample. 

  {As evident from Table \ref{table: SB-Mass Results}, we could not resolve the core radii of AXES-2MRS groups with the one-dimensional beta model, which is a common issue in galaxy groups and has also been observed in the X-GAP group sample (Dominique Eckert, private communication 2024). The emissivity profiles of galaxy groups tend to rise, following a power law, up to the groups' centre without a well-defined core. As mentioned in Sect. \ref{sectionsb}, we are only interested in the power-law behaviour of the beta model at large radii; however, we conclude that the one-dimensional beta model is not an ideal choice at the galaxy group scale.}

Using \texttt{linmix} with intrinsic scatter, the fitting parameters come out to be consistent with our results using orthogonal distance regression. However, the reported uncertainties {of {\tt linmix}} are larger by a factor of 1.5-2. This might help resolve the previously found tensions in the fit parameters obtained without the intrinsic scatter. As we have demonstrated, the scatter of the scaling relations is a meaningful parameter that allows one to access the physics of galaxy groups.

As discussed in this paper, the differences in the literature results on the scaling relations might be associated with the selection of the sample. 
Our study has employed the largest angular scales ever considered in X-ray source identification. Our results on the scaling relations are within the difference between the literature results, which limits the scope of contribution from the `known unknowns'.
The number of AXES sources combined with the spatial scales employed in searching for the X-ray emission approaches the source confusion limit. While future detailed studies of AXES sources will be beneficial, deeper surveys will have to use smaller spatial scales when searching for X-ray emission, which will lead to different selection effects. With the advances in hydrodynamical simulations, modelling of the X-ray emission from the outskirts of groups and clusters of galaxies has become reliable, which makes AXES the most suitable sample for comparison to simulations.

\begin{acknowledgements}
We thank the anonymous referee for the useful comments on the manuscript. ET acknowledges the Estonian Research Council grant PRG1006 and the Centre of Excellence ‘Foundations of the Universe’ (TK202) funded by the Ministry of Education and Research. The authors thank an anonymous referee for useful comments, which helped improve the manuscript. AF and KH thank Dominique Eckert for the discussions. AF thanks Lorenzo Lovisari, Fabio Gastaldello and Mariachiara Rossetti for organising the conference on "GALAXY GROUPS IN THE ERA OF EROSITA AND EUCLID: A MULTIWAVELENGTH VIEW", where a revision of this paper was refined.
\end{acknowledgements}

\bibliographystyle{aa} 
\bibliography{ref} 

\begin{appendix}

\section{Details of the \textit{XMM-Newton} observations}
    
Table \ref{table: Optical Results} lists the optical parameters of our \textit{XMM-Newton} subsample. The positions (RA and Dec) correspond to the X-ray peak emission centres taken as a centroid of a wavelet reconstruction of the 0.5--2 keV image on scales 0.5-4 arcmin. Also shown are the median group redshifts, the optical line-of-sight velocity dispersions determined using the Gapper method, and the number of member galaxies.

\begin{table}[hbt!]
\caption{\textit{XMM-Newton} subsample of AXES-2MRS groups.}             
\label{table: Optical Results}
\centering                          
\begin{tabular}{c c c l l c}        
\hline\hline              
Group ID\!\!\!\! &\multicolumn{1}{c}{RA} & \multicolumn{1}{c}{Dec} & \multicolumn{1}{c}{$z_{\rm{med}}$} & \multicolumn{1}{c}{$\sigma$$_{v}$} & $N_{\rm{gal}}$\\
 {\footnotesize 2MRS} & {\footnotesize (J2000)} & {\footnotesize (J2000)} & & \multicolumn{1}{c}{\footnotesize (km s$^{-1}$)} &\\
\hline                  
   361 & 16.853 & 32.399 & 0.016 & 370 $\pm$ 83 & 18\\
   505 & 21.445 & --1.395 & 0.0171 & 438 $\pm$ 76 & 29\\
   827 & 35.780 & 42.986 & 0.0197 & 510 $\pm$ 117 & 17\\
   859 & 36.401 & 36.961 & 0.0353 & 480 $\pm$ 167 & 8\\
   1571 & 61.642 & 30.379 & 0.0179 & 370 $\pm$ 103 & 12\\ 
   1830 & 74.732 & --0.484 & 0.0144 & 320 $\pm$ 85 & 13\\
   2009 & 86.370 & --25.936 & 0.0388 & 925 $\pm$ 201 & 19\\
   2161 & 96.162 & --37.337 & 0.0329 & 569 $\pm$ 151 & 13\\
   2533 & 315.436 & --13.311 & 0.0278 & 409 $\pm$ 125 & 10\\
   2541 & 117.844 & 50.202 & 0.0229 & 521 $\pm$ 170 & 9\\
   2657 & 125.154 & 21.072 & 0.017 & 373 $\pm$ 104 & 12\\
   3551 & 164.543 & 1.612 & 0.0405 & 339 $\pm$ 110 & 9\\
   3718 & 170.612 & 24.296 & 0.027 & 478 $\pm$ 166 & 8\\
   4050 & 182.018 & 25.239 & 0.023 & 319 $\pm$ 98 & 10\\
   4808 & 202.351 & 11.765 & 0.0239 & 347 $\pm$ 101 & 11\\
   5089 & 210.908 & --33.983 & 0.0139 & 238 $\pm$ 66 & 12\\
   5841 & 244.338 & 34.903 & 0.0303 & 313 $\pm$ 96 & 10\\
   5914 & 247.417 & 40.826 & 0.0318 & 591 $\pm$ 111 & 25\\
   6015 & 254.498 & 27.858 & 0.0345 & 309 $\pm$ 108 & 8\\
   6116 & 260.202 & --1.039 & 0.0286 & 538 $\pm$ 137 & 14\\
   6407 & 281.827 & --63.332 & 0.015 & 471 $\pm$ 83 & 28\\
   6666 & 304.458 & --70.819 & 0.0131 & 420 $\pm$ 137 & 9\\
   6916 & 316.840 & --25.459 & 0.0359 & 577 $\pm$ 201 & 8\\
   7427 & 348.942 & --2.389 & 0.0234 & 473 $\pm$ 145 & 19\\
   7727 & 181.04 & 20.293 & 0.0248 & 445 $\pm$ 94 & 20\\
\hline
\end{tabular}
\tablefoot{
RA and Dec are coordinates of the peak X-ray emission centres.
}
\end{table}

In Table \ref{table: Observations}, we provide a detailed summary of the observations for our \textit{XMM-Newton} subsample. The table includes Group ID (2MRS group identifier from \citet{tempel18}), OBS-ID (\textit{XMM-Newton} observation identifier), DATE-OBS (date of the observation), Clean-EXP (clean exposure time), kT (X-ray gas temperature), and $L_{\rm X}$ (0.1-2.4 keV X-ray luminosity) measured by \textit{XMM-Newton} (for $L_{\rm X}$ measured by ROSAT, refer to Fig. \ref{lxxmmrass}). It also includes a$_{spec}$, b$_{spec}$, and $\theta$ which are the semi-major axis, semi-minor axis, and position angle of the elliptical extraction regions used in the spectral analysis, respectively (see the green regions in Fig. \ref{zoo1}).

Table \ref{table: SB-Mass Results} complements Table \ref{table: Observations} with more X-ray properties obtained from the \textit{XMM-Newton} observations. In particular, it contains the details of the surface brightness profile fitting using a single $\beta$-model and mass estimates. The columns are $\beta$ (slope of the surface brightness profile), R$_{kT}$ (the outer radius of the initial mass estimate), $\Delta_{kT}$ (the overdensity of the measurement at R$_{kT}$), M$_{\Delta_{R_{kT}}}$ (the mass estimate at the initial overdensity $\Delta_{kT}$), R$_{C}$ (the core radius of the $\beta$ model), DS (the distance scale),  M$_{10000}$ (mass estimate at the overdensity covered by the data), and c$_{200}$ (halo concentration).

\begin{table*}[b]
\caption{XMM observations and basic properties of AXES-2MRS.}             
\label{table: Observations}
\centering                          
\begin{tabular}{l c c c c c c c r}        
\hline\hline                
\!\!\!\!Group ID & OBS-ID & DATE-OBS & \multicolumn{1}{c} {Clean-EXP}\!\!\!\!\!\! &  $k\,T$ & $L_{\rm X}$ & $a_{\rm spec}$ & $b_{\rm spec}$& $\theta$\ \ \ \ \\
 2MRS & \textit{XMM-Newton} & (UTC) & \multicolumn{1}{c}{(ks)}\!\! & (keV) & (10$^{43}$ erg s$^{-1}$) & (arcmin) & (arcmin) & \!\!\!(deg) \\
\hline                  
   361\tablefootmark{M} & 0551720101 & 2008-07-01 & 24.1 & 1.898 $\pm$ 0.119 & 0.716 $\pm$ 0.004 & 4.5 & 3.5 & 82.9\\ 
   505\tablefootmark{OS} & 0743700201 & 2015-01-09 & 69.4 & 1.417 $\pm$ 0.077 & 0.344 $\pm$ 0.004 & 2.7 & 1.4 & 33.9\\
   827 & 0002970201 & 2002-02-05 & 13.5 & 2.159 $\pm$ 0.121 & 1.027 $\pm$ 0.009 & 4.7 & 3.9 & 277.7\\
   859 & 0863880401 & 2020-07-21 & 15.1 & 1.708 $\pm$ 0.110 & 0.769 $\pm$ 0.02 & 2.8 & 2.2 & 278.4\\
   1571 & 0883620101 & 2021-09-13 & 7.2 & 1.490 $\pm$ 0.251 & 0.355 $\pm$ 0.011 & 2.9 & 1.9 & 7.9\\ 
   1830 & 0673180301 & 2012-02-24 & 3.1 & 1.170 $\pm$ 0.005 & 0.573 $\pm$ 0.01 & 3.2 & 2.6 & 309.7\\
   2009 & 0302030101 & 2006-02-17 & 27.2 & 3.132 $\pm$ 0.156 & 1.956 $\pm$ 0.014 & 4.8 & 3.6 & 341.7\\
   2161 & 0800761301 & 2017-10-11 & 14.1 & 1.651 $\pm$ 0.045 & 1.284 $\pm$ 0.013 & 3.0 & 2.5 & 344.4\\
   2533 & 0864052501 & 2021-04-23 & 7.2 & 1.455 $\pm$ 0.046 & 1.241 $\pm$ 0.016 & 4.9 & 3.6 & 272.4\\
   2541\tablefootmark{OS} & 0800761001 & 2018-04-19 & 8.4 & 1.477 $\pm$ 0.039 & 1.115 $\pm$ 0.018 & 2.9 & 2.1 & 29.5\\
   2657 & 0108860501 & 2001-10-15 & 15.8 & 1.516 $\pm$ 0.054 & 0.343 $\pm$ 0.005 & 3.3 & 3.0 & 4.3\\
   3551 & 0601930101 & 2009-05-26 & 18.1 & 2.090 $\pm$ 0.115 & 1.218 $\pm$ 0.01 & 4.6 & 3.1 & 334.2\\
   3718 & 0112270301 & 2001-12-02 & 6.5 & 1.563 $\pm$ 0.057 & 1.156 $\pm$ 0.013 & 3.7 & 3.1 & 57.1\\
   4050 & 0151400201 & 2003-05-26 & 8.2 & 1.361 $\pm$ 0.179 & 0.373 $\pm$ 0.008 & 3.3 & 2.1 & 291.6\\
   4808 & 0041180801 & 2001-12-30 & 13.7 & 1.255 $\pm$ 0.073 & 0.56 $\pm$ 0.008 & 3.5 & 1.6 & 283.7\\
   5089\tablefootmark{M} & 0741930101 & 2014-07-25 & 90.5 & 2.271 $\pm$ 0.060 & 0.86 $\pm$ 0.004 & 4.8 & 3.4  & 313.0\\
   5841 & 0800761701 & 2018-01-16 & 7.1 & 1.687 $\pm$ 0.179 & 0.742 $\pm$ 0.016 & 2.7 & 1.8 & 282.0\\
   5914 & 0203710201 & 2004-09-07 & 3.5 & 1.175 $\pm$ 0.037 & 0.856 $\pm$ 0.02 & 3.0 & 2.9 & 0.0\\
   6015 & 0654800201 & 2010-08-26 & 38.4 & 1.877 $\pm$ 0.097 & 2.14 $\pm$ 0.014 & 4.7 & 3.8 & 291.7\\
   6116 & 0400930101 & 2006-08-25 & 23.1 & 2.265 $\pm$ 0.210 & -- & 6.8 & 5.2 & 321.3\\
   6407\tablefootmark{M} & 0405550401 & 2006-09-07 & 17.1 & 1.168 $\pm$ 0.041 & 0.513 $\pm$ 0.008 & 2.8 & 1.9 & 23.9\\
   6666\tablefootmark{OS} & 0022340101 & 2002-03-31 & 8.8 & 1.094 $\pm$ 0.051 & 0.475 $\pm$ 0.009 & 2.8 & 1.7 & 40.4\\
   6916 & 0741581601 & 2014-10-22 & 5.1 & 2.213 $\pm$ 0.308 & 1.719 $\pm$ 0.025 & 4.7 & 5.2 & 77.0\\
   7427 & 0501110101 & 2007-11-22 & 24.5 & 1.635 $\pm$ 0.047 & 1.631 $\pm$ 0.009 & 4.6 & 3.7 & 41.0\\
   7727\tablefootmark{OS} & 0112270601 & 2003-01-02 & 2.43 & 1.055 $\pm$ 0.119 & 0.385 $\pm$ 0.013 & 2.6 & 2.3 & 275.9\\
\hline                                   
\end{tabular}

\tablefoot{
\tablefoottext{M}{Groups showing merging behaviour.} 
\tablefoottext{OS}{Over-split groups. Refer to Sect. \ref{sectionxmm} for more details.} 
}
\end{table*}

\begin{table*}[t]
\caption{Surface brightness profiles, mass estimates, and halo concentrations of our \textit{XMM-Newton} subsample.}             
\label{table: SB-Mass Results}
\centering                          
\begin{tabular}{c c r l c c r c c c c}        
\hline\hline                
Group ID & $\beta$ & $R_{\rm kT}$\ \  & \ \ $R_{\rm C}$ & DS & $M_{\Delta_{R_{\rm kT}}}$ & $\Delta_{R_{\rm kT}}$\ \  &  $M_{10000}$ & $c_{200}$\\

 {\footnotesize 2MRS} & & (kpc)\ & \ \ ($''$) & (kpc/$''$) & (10$^{13}$ M$_{\odot}$) & & (10$^{13}$ M$_{\odot}$) & \\
\hline                  
   361 & 0.337 $\pm$ 0.042 & 83.7 & 0.001 & 0.346 & 0.65 $\pm$ 0.12 & 19161 & 0.72 $\pm$ 0.19 & 1.38 $\pm$ 0.99\\ 
   505 & 0.363 $\pm$ 0.01 & 47.2 & 0.0391 & 0.366 &  0.3 $\pm$ 0.07 & 48373 & 0.37 $\pm$ 0.12 &   0.395 $\pm$ 0.205\\
   827 & 0.337 $\pm$ 0.008 & 130.1 & 0.001 & 0.502 & 1.16 $\pm$ 0.09 & 8988 & 1.14 $\pm$ 0.13 & 0.67 $\pm$ 0.42\\
   859 & 0.354 $\pm$ 0.008 & 107.9 & 0.001 & 0.714 & 0.8 $\pm$ 0.08 & 10710 & 0.8 $\pm$ 0.12 & 0.925 $\pm$ 0.745\\
   1571 & 0.313 $\pm$ 0.017 & 54.7 & 0.00002 & 0.372 & 0.31 $\pm$ 0.1 & 32663 & 0.37 $\pm$ 0.17 & 0.83 $\pm$ 0.63\\ 
   1830 & 0.363 $\pm$ 0.001 & 51.8 & 0.039 & 0.296 & 0.27 $\pm$ 0.05 & 33594 & 0.32 $\pm$ 0.08 & 0.98 $\pm$ 0.7\\
   2009 & 0.374 $\pm$ 0.099 & 202.4 & 0.003 & 0.795 & 2.89 $\pm$ 0.37 & 5866 & 2.66 $\pm$ 0.47 & 0.265 $\pm$ 0.165\\
   2161 & 0.390 $\pm$ 0.102 & 107.5 & 0.003 & 0.649 & 0.85 $\pm$ 0.2 & 11512 & 0.86 $\pm$ 0.29 & 0.46 $\pm$ 0.33\\
   2533 & 0.334 $\pm$ 0.019 & 149.1 & 0.001 & 0.578 & 0.89 $\pm$ 0.06 & 4537 & 0.78 $\pm$ 0.08 & 1.185 $\pm$ 0.965\\
   2541 & 0.484 $\pm$ 0.009 & 69.1 & 0.019 & 0.455 & 0.6 $\pm$ 0.08 & 31206 & 0.71 $\pm$ 0.13 & 0.625 $\pm$ 0.455\\
   2657 & 0.371 $\pm$ 0.006 & 62.1 & 0.003 & 0.328 & 0.43 $\pm$ 0.07 & 30624 & 0.5 $\pm$ 0.12 & 1.03 $\pm$ 0.75\\
   3551 & 0.301 $\pm$ 0.014 & 183.1 & 0.0002 & 0.778 & 1.41 $\pm$ 0.09 & 3867 & 1.21 $\pm$ 0.11 & 50.05 $\pm$ 49.95\\
   3718 & 0.334 $\pm$ 0.01 & 106.7 & 0.016 & 0.521 & 0.68 $\pm$ 0.06 & 9538 & 0.68 $\pm$ 0.08 & 0.72 $\pm$ 0.57\\
   4050 & 0.331 $\pm$ 0.011 & 75.5 & 0.001 &  0.455 & 0.42 $\pm$ 0.08 & 16473 & 0.45 $\pm$ 0.12 & 2.055 $\pm$ 1.755\\
   4808 & 0.435 $\pm$ 0.137 & 76.4 & 0.01 & 0.468 &  0.51 $\pm$ 0.2 & 19484 & 0.56 $\pm$ 0.31 & 2.475 $\pm$ 2.225\\
   5089\textsuperscript{*} & 0.268 $\pm$ 0.004 & 68.4 & 0.001 & 0.274 & 0.51 $\pm$ 0.07 & 27368 & 0.58 $\pm$ 0.14 & --\\
   5841 & 0.386 $\pm$ 0.01 & 84.5 & 0.001 & 0.614 &  0.67 $\pm$ 0.11 & 18874 & 0.74 $\pm$ 0.17 & 12.91 $\pm$ 12.39\\
   5914 & 0.457 $\pm$ 0.169 & 112.2 & 0.264 & 0.634 & 0.74 $\pm$ 0.22 & 8824 & 0.72 $\pm$ 0.31 & 0.275 $\pm$ 0.175\\
   6015 & 0.307 $\pm$ 0.001 & 175.7 & 0.006 & 0.685 & 1.24 $\pm$ 0.07 & 3872 & 1.07 $\pm$ 0.08 & --\\
   6116\textsuperscript{\textdagger} & -- & -- & -- & -- & -- & -- & -- & --\\
   6407 & 0.319 $\pm$ 0.044 & 43.9 & 0.0004 & 0.306 & 0.2 $\pm$ 0.07 & 40579 & 0.24 $\pm$ 0.12 & 0.24 $\pm$ 0.14\\
   6666 & 0.527 $\pm$ 0.006 & 42.7 & 0.02 & 0.307 & 0.3 $\pm$ 0.07 & 66548 & 0.38 $\pm$ 0.17 & 0.715 $\pm$ 0.535\\
   6916 & 0.346 $\pm$ 0.031 & 222.4 & 0.001 & 0.748 & 2.08 $\pm$ 0.16 & 3187 & 1.73 $\pm$ 0.19 & 0.67 $\pm$ 1.0\\
   7427 & 0.418 $\pm$ 0.066 & 127.2 & 0.007 & 0.508  & 1.06 $\pm$ 0.14 & 8774 & 1.04 $\pm$ 0.2 & 1.16 $\pm$ 0.93\\
   7727 & 0.349 $\pm$ 0.072 & 70.4 & 0.022 & 0.478 & 0.32 $\pm$ 0.1 & 15462 & 0.34 $\pm$ 0.15 & 0.335 $\pm$ 0.215\\
\hline      
\end{tabular}

\tablefoot{
\tablefoottext{\textdagger}{Has a large offset relative to the \textit{XMM-Newton} pointing.}
\tablefoottext{*}{Shows a significant merging behaviour (see Sect. \ref{sectionxmm})}. Refer to Sect. \ref{conclusion} for a discussion about the artificially small  R$_{\rm C}$ values.}
\end{table*}

\section{\textit{XMM-Newton} images of AXES-2MRS groups}

The X-ray images for the XMM-subsample used in this work are shown in Fig. \ref{zoo1}. The size of the spectral extraction regions (green dashed ellipses) are listed in Table \ref{table: Observations}.

\begin{figure*}
    \centering
    \begin{subfigure}[b]{0.33\textwidth}
        \includegraphics[width=\textwidth]{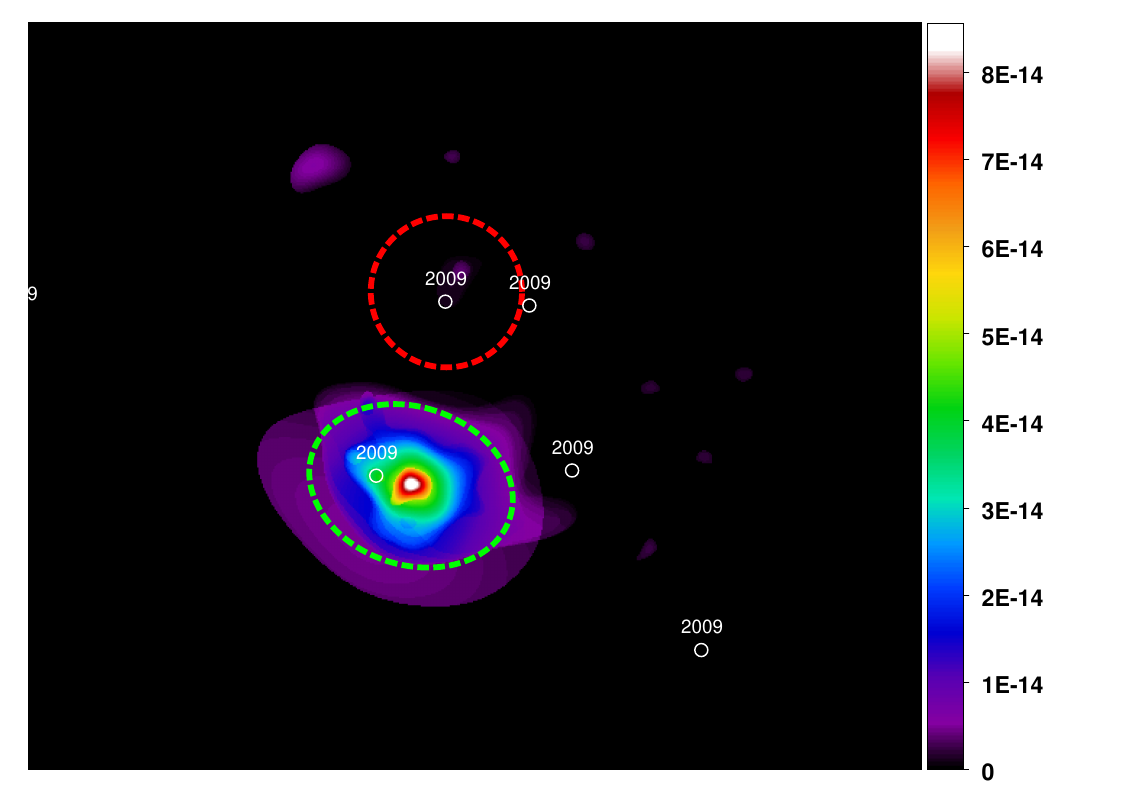}
    \end{subfigure}
    \begin{subfigure}[b]{0.33\textwidth}
        \includegraphics[width=\textwidth]{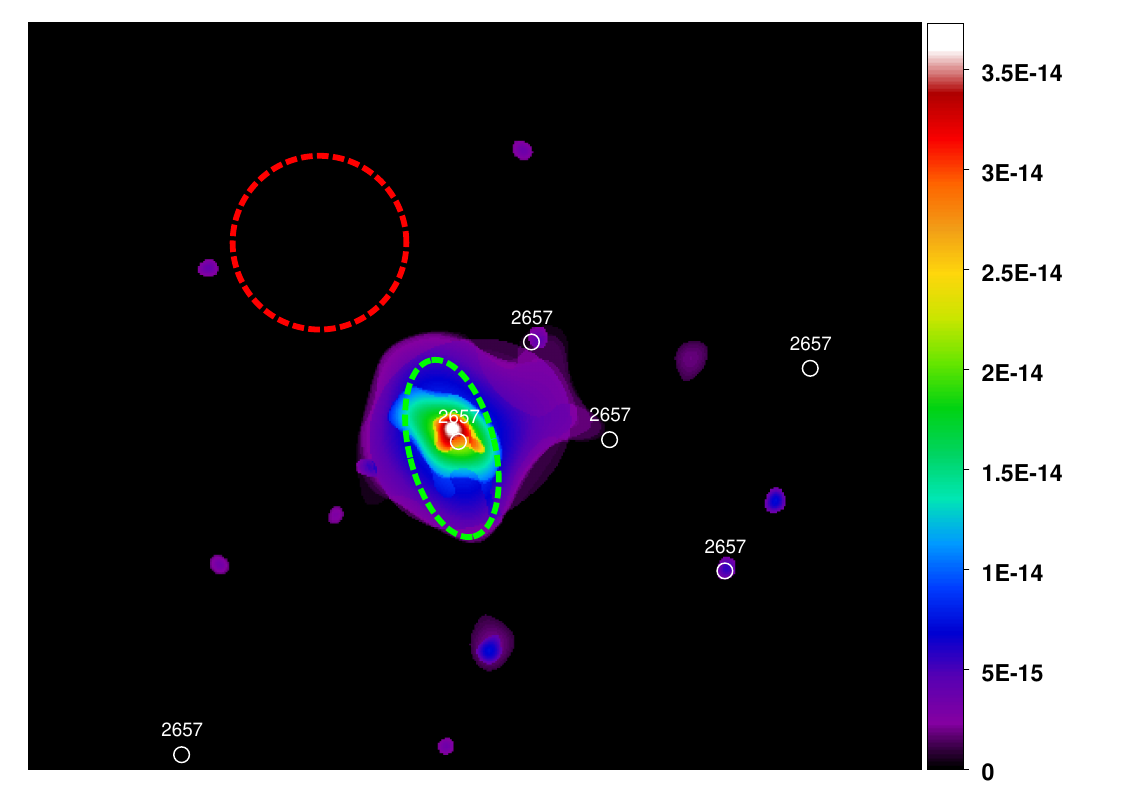}
    \end{subfigure}
    \begin{subfigure}[b]{0.33\textwidth}
        \includegraphics[width=\textwidth]{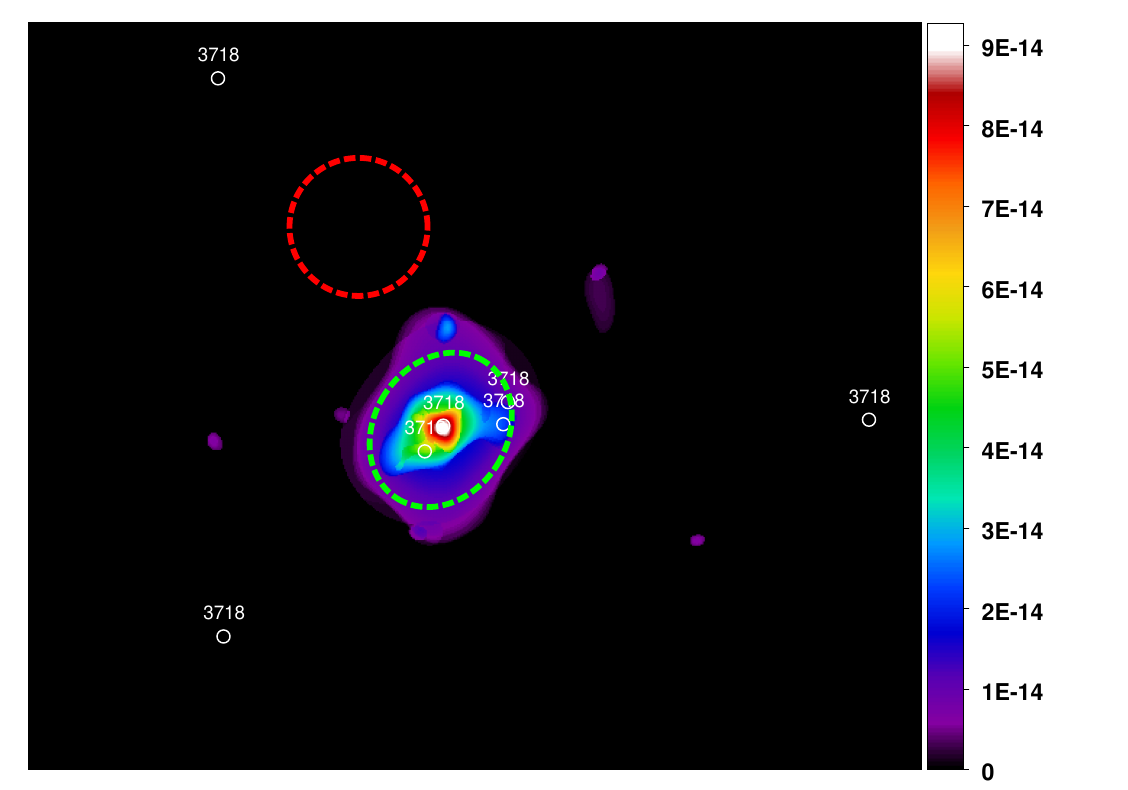}
    \end{subfigure}
    \begin{subfigure}[b]{0.33\textwidth}
         \includegraphics[width=\textwidth]{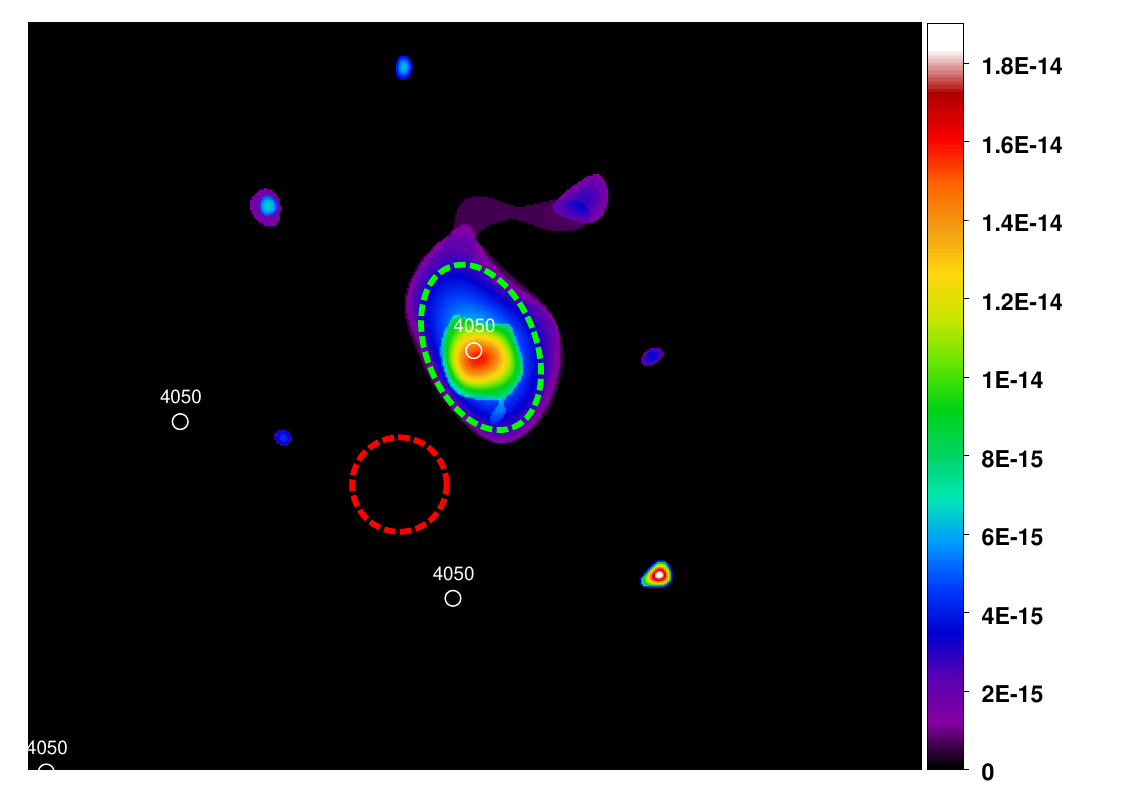}
    \end{subfigure}
    \begin{subfigure}[b]{0.33\textwidth}
        \includegraphics[width=\textwidth]{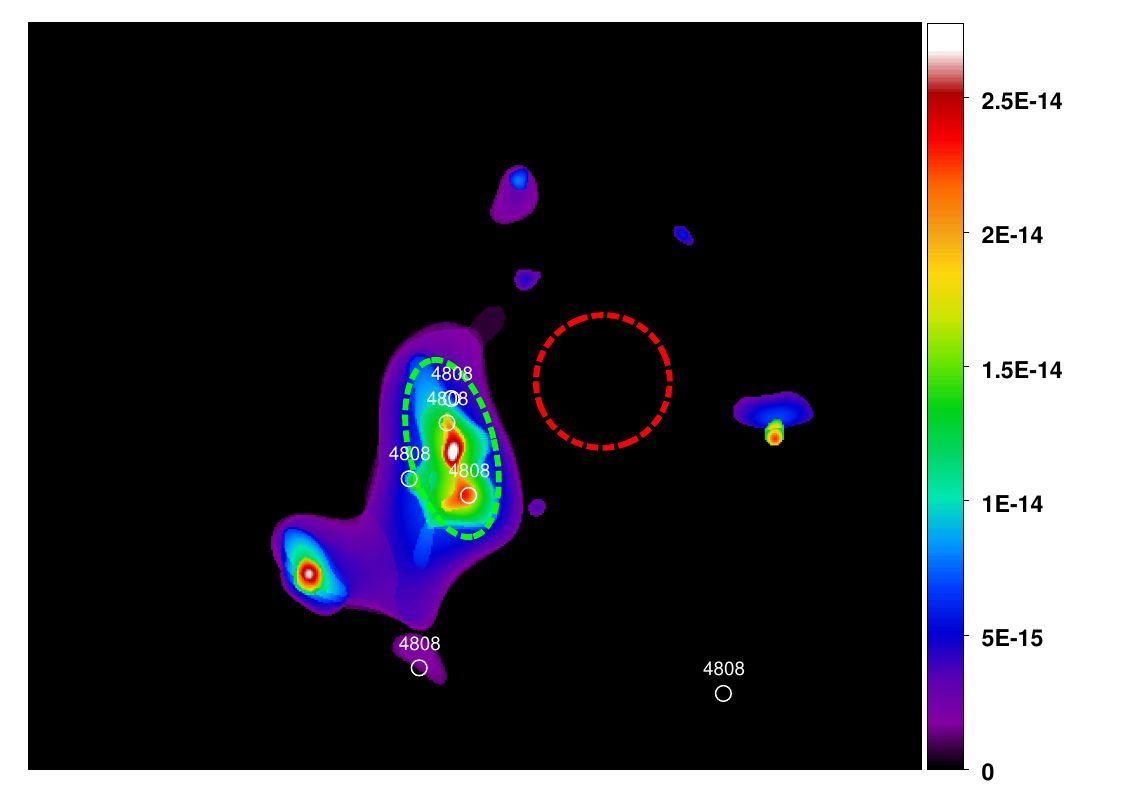}
    \end{subfigure}
    \begin{subfigure}[b]{0.33\textwidth}
        \includegraphics[width=\textwidth]{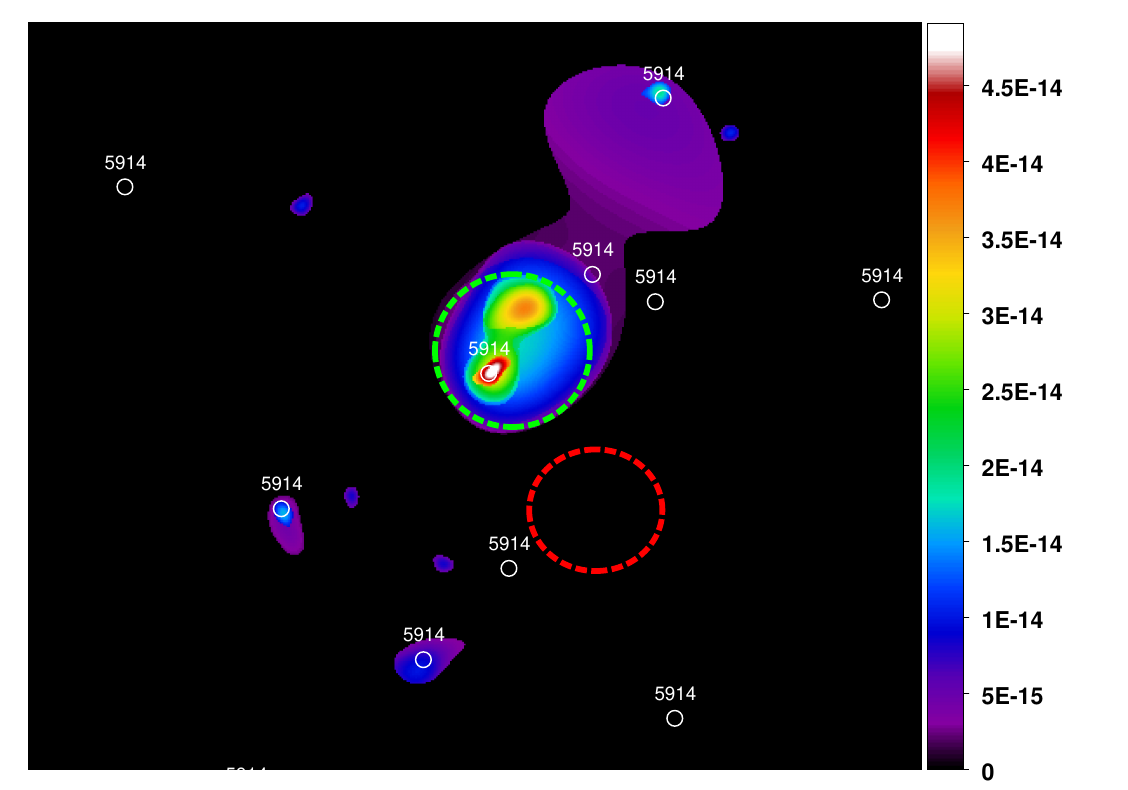}
    \end{subfigure}
    \begin{subfigure}[b]{0.33\textwidth}
        \includegraphics[width=\textwidth]{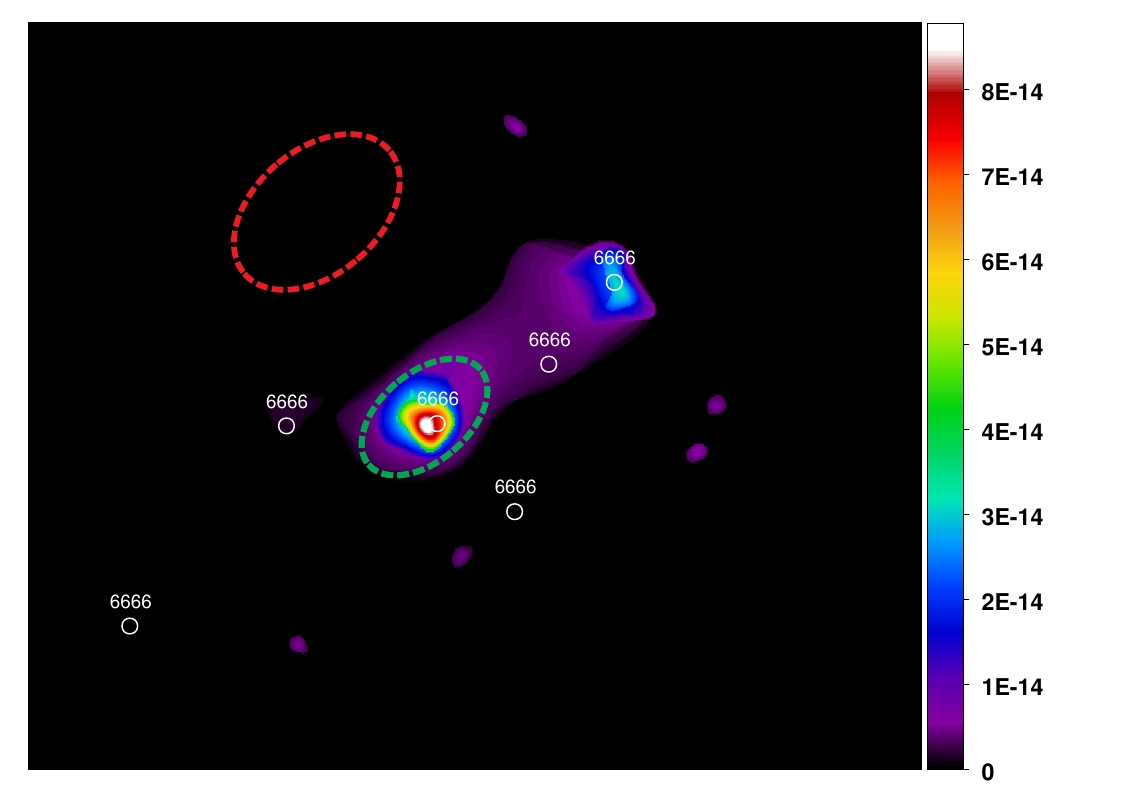}
    \end{subfigure}
    \begin{subfigure}[b]{0.33\textwidth}
        \includegraphics[width=\textwidth]{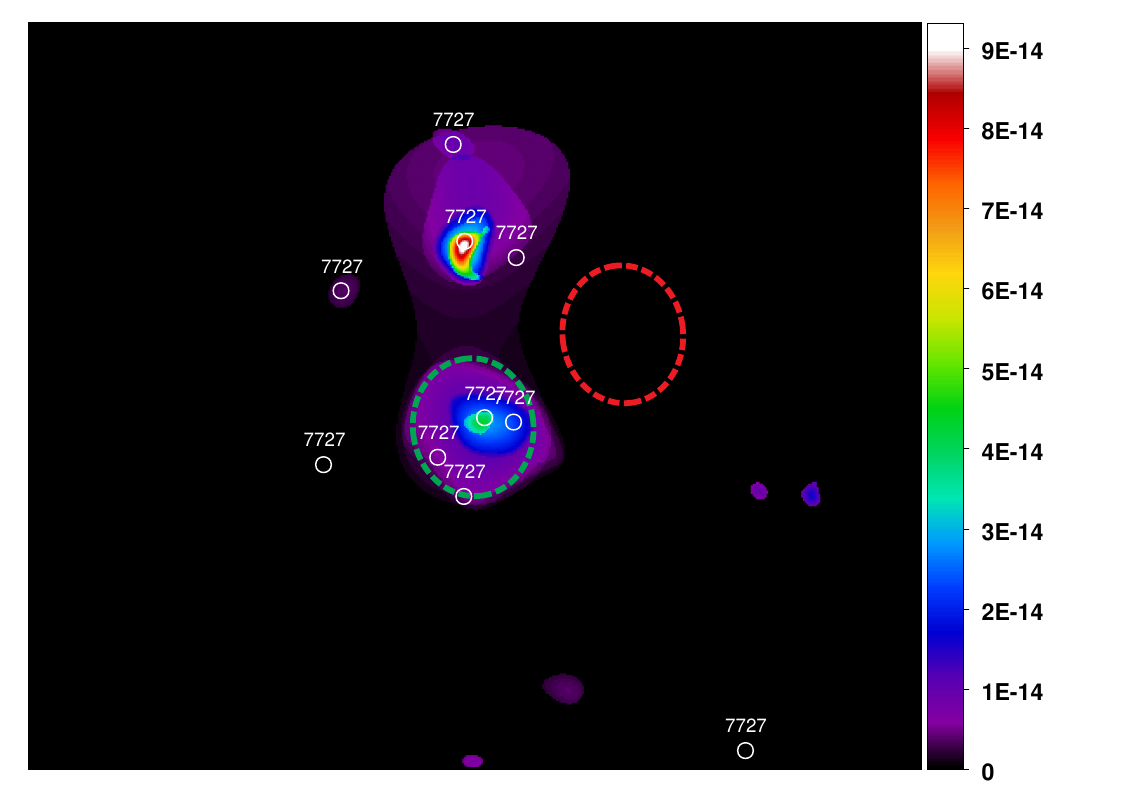}
    \end{subfigure}
    \begin{subfigure}[b]{0.33\textwidth}
        \includegraphics[width=\textwidth]{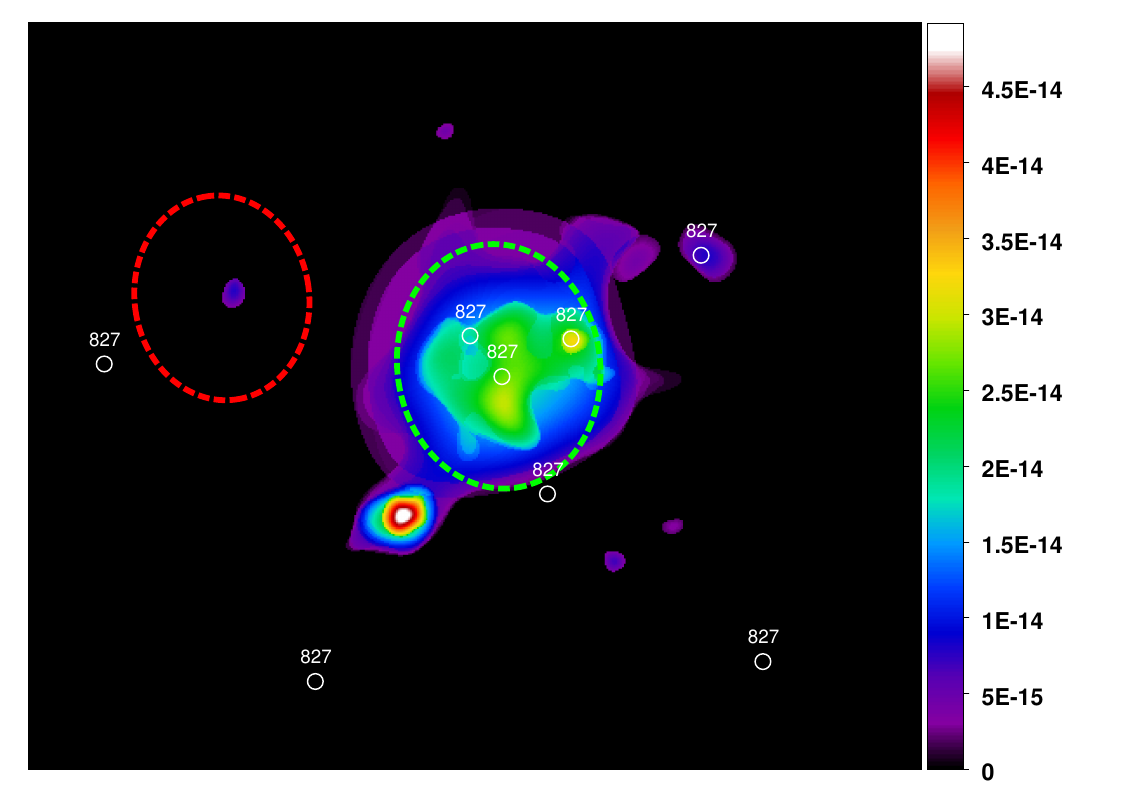}
    \end{subfigure}
    \begin{subfigure}[b]{0.33\textwidth}
        \includegraphics[width=\textwidth]{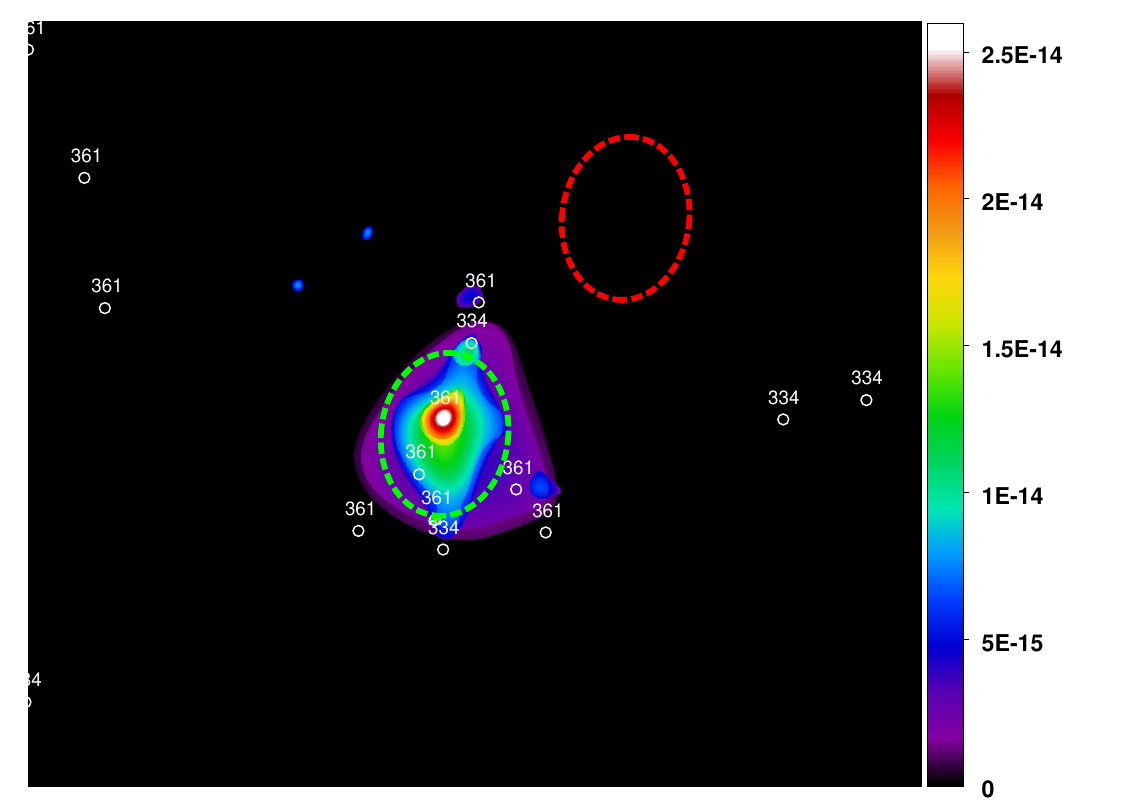}
    \end{subfigure}
    \begin{subfigure}[b]{0.33\textwidth}
        \includegraphics[width=\textwidth]{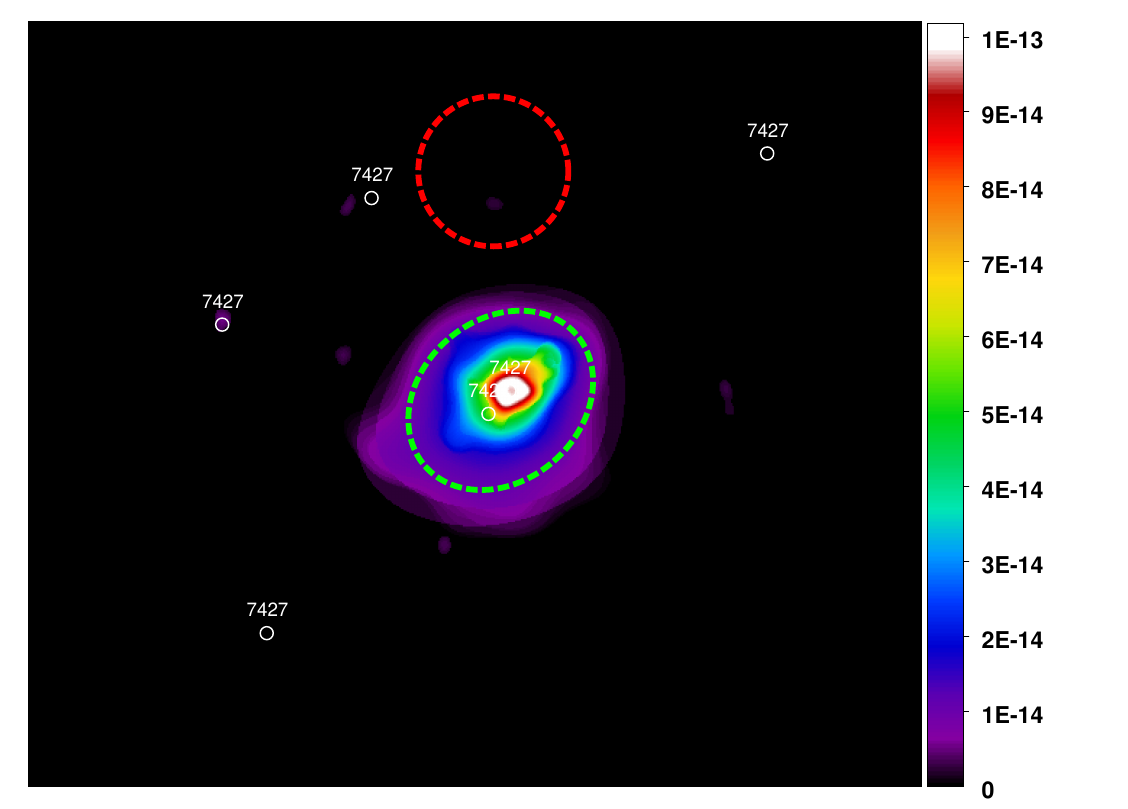}
    \end{subfigure}
    \begin{subfigure}[b]{0.33\textwidth}
        \includegraphics[width=\textwidth]{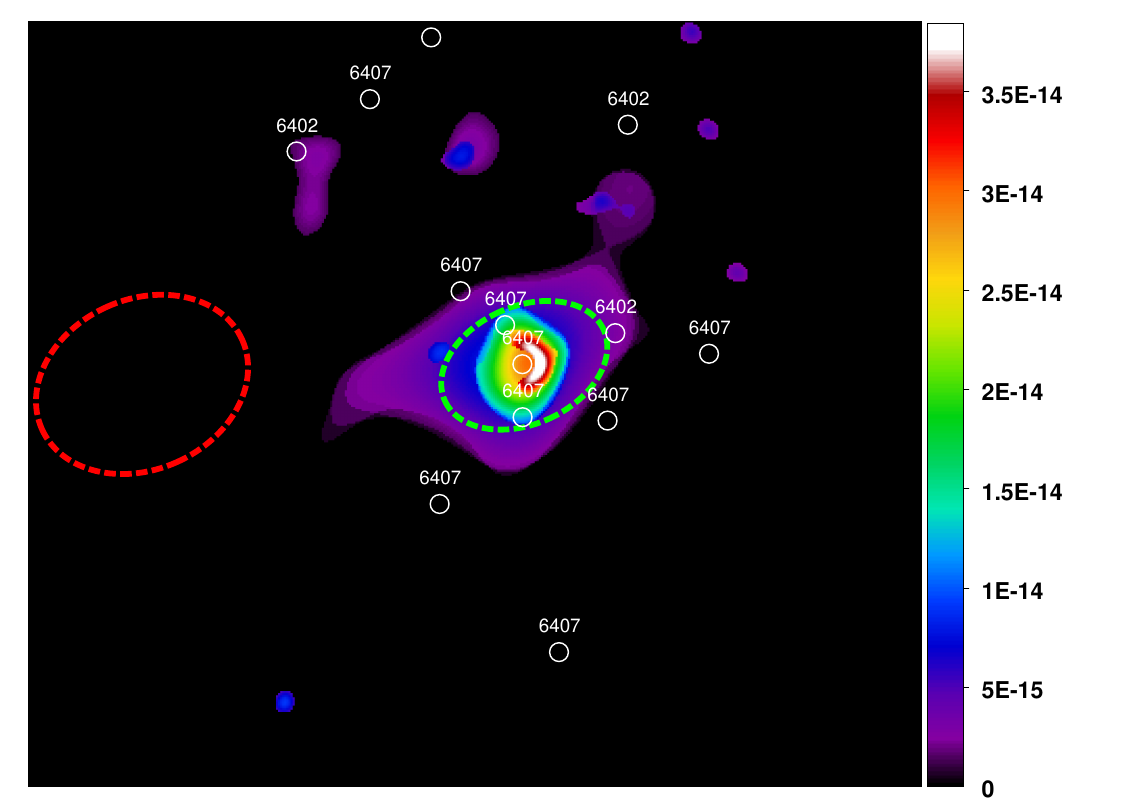}
    \end{subfigure}
    \begin{subfigure}[b]{0.33\textwidth}
        \includegraphics[width=\textwidth]{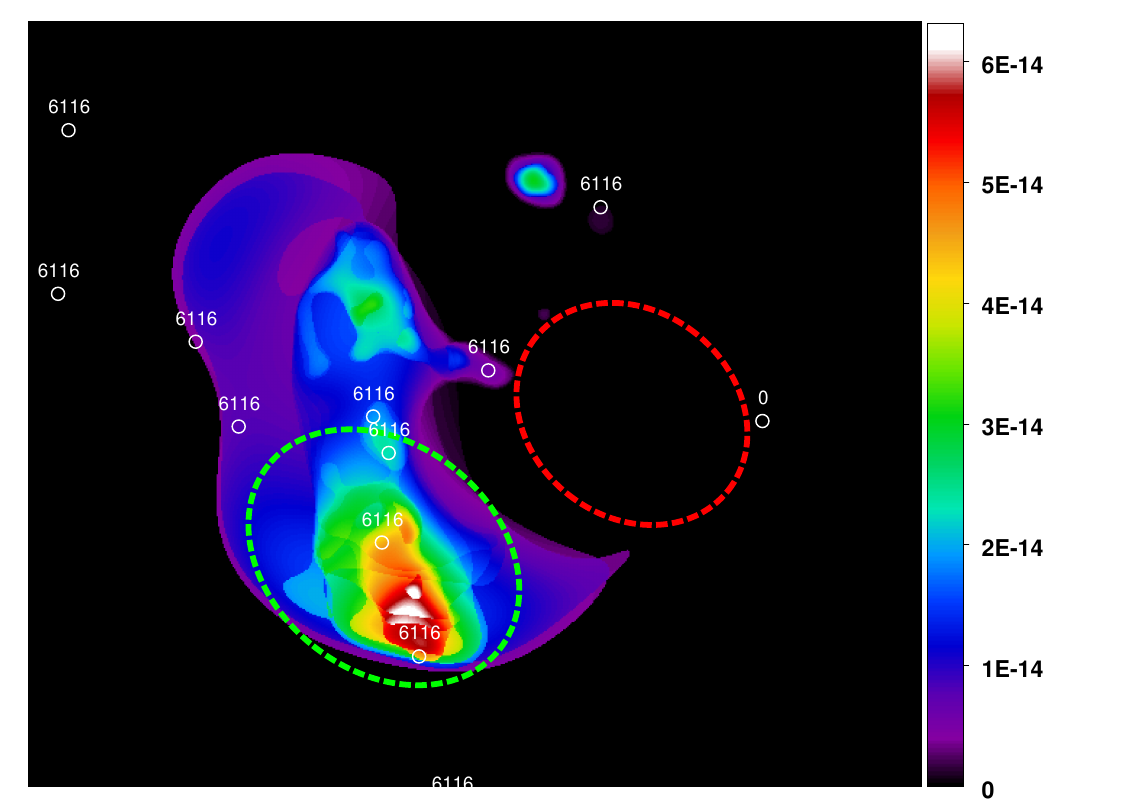}
    \end{subfigure}
    \begin{subfigure}[b]{0.33\textwidth}
        \includegraphics[width=\textwidth]{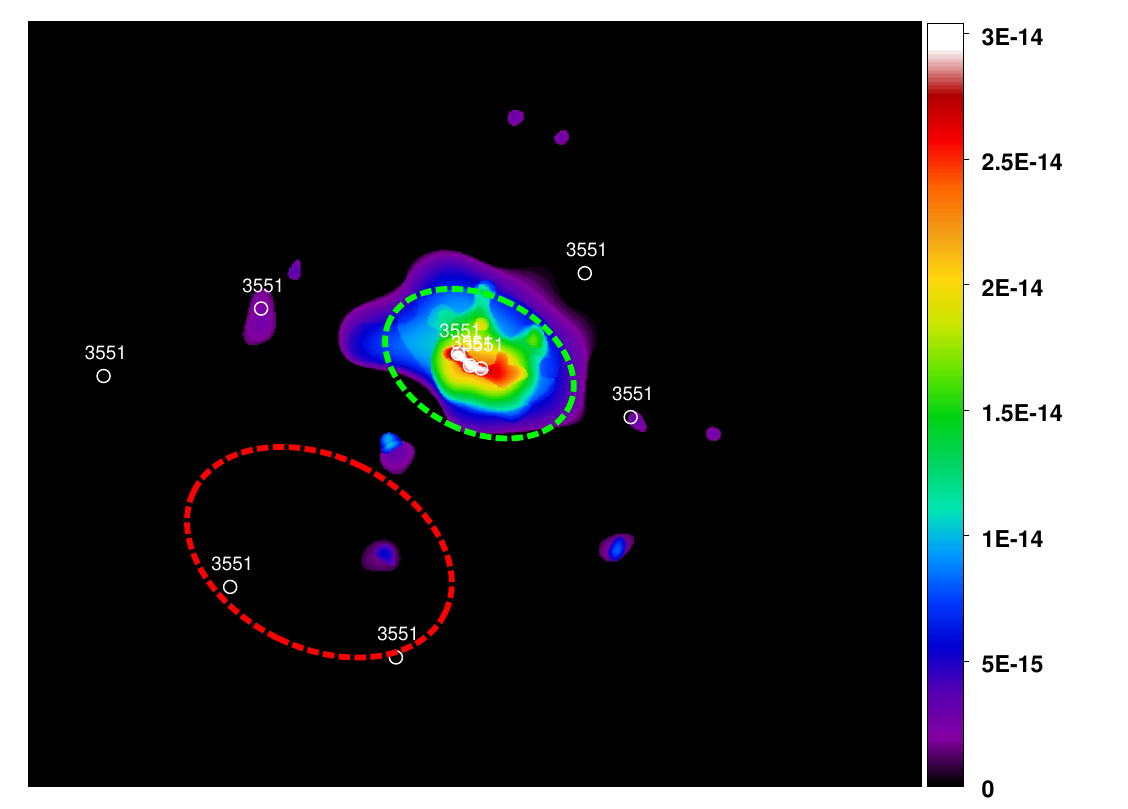}
    \end{subfigure}
    \begin{subfigure}[b]{0.33\textwidth}
        \includegraphics[width=\textwidth]{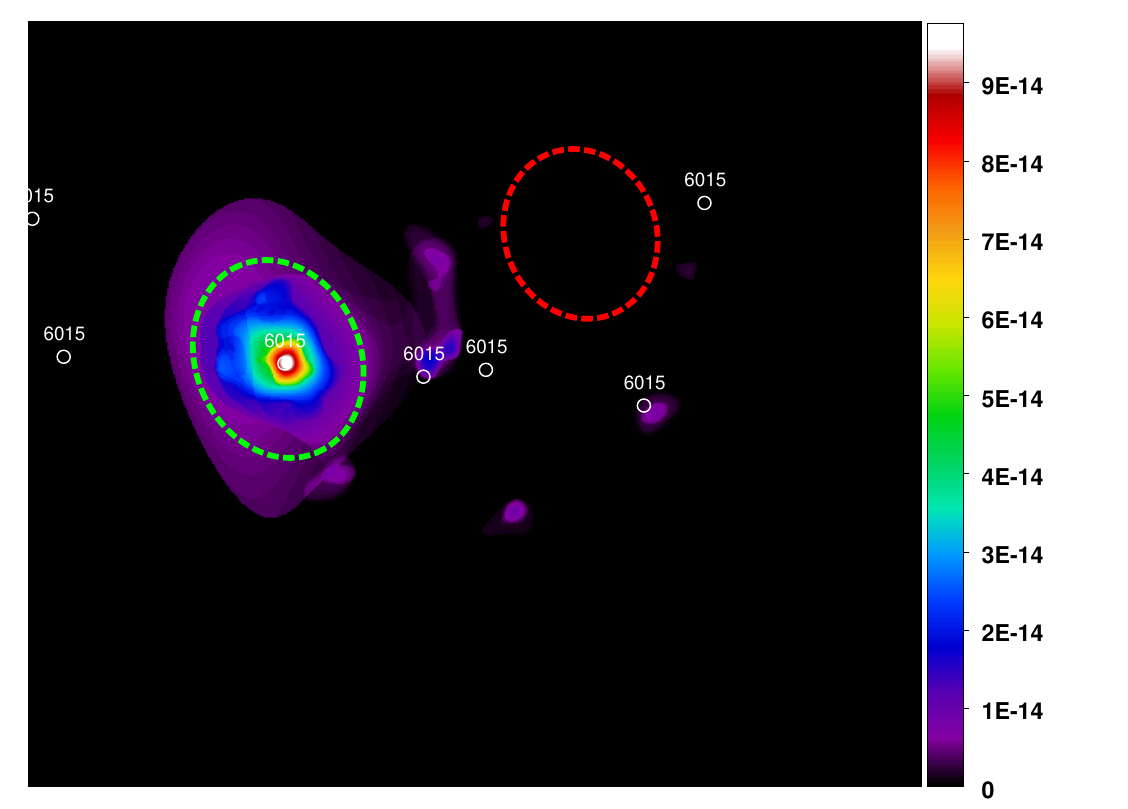}
    \end{subfigure}
        \caption{Twenty-five \textit{XMM-Newton} X-ray pointings for the full sample used in this work. Green dashed ellipses are spectral extraction regions. Red dashed ellipses are background extraction regions. Small white circles represent member galaxies in each group, each denoted with the respective group ID. The colour bar scale is given in counts per second per square centimetre per square arcminute.}
    \label{zoo1}
\end{figure*}

\begin{figure*}[h]
    \ContinuedFloat
    \centering
    \begin{subfigure}[b]{0.33\textwidth}
        \includegraphics[width=\textwidth]{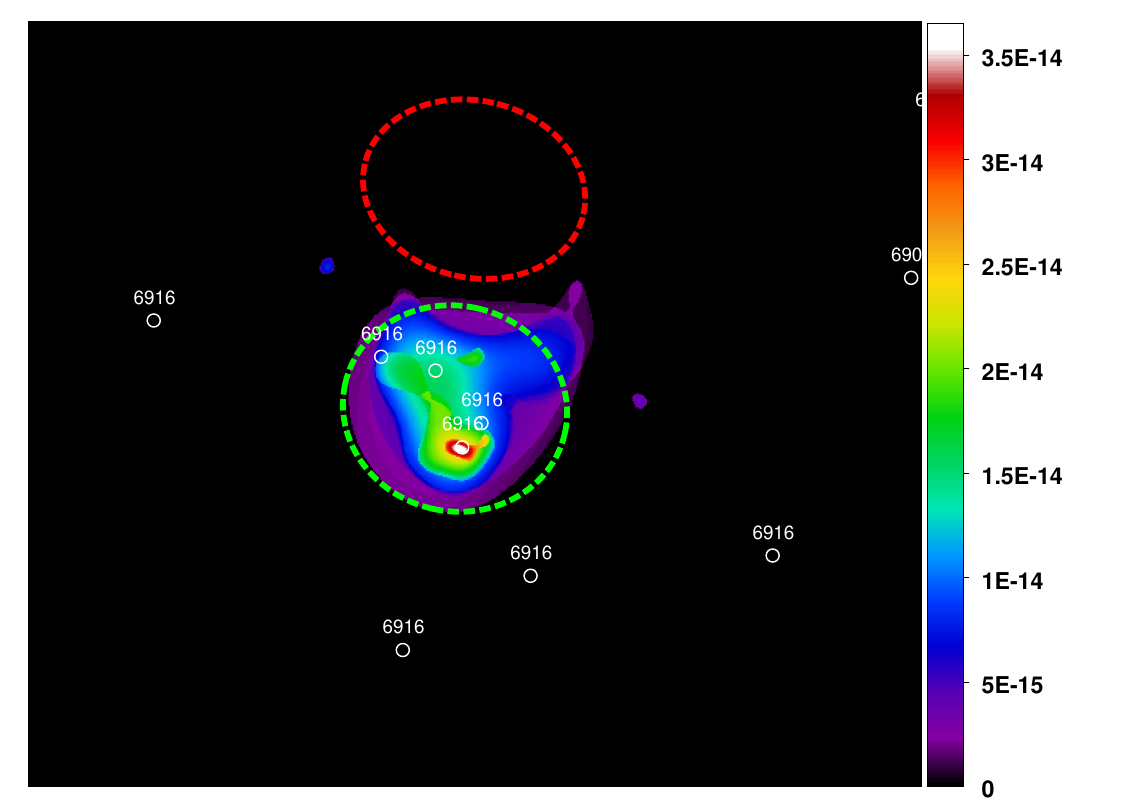}
    \end{subfigure}
    \begin{subfigure}[b]{0.33\textwidth}
        \includegraphics[width=\textwidth]{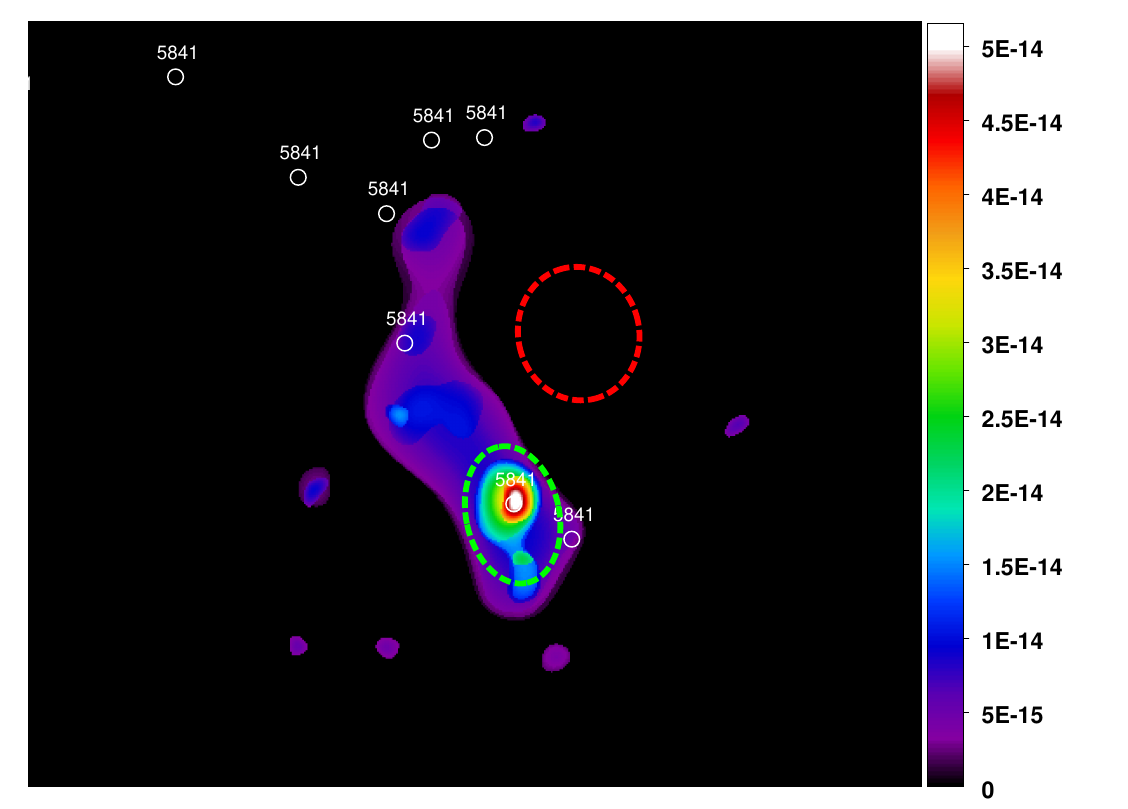}
    \end{subfigure}
    \begin{subfigure}[b]{0.33\textwidth}
        \includegraphics[width=\textwidth]{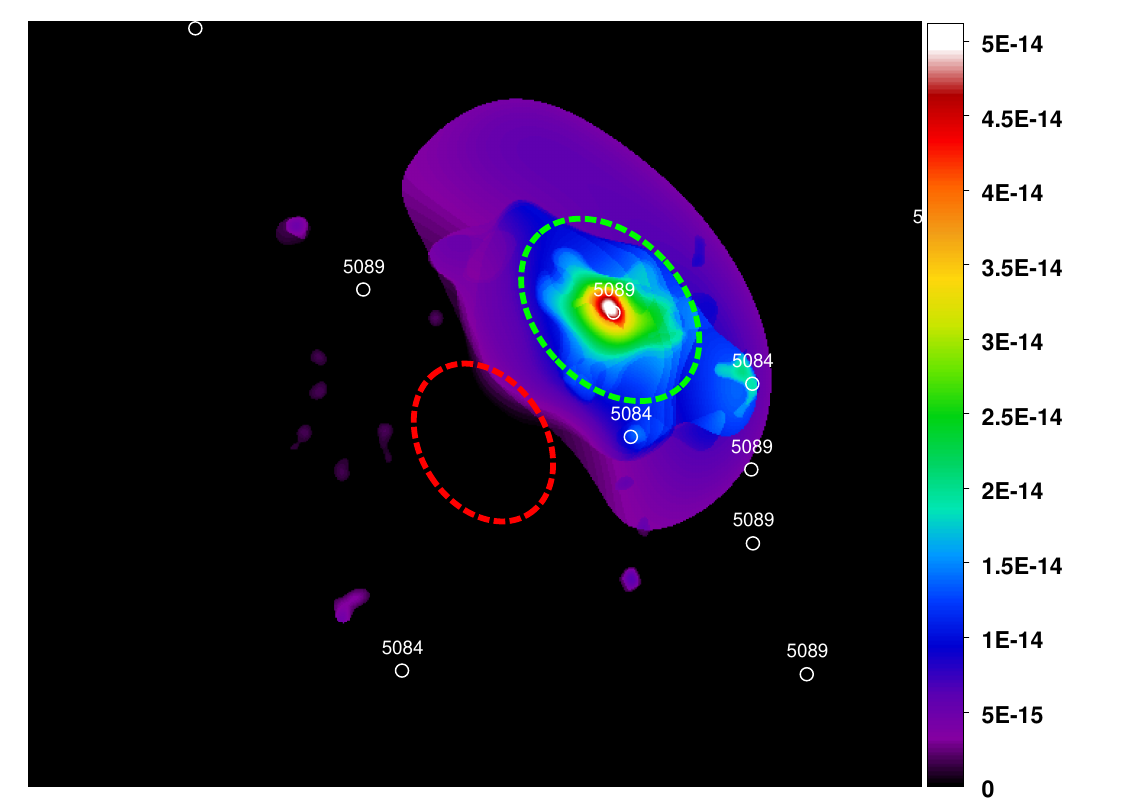}
        \end{subfigure}
        \begin{subfigure}[b]{0.33\textwidth}
        \includegraphics[width=\textwidth]{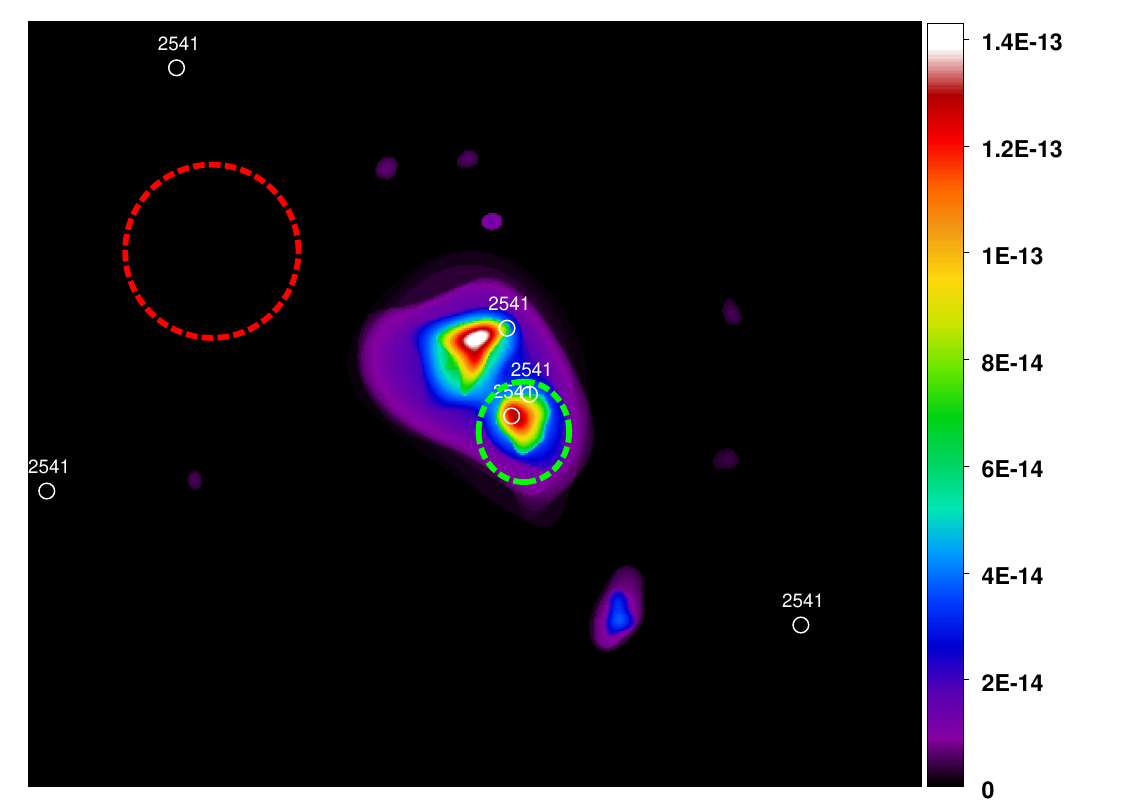}
    \end{subfigure}
    \begin{subfigure}[b]{0.33\textwidth}
        \includegraphics[width=\textwidth]{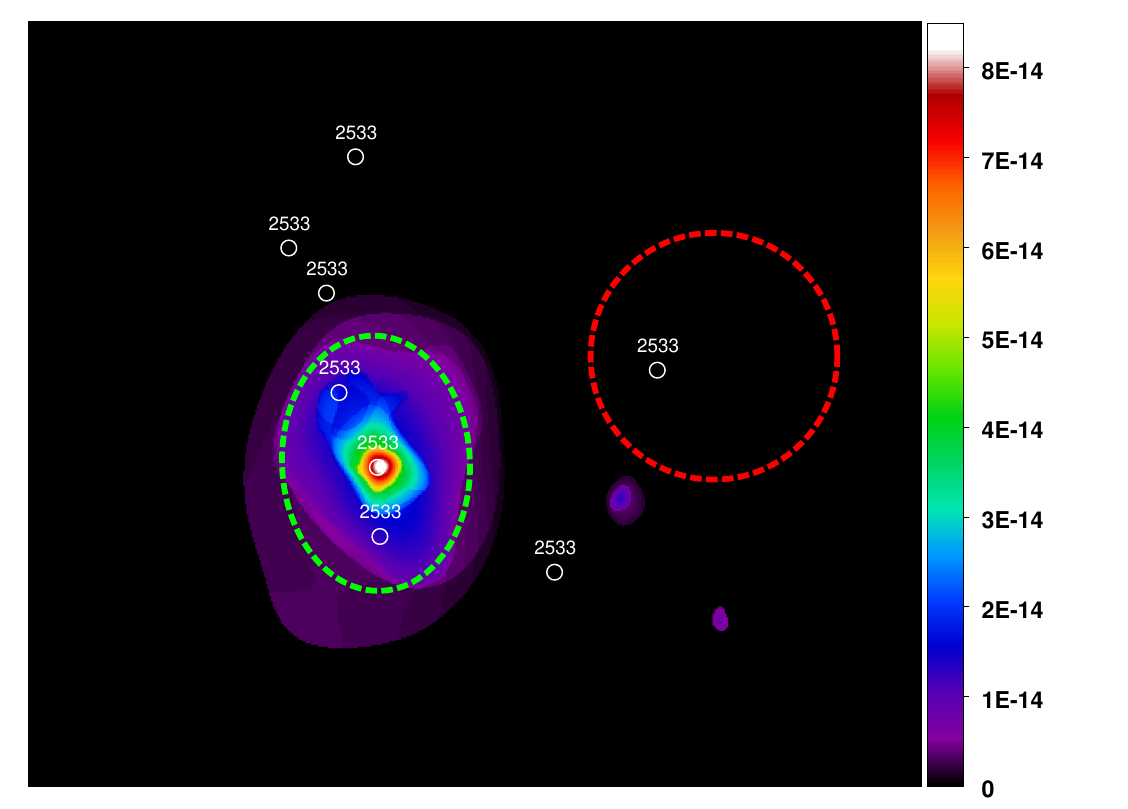}
    \end{subfigure}
    \begin{subfigure}[b]{0.33\textwidth}
        \includegraphics[width=\textwidth]{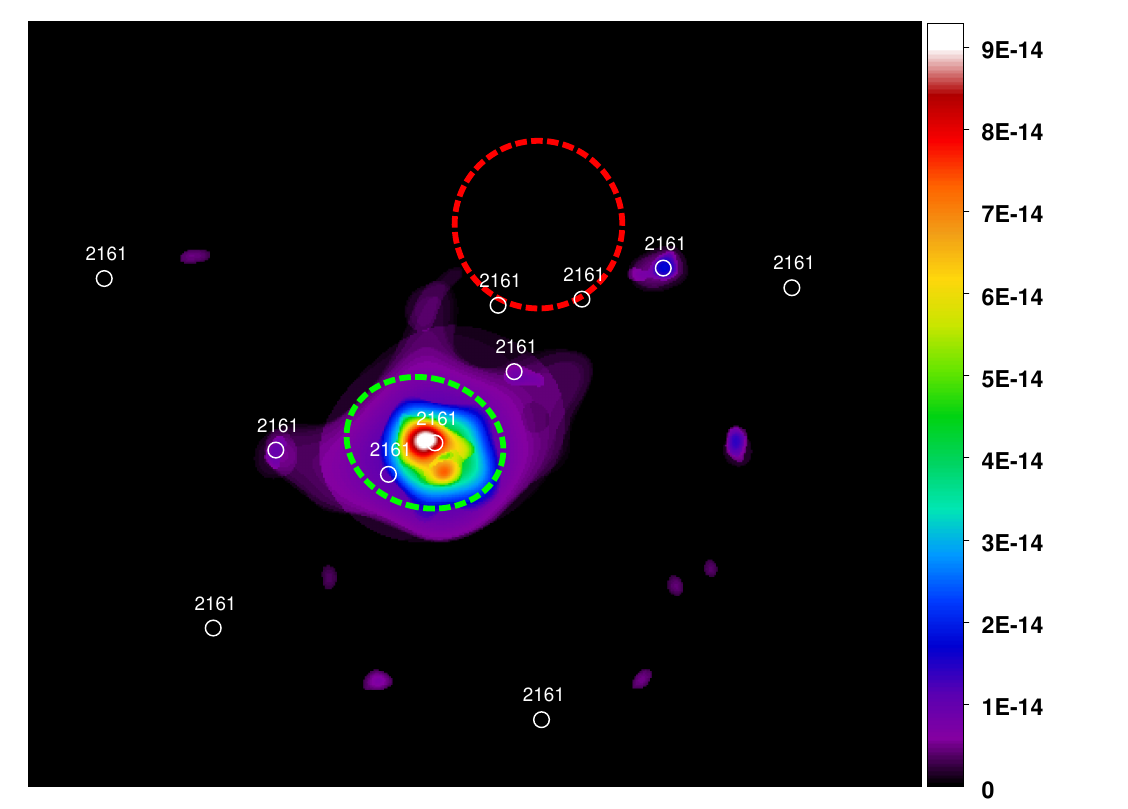}
        \end{subfigure}
        \begin{subfigure}[b]{0.33\textwidth}
        \includegraphics[width=\textwidth]{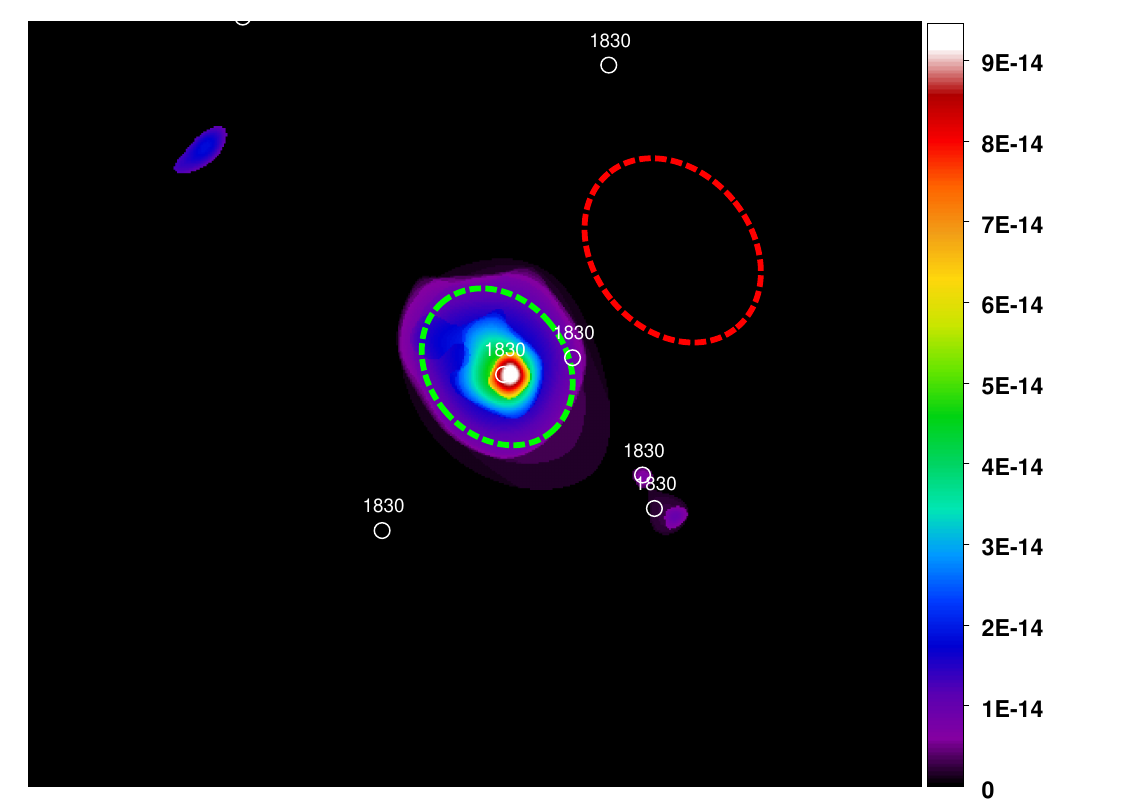}
    \end{subfigure}
    \begin{subfigure}[b]{0.33\textwidth}
        \includegraphics[width=\textwidth]{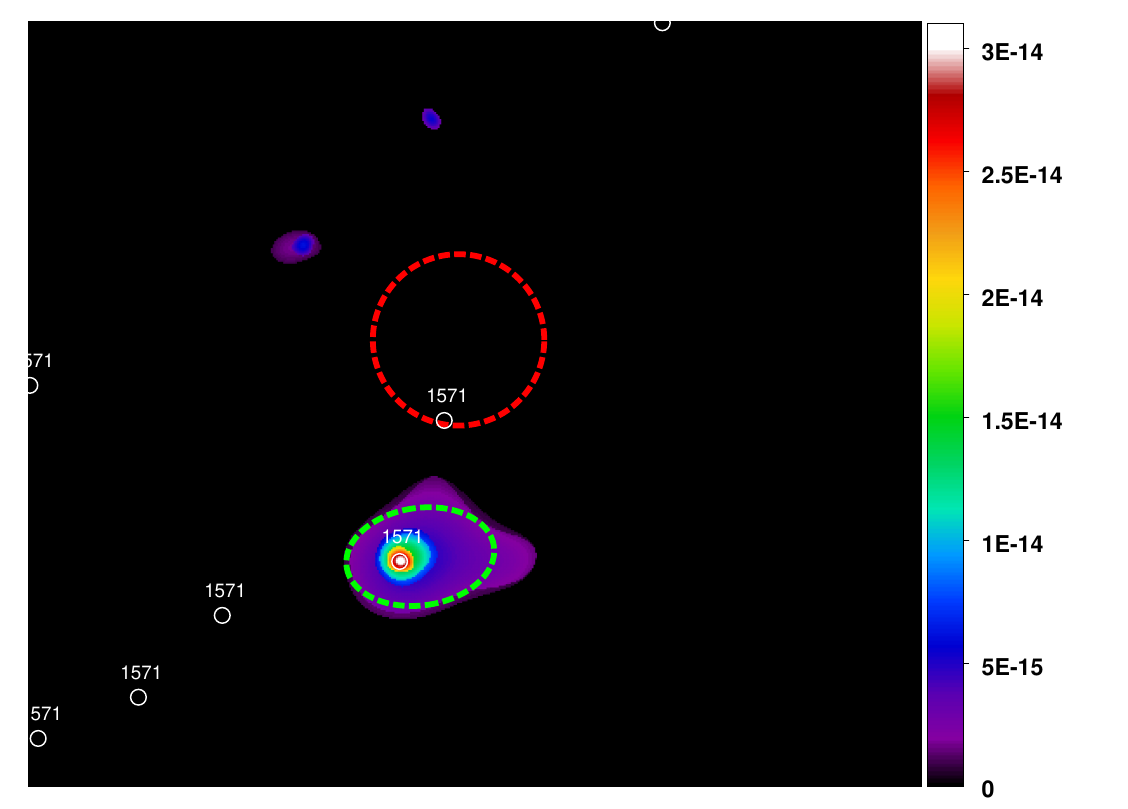}
    \end{subfigure}
    \begin{subfigure}[b]{0.33\textwidth}
        \includegraphics[width=\textwidth]{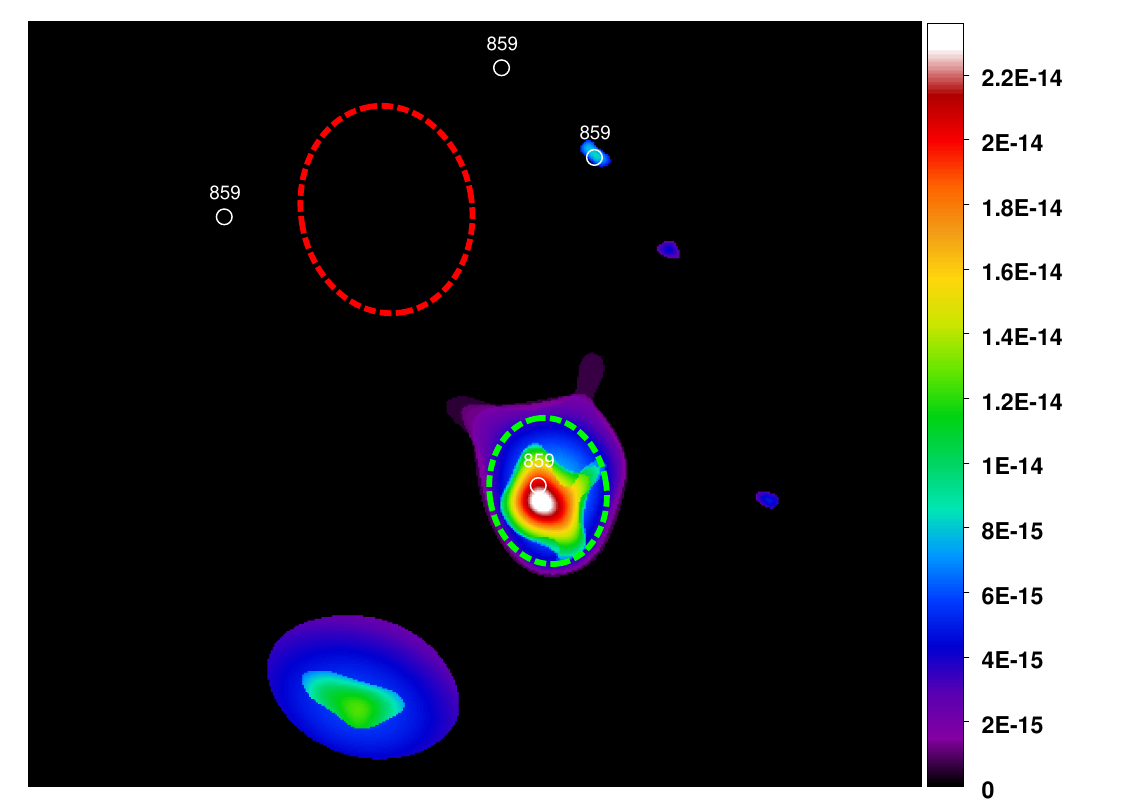}
        \end{subfigure}
    \begin{subfigure}[b]{0.33\textwidth}
        \includegraphics[width=\textwidth]{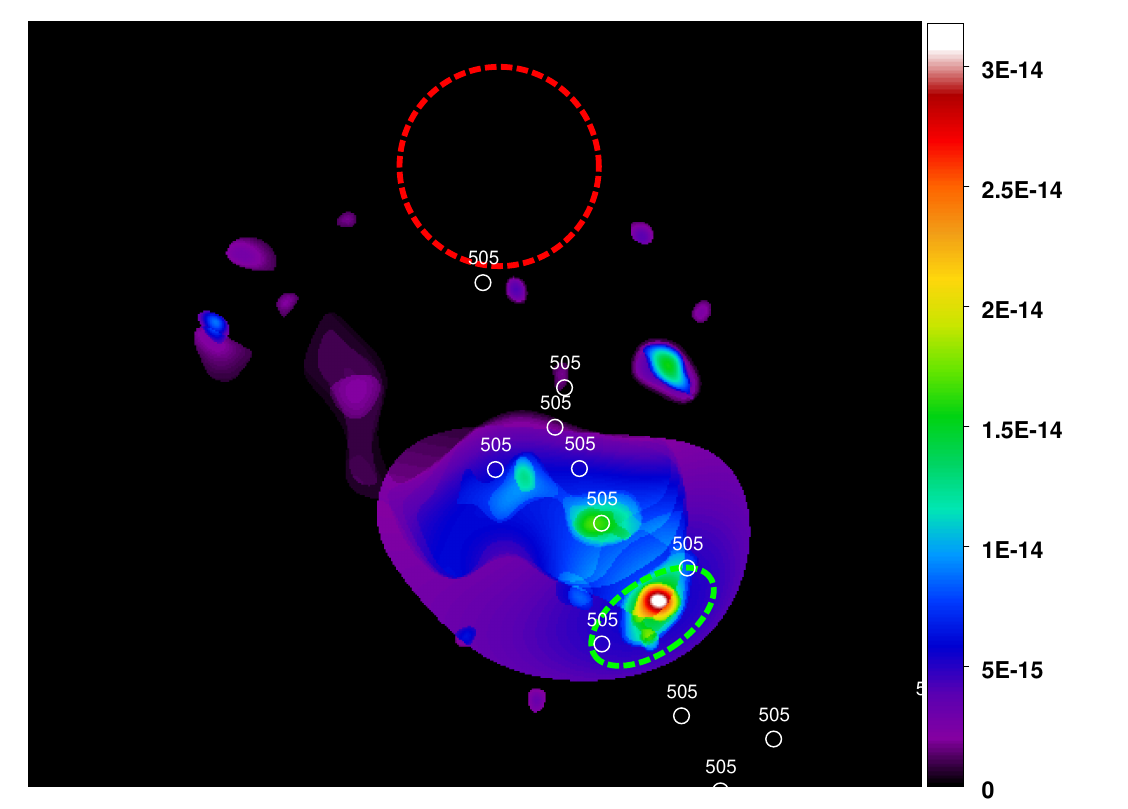}
    \end{subfigure}
    \caption{Continued.}
    \label{zoo2}
\end{figure*}

\section{AXES-2MRS group catalogue}

In Table\ref{tab:catalogue_columns} we describe the X-ray properties of the Full AXES-2MRS Group catalogue. Source flux and luminosities are based on RASS data. The optical properties of the groups are calculated only for groups with at least 5 clean members. The redshifts are reported in the CMB frame, using the catalogues of \cite{tempel18}.   {The catalogues described in Table C.1 are only available in electronic form at the CDS via anonymous ftp to cdsarc.u-strasbg.fr (130.79.128.5) or via http://cdsweb.u-strasbg.fr/cgi-bin/qcat?J/A+A/ .}

\begin{table*}

    \caption{Descriptions of the columns of the AXES-2MRS catalogue.}
    \label{tab:catalogue_columns}
    \centering
    \begin{tabular*}{\textwidth}
    {l@{\extracolsep{\fill}}llr}
        \hline\hline
        \multicolumn{1}{l}{Column} & \multicolumn{1}{l}{Unit} & \multicolumn{1}{c}{Description} & \multicolumn{1}{r}{Example} \\
        \hline
        \texttt{GROUP\_ID} (1) & &  2MRS group  identification number from \cite{tempel18}& 2150 \\
        \texttt{AXES\_ID} (2) & & Extended X-ray source ID in the AXES catalogue  & 93280903 \\
        \texttt{RA} (3) & deg & X-ray detection right ascension (J2000) & 95.56809 \\
        \texttt{DEC} (4) & deg & X-ray detection declination (J2000) & 
        --64.67731 \\
        \texttt{NMEM} (5) & & Number of spectroscopic members in 2MRS group catalogue  & 23 \\
        \texttt{NMEM\_CLEAN} (6) & & Number of spectroscopic members after the cleaning  & 23 \\
        \texttt{ZSPEC} (7) & & 2MRS group redshift & 0.0281 \\
        \texttt{ZSPEC\_CLEAN} (8) & & Group redshift, assigned using median value of clean members & 0.0281 \\
        \texttt{CLUVDISP\_GAP} (9) & km s$^{-1}$ & Gapper estimate of the cluster velocity dispersion & 582.223 \\
        \texttt{GAUSSIANITY} (10) &  & Gaussianity, based on the substructure analysis & G \\
        \texttt{LX0124} (11) & ergs s$^{-1}$ & Luminosity in the (0.1-2.4) keV band of the cluster, aperture $R_\text{500c}$ & $2.3\times10^{43}$ \\
        \texttt{ELX} (12) & ergs s$^{-1}$ & Uncertainty on \texttt{LX0124} & $7.59\times10^{41}$ \\
        \texttt{FLUX052} (13) & ergs s$^{-1}$ cm$^{-2}$ & Galaxy cluster X-ray flux in the 0.5-2.0 keV band & $7.25\times10^{-12}$ \\
        \texttt{EFLUX052} (14) & ergs s$^{-1}$ cm$^{-2}$ & Uncertainty on \texttt{FLUX052} & $2.39\times10^{-13}$ \\
        \texttt{R\_E} (15) & arcmin & Apparent radial extent of X-ray emission  & 16.8 \\
        \texttt{R\_500} (15) & arcmin & Estimated $R_{500}$ radius & 48.6 \\
        \hline
    \end{tabular*}
\end{table*}

\end{appendix}
\end{document}